\pdfoutput=1
\documentclass[11pt]{article}

\usepackage{epsfig,latexsym,amsfonts,amsmath,amsthm,amssymb,amsbsy,multirow,slashed,wasysym,textcomp,cite,comment,subfigure,wrapfig,datetime,mathtools,cancel,mathrsfs}
\usepackage{datetime2}
\usepackage[font={footnotesize,sl},bf]{caption}

\usepackage{setspace}
\usepackage{tikz}
\usetikzlibrary{calc}
\usetikzlibrary{decorations.pathreplacing}
\usetikzlibrary{decorations.pathmorphing}
\usetikzlibrary{decorations.markings}
\usetikzlibrary{arrows.meta}
\usetikzlibrary{positioning}

\usepackage{xcolor}
\definecolor{oldmauve}{rgb}{0.4, 0.19, 0.28}
\definecolor{pansypurple}{rgb}{0.47, 0.09, 0.29}
\definecolor{burgundy}{rgb}{0.5, 0.0, 0.13}
\definecolor{carminepink}{rgb}{0.92, 0.3, 0.26}
\definecolor{blue(pigment)}{rgb}{0.2, 0.2, 0.6}
\definecolor{darkseagreen}{rgb}{0.56, 0.74, 0.56}
\definecolor{darkspringgreen}{rgb}{0.09, 0.45, 0.27}
\definecolor{ceruleanblue}{rgb}{0.16, 0.32, 0.75}

\definecolor{navyblue}{RGB}{0,0,128}

\numberwithin{equation}{section}

\def\bea{\begin{eqnarray}}
\def\eea{\end{eqnarray}}
\newcommand{\beq}{\begin{eqnarray}}
\newcommand{\eqq}{\end{eqnarray}}
 \newcommand{\badat}{\begin{alignedat}}
 \newcommand{\eadat}{\end{alignedat}}

\newcommand{\eal}[1]{\be \begin{aligned} #1 \end{aligned}\end{equation}} 
\newcommand{\eqn}[1]{\be #1 \end{equation}} 
\newcommand{\eqa}[1]{\bea  #1\end{eqnarray}}

\newcommand{\bz}{\bar{z}}

\newcommand{\scri}{\mathscr{I}}
\newcommand{\sq}[1]{[#1]}
\newcommand{\an}[1]{\left\langle#1\right\rangle}
\newcommand{\mo}{\mathcal{O}}
\newcommand{\mc}{\mathcal{C}}
\newcommand{\mA}{\mathcal{A}}

\newcommand{\zb}{\bar{z}}
\newcommand{\Zb}{\bar{Z}}

\newcommand{\hypf}{{}_2F_{1}}

\newcommand{\D}{\Delta}
\newcommand{\e}{\epsilon}
\newcommand{\om}{\omega}

\long\def\new#1\endnew{{\bf #1}}		
\long\def\del#1\enddel{}

\def\del{\partial}

\newcommand{\be}{\begin{eqnarray}}
\newcommand{\en}{\end{eqnarray}}

\def\bz{{\bar z}}

\numberwithin{equation}{section} % equation numbers follow sections

\author{}
\numberwithin{equation}{section} % equation numbers follow sections
\usepackage{hyperref}
\usepackage{orcidlink}

\begin{document}

\begin{titlepage}

  \thispagestyle{empty}

 \begin{flushright}
%NORDITA 2022-084 \\
% UUITP-55/22 \\
 \end{flushright}

%\vskip2cm

  \begin{center}  
%{\LARGE\textbf{Carrollian Amplitudes from the Flat}}
%\vskip0.2cm
%{\LARGE\textbf{Limit of AdS Witten Diagrams}}

{\LARGE\textbf{Carrollian Amplitudes from Holographic Correlators}}

\vskip1cm
Luis F. Alday\footnote{\fontsize{8pt}{10pt}\selectfont\ \href{mailto:luis.alday@maths.ox.ac.uk}
{luis.alday@maths.ox.ac.uk}},
Maria Nocchi\footnote{\fontsize{8pt}{10pt}\selectfont\ \href{mailto:maria.nocchi@maths.ox.ac.uk}
{maria.nocchi@maths.ox.ac.uk}}, 
Romain Ruzziconi\footnote{\fontsize{8pt}{10pt}\selectfont\ \href{mailto:Romain.Ruzziconi@maths.ox.ac.uk}{romain.ruzziconi@maths.ox.ac.uk}}, 
Akshay Yelleshpur Srikant\footnote{\fontsize{8pt}{10pt}\selectfont \ \href{mailto:Akshay.YelleshpurSrikant@maths.ox.ac.uk}{yelleshpursr@maths.ox.ac.uk}}
\vskip0.5cm

\normalsize
\medskip

\textit{Mathematical Institute, University of Oxford, \\ Andrew Wiles Building, Radcliffe Observatory Quarter, \\
Woodstock Road, Oxford, OX2 6GG, UK}

\end{center}

\vskip0.5cm

\begin{abstract}
Carrollian amplitudes are flat space amplitudes written in position space at null infinity which can be re-interpreted as correlators in a putative dual Carrollian CFT. We argue that these amplitudes are the natural objects obtained in the flat space limit of AdS Lorentzian boundary correlators. The flat limit is taken entirely in position space by introducing Bondi coordinates in the bulk. From the bulk perspective, this procedure makes it manifest that the flat limit of any Witten diagram is the corresponding flat space Feynman diagram. It also makes explicit the fact that the flat limit in the bulk is implemented
by a Carrollian limit at the boundary. We systematically analyse tree-level two, three and four-point correlators. Familiar features such as the distributional nature of Carrollian amplitudes and the presence of a bulk point singularity arise naturally as a consequence of requiring a finite and non-trivial Carrollian
limit.
\end{abstract}

\end{titlepage}
\setcounter{page}{2}

\setcounter{tocdepth}{2}
\tableofcontents

\section{Introduction}

\begin{figure}[ht!]
\centering
\begin{tikzpicture}[scale=1]
	\def\h{3};\def\l{2};\def\dy{0.5};\def\ofx{1.7};\def\ofy{0.2};\def\ec{1.6};\def\dh{0.2};
	\coordinate (bl) at (-\l,-\h);
	\coordinate (br) at ( \l,-\h);
	\coordinate (cl) at (-\l,  0);
	\coordinate (cr) at ( \l,  0);
	\coordinate (cc) at (  0,  0);
	\coordinate (tl) at (-\l, \h);
	\coordinate (tr) at ( \l, \h);
	\coordinate (bbl) at (-\l,-2*\h);
	\coordinate (bbr) at ( \l,-2*\h);
	\coordinate (bil) at ($(cl)-(0,\ec)$);
	\coordinate (bir) at ($(cr)-(0,\ec)$);
	\coordinate (biil) at ($(bil)+(0,\dh)$);
	\coordinate (biir) at ($(bir)+(0,\dh)$);
	\coordinate (biiil) at ($(cl)+(0,\ec)$);
	\coordinate (biiir) at ($(cr)+(0,\ec)$);
	\coordinate (bivl) at ($(biiil)-(0,\dh)$);
	\coordinate (bivr) at ($(biiir)-(0,\dh)$);
	\coordinate (cb) at ($(bil)!0.5!(biir)$);
	\coordinate (ch) at ($(biiil)!0.5!(bivr)$);
	\draw[red,opacity=0] ($(cc)+(-4,-4)$) -- ($(cc)+(-4,4)$) -- ($(cc)+(4,4)$) -- ($(cc)+(4,-4)$) -- cycle;
	\draw[] (bl) -- (cl) -- (tl);
	\draw[] (br) -- (tr)node[anchor=north west]{$\mathscr I_{\text{AdS}}$};
	\draw[] (bl) arc[x radius=\l, y radius=\dy, start angle=-180, end angle=0];
	\draw[densely dashed] (bl) arc[x radius=\l, y radius=\dy, start angle=180, end angle=0];
	\draw[] (tl) arc[x radius=\l, y radius=\dy, start angle=-180, end angle=180];
	\draw[orange] (bil) arc[x radius=\l, y radius=\dy, start angle=-180, end angle=0];
	\draw[orange,densely dashed] (bil) arc[x radius=\l, y radius=\dy, start angle=180, end angle=0];
	\draw[orange] (biil) arc[x radius=\l, y radius=\dy, start angle=-180, end angle=0];
	\draw[orange,densely dashed] (biil) arc[x radius=\l, y radius=\dy, start angle=180, end angle=0];
	\draw[navyblue] (bivl) arc[x radius=\l, y radius=\dy, start angle=-180, end angle=0];
	\draw[navyblue,densely dashed] (bivl) arc[x radius=\l, y radius=\dy, start angle=180, end angle=0];
	\fill[navyblue,opacity=0.2] (biiil) to[in=100,out=80,looseness=0.43] (biiir) -- (bivr) to[in=80,out=100,looseness=0.43] (bivl) -- cycle;
	\fill[orange,opacity=0.2] (bil) to[in=100,out=80,looseness=0.43] (bir) -- (biir) to[in=80,out=100,looseness=0.43] (biil) -- cycle;
	\fill[orange,opacity=0.5] (bil) to[in=-100,out=-80,looseness=0.43] (bir) -- (biir) to[in=-80,out=-100,looseness=0.43] (biil) -- cycle;
	\def\sq{0.45};
	\coordinate (T) at ($(cc)+(0,\sq)$);
	\coordinate (L) at ($(cc)-(\sq,0)$);
	\coordinate (R) at ($(cc)+(\sq,0)$);
	\coordinate (B) at ($(cc)-(0,\sq)$);
	\draw (L) -- (B) -- (R) -- (T) -- cycle;
	\draw[] (L) to[out=-20,in=-160] (R);
	\draw[densely dashed] (L) to[out=20,in=160] (R);
	\fill[orange,opacity=0.3] (L) to[out=20,in=160] (R) -- (B) -- (L);
	\fill[navyblue,opacity=0.3] (L) to[out=-20,in=-160] (R) -- (T) -- (L);
	\node[circle,inner sep=1pt,fill] at (cc) {};
	\node[left] at (tl) {$\tau=+\pi$};
	\node[left] at (bl) {$\tau=-\pi$};
	\node[left] at (cl) {$\tau=0$};
	\node[left,navyblue] at ($(biiil)!0.5!(bivl)$) {$\tau=+\frac{\pi}{2}$};
	\node[left,orange] at ($(bil)!0.5!(biil)$) {$\tau=-\frac{\pi}{2}$};
	\node[circle,inner sep=1pt,fill,outer sep=3pt] (one) at ($(cb)+({\l*cos(-135)},{\dy*sin(-135)})$) {};
	\node[circle,inner sep=1pt,fill,outer sep=3pt] (two) at ($(cb)+({\l*cos(-45)},{\dy*sin(-45)})$) {};
	\node[circle,inner sep=1pt,fill,outer sep=3pt] (thr) at ($(ch)+({\l*cos(60)},{\dy*sin(60)})$) {};
	\node[circle,inner sep=1pt,fill,outer sep=3pt] (fou) at ($(ch)+({\l*cos(120)},{\dy*sin(120)})$) {};
	\node[circle,inner sep=1pt,fill,outer sep=3pt] (int) at (cc) {};
	\draw[red!75!black,decoration={markings,mark=at position 0.3 with {\arrow{>}}},postaction={decorate}] (one) -- (int);
	\draw[red!75!black,decoration={markings,mark=at position 0.3 with {\arrow{>}}},postaction={decorate}] (two) -- (int);
	\draw[red!75!black,decoration={markings,mark=at position 0.3 with {\arrow{<}}},postaction={decorate}] (thr) -- (int);
	\draw[red!75!black,decoration={markings,mark=at position 0.3 with {\arrow{<}}},postaction={decorate}] (fou) -- (int);
	\draw[] (cl) arc[x radius=\l, y radius=\dy, start angle=-180, end angle=0];
	\draw[densely dashed] (cl) arc[x radius=\l, y radius=\dy, start angle=180, end angle=0];
	\draw[navyblue] (biiil) arc[x radius=\l, y radius=\dy, start angle=-180, end angle=0];
	\draw[navyblue,densely dashed] (biiil) arc[x radius=\l, y radius=\dy, start angle=180, end angle=0];
	\fill[navyblue,opacity=0.5] (biiil) to[in=-100,out=-80,looseness=0.43] (biiir) -- (bivr) to[in=-80,out=-100,looseness=0.43] (bivl) -- cycle;
\end{tikzpicture}
\caption{Usual set-up to take the flat limit of holographic correlators in a global patch of AdS. Here $\tau$ represents the global time coordinate (see Equation \eqref{global coordinate}). The boundary operators are inserted in little strips (blue strip for outgoing operators and orange strip for incoming operators), such that the scattering occurs in a small region in the center of AdS, which can be approximated by flat space. Even if this procedure has been shown to be very efficient in practice, the precise operation one has to implement in the dual theory is not completely clear. One of the objectives of this paper is to present a flat space limit that does not require this restriction in the operator insertions a priori, and implement clearly the Carrollian limit in the dual theory, see Figure \ref{fig:flatlimit}.}
\label{fig:centerAdS}
\end{figure}
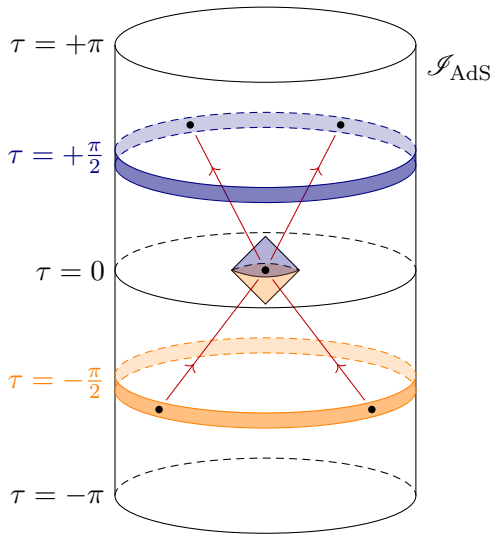

The holographic principle has been remarkably successful in spacetimes with a negative cosmological constant \cite{Maldacena:1997re,Witten:1998qj,Aharony:1999ti}. Building on this progress, early efforts have been made to extend this understanding to obtain a holographic description of gravity in asymptotically flat spacetimes \cite{Susskind:1998vk,Polchinski:1999ry,Giddings:1999jq}. The flat limit is typically implemented by making specific kinematic choices and zooming into the center of AdS \footnote{References \cite{Gadde:2022ghy, Marotta:2024sce} implement the flat limit differently.} ($\ell\gg r$, where $\ell$ is the AdS radius and $r$ is the radial distance measured from the centre, see Figure \ref{fig:centerAdS}). This approach has significantly advanced our comprehension of how flat space physics can be derived from AdS \cite{Gary:2009ae,Penedones:2010ue,Fitzpatrick:2011hu,Alday:2017vkk,Hijano:2019qmi,Gadde:2022ghy, Marotta:2024sce}.  Nonetheless, it remains challenging to track the implications of this bulk flat limit within the dual theory at the boundary.

In parallel, important progress has been made in the understanding of flat space holography, and what properties should be expected for a putative dual theory in that context. Two main proposals have been pushed forward. On the one hand, celestial holography suggests that the dual theory to gravity in four-dimensional asymptotically flat spacetime is a two-dimensional CFT living on the celestial sphere \cite{deBoer:2003vf,He:2015zea,Pasterski:2016qvg,Cheung:2016iub,Pasterski:2017kqt,Strominger:2017zoo,Pasterski:2017ylz}. On the other hand, Carrollian holography proposes that the putative dual theory is instead a three-dimensional Carrollian CFT living at null infinity \cite{Arcioni:2003xx,Dappiaggi:2005ci,Barnich:2006av,Barnich:2010eb,Bagchi:2010zz,Barnich:2012xq,Barnich:2012rz,Bagchi:2012xr,Bagchi:2014iea,Bagchi:2015wna,Bagchi:2016bcd,Ciambelli:2018wre,Donnay:2022aba,Bagchi:2022emh,Donnay:2022wvx,Saha:2023hsl}. While these two approaches look a priori completely different, they have been shown to be equivalent \cite{Donnay:2022aba,Bagchi:2022emh,Donnay:2022wvx,Mason:2023mti}, offering two complementary roads towards flat space holography. It is possible to engineer the flat limit procedure such as to land either on the celestial CFT \cite{deGioia:2022fcn,deGioia:2023cbd,deGioia:2024yne} or on the Carrollian CFT \cite{Bagchi:2023fbj,Bagchi:2023cen}, which has led to interesting progress. However, the precise understanding of the effect of this limit at the level of the CFT dual to AdS remains to be clarified. For instance, what does it mean to select a little strip in the boundary CFT from an intrinsic perspective? (see Figure \ref{fig:centerAdS})

In a different set-up, earlier works \cite{Barnich:2012aw,Poole:2018koa,Compere:2019bua,Compere:2020lrt} have shown that asymptotically AdS Einstein metrics can be reduced to asymptotically flat metrics by taking a careful flat limit in Bondi gauge ($r \gg \ell \to \infty$), which is conceptually different from the limit usually considered in the scattering amplitude context (compare Figures \ref{fig:centerAdS} and \ref{fig:flatlimit}). Indeed, while Bondi gauge was originally introduced in asymptotically flat spacetimes \cite{Bondi:1962px,Sachs:1962zza} to study outgoing radiation using a null foliation, it also exists in AdS and has the desirable property to admit a smooth flat limit \cite{Barnich:2012aw,Poole:2018koa,Compere:2019bua,Compere:2020lrt,Ciambelli:2018wre,Campoleoni:2018ltl,Ciambelli:2020eba,Ciambelli:2020ftk,Ruzziconi:2020wrb,Geiller:2021vpg,Geiller:2022vto,Campoleoni:2023fug,Ciambelli:2024kre}. Remarkably, this flat limit in the bulk can be directly re-interpreted as a Carrollian limit \cite{Levy1965} at the boundary, which formally consists in taking the speed of light to zero, i.e. $c\to 0$. In particular, the conformal geometry at the AdS boundary reduces to the conformal Carrollian geometry at null infinity \cite{Henneaux:1979vn,1977asst.conf....1G,Duval:2014uva,Ashtekar:2014zsa}, offering a potential understanding of what precisely happens at the level of the dual theory in the limit. This interpretation works at the level of the asymptotic symmetries, the dynamics and the phase space of Einstein gravity, and constitutes probably the most appealing advantage for the Carrollian approach to flat space holography. 

In this work, we exploit this powerful property of AdS Bondi coordinates to take the flat limit of Lorentzian boundary correlators, computed via Witten diagrams \cite{Witten:1998qj,Freedman:1998tz,Liu:1998bu,DHoker:1998ecp,DHoker:1998bqu}, upon analytic continuation. This procedure has the following advantages with respect to the other approaches: $(i)$ the background AdS metric reduces smoothly to the Minkowski metric in the flat limit ($\ell \to \infty)$ without further rescalings, by contrast with other coordinate systems such as Poincaré coordinates. As a consequence, all the building blocks for computing holographic correlators, such as bulk-to-bulk or bulk-to-boundary propagators, reduce to those of flat space in the limit. $(ii)$ The Bondi coordinates relate the AdS boundary to null infinity in the limit, which is particularly suitable to describe a massless scattering in flat space. This contrasts with the global coordinates which are more tied to spatial infinity. $(iii)$ As mentioned above, the flat limit in the bulk induces the Carrollian limit in the boundary theory by simply identifying the inverse bulk AdS radius $\ell^{-1}$ with the speed of light $c$ at the boundary. $(iv)$ The limit is taken purely in position space, which yields a very clean interpretation. 

Using this set-up, we show that the flat limit of AdS Witten diagrams leads naturally to Carrollian amplitudes \cite{Donnay:2022aba,Bagchi:2022emh,Donnay:2022wvx,Nguyen:2023miw,Mason:2023mti,Liu:2024nfc,Stieberger:2024shv}. The latter are massless scattering amplitudes written in position space at $\mathscr I$, which have a direct interpretation as correlators in a boundary Carrollian CFT. We apply this procedure to two, three and four-point functions for scalar fields, including contact and exchange diagrams. We also study the Carrollian limit directly at the level of the CFT correlators at the boundary of AdS, and show that they reproduce the desired Carrollian correlators at null infinity in the limit. In particular, we explain how the familiar features like the distributional nature of Carrollian correlators and an intimate connection to the bulk point singularity emerge naturally from the Carrollian limit. More generally, our procedure provides a systematic way to reduce CFT correlators to Carrollian CFT correlators, which constitutes an important path toward the connection between AdS/CFT and flat space holography.

This paper is organised as follows. In Section \ref{sec:Carrollian amplitudes in flat spacetime}, we review the notion of Carrollian amplitudes and discuss their relation with the usual Feynman rules. In Section \ref{sec:Flat limit of AdS Witten diagrams in Bondi coordinates}, we show that the building blocks of the Witten diagrams admit a well-defined flat limit in Bondi coordinates, so that the holographic correlators reduce to the Carrollian amplitudes in the limit. In Section \ref{sec:Carrollian limit of boundary CFT correlators}, we take the Carrollian limit directly on the expression of the boundary CFT correlators and show that they reduce correctly to the Carrollian correlators encoding the bulk $\mathcal{S}$-matrix. The core of the paper is also supplemented with some appendices: Appendix \ref{sec:Propagators in flat space} contains further conventions and details on the flat space propagators and Appendix \ref{sec:dbar1112} contains some details on the computation of $D$-functions.

\section{Carrollian amplitudes in flat spacetime}
\label{sec:Carrollian amplitudes in flat spacetime}

\subsection{Definition and properties}
\label{sec:Definition and properties}

\begin{figure}[h!]
\centering
\includegraphics[scale=1.2]{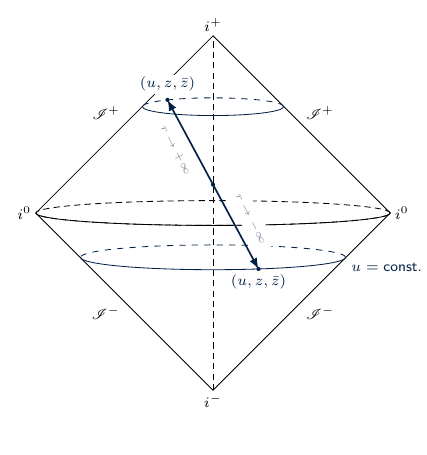}
\caption{There is a natural geometric identification between $\mathscr I^-$ and $\mathscr I^+$. Starting from a point at $\mathscr I^-$ and following a null geodesic, we end up on a point at $\mathscr I^+$ with the same $(u,z,\bar z)$ coordinates.}
    \label{FigureBondi}
\end{figure}

In this section, we briefly review the definition and properties of Carrollian amplitudes following the notations and conventions of \cite{Donnay:2022wvx,Mason:2023mti}. In Minkowski space, the (planar) Bondi coordinates $( u,r,z,\bar z)$, with $u,r \in \mathbb{R}$, $z \in \mathbb{C}$, are related to the Cartesian coordinates $X^\mu = (X^0,X^1,X^2,X^3)$ via
\begin{equation}
    X^\mu = u\,\partial_z\partial_{\bar z}q^\mu(z,\bar z) + r\, q^\mu(z,\bar z) \ , \label{Retarded flat BMS coordinates}
\end{equation}
where 
\begin{equation}
    q^\mu(z,\bar z) \equiv \frac{1}{\sqrt{2}} \Big(1+z\bar z,z+\bar z,-i(z-\bar z),1-z\bar z\Big) \label{q in terms of z bar z future}
\end{equation} is a null vector. The Minkowski line element reads\footnote{The relation of the line element \eqref{Minkowski metric} in planar Bondi coordinates with the more standard form of the Bondi line element $ds^2 = -du^2 - 2 du dr + 4 r^2 (1+z \bar z)^{-2} dz d\bar z$ is discussed in Appendix A of \cite{Donnay:2022wvx}.}
\begin{equation}
    d s^2 = -2 d u d  r+2r^2 dz d\bar z \ .
    \label{Minkowski metric}
\end{equation} Here, future null infinity $\mathscr I^+$ is reached in the limit $r\to + \infty$ while past null infinity $\mathscr I^-$ is obtained in the limit $r\to - \infty$ (see Figure \ref{FigureBondi}). In particular, there is a natural antipodal identification between points at $\mathscr I^-$ and points at $\mathscr I^+$ which have the same boundary coordinates $x^a = (u,z,\bar z)$. We will use the same notation $\mathscr I$ for $\mathscr I^+$ and $\mathscr I^-$. 

The geometric structure induced at the boundary via conformal compactification \cite{Penrose:1962ij,Penrose:1965am} is a (conformal) Carrollian structure \cite{Henneaux:1979vn,1977asst.conf....1G,Duval:2014uva,Ashtekar:2014zsa}, which consists of a degenerate metric $q_{ab} = 0 du^2 + 2 dz d\bar z$ and a vector field $n^a \partial_a = \partial_u$ in the kernel of the metric, i.e. $q_{ab} n^b = 0$. The boundary action of the bulk Poincaré symmetries is generated by vector fields 
\begin{equation}
 \xi = (\mathcal{T} + {u} \alpha ) \partial_u  + \mathcal{Y}\partial_z + \bar{\mathcal{Y}}\partial_{\bar{z}} \ , \qquad \alpha = \frac{1}{2}(\partial_z \mathcal{Y} + \partial_{\bar{z}} \bar{\mathcal{Y}} ) \ ,
    \label{BMS vectors}
\end{equation} where $\mathcal{T}(z, \bar z) = 1,z,\bar z, z \bar z$ are the four translations, and $\mathcal{Y}(z) = 1, z, z^2$ and $\bar{\mathcal{Y}}(\bar{z}) = 1, \bar z,  {\bar{z}}^2$ are the six Lorentz transformations satisfying the conformal Killing equation in two dimensions, i.e. $\partial_z \bar{\mathcal{Y}}=\partial_{\bar z} \mathcal{Y}=0$. These correspond to the (global part of the) conformal symmetries of the Carrollian structure at $\mathscr I$:
\begin{equation}
    \mathcal{L}_{\xi} q_{ab} = 2 \alpha q_{ab} \ , \qquad \mathcal{L}_{\xi} n^a = - \alpha n^a \ .
\end{equation} A Carrollian primary operator $\Phi_{(k,\bar{k})}(u,z,\bar z)$ of Carrollian weights $(k,\bar{k})$ is defined as an operator transforming as 
\begin{equation}    \delta_{(\mathcal{T},\mathcal{Y}, \bar{\mathcal{Y}})} \Phi_{(k,\bar{k})} = \Big[\Big(\mathcal{T} + \frac{u}{2}(\partial_z \mathcal{Y} + \partial_{\bar{z}}\bar{\mathcal{Y}}) \Big)  \partial_u + \mathcal{Y} \partial_z + \bar{\mathcal{Y}} \partial_{\bar z} + k \partial_z \mathcal{Y} + \bar{k}  \partial_{\bar z} \bar{\mathcal{Y}} \Big] \Phi_{(k,\bar{k})} 
\label{carrollian primary}
\end{equation} under the action of conformal Carrollian symmetries. Notice that, if $\Phi_{(k,\bar k)}$ is a conformal Carrollian primary, then the descendant $\partial_u^m \Phi_{(k,\bar k)}$ ($m \in \mathbb{N}$) is also a conformal Carrollian primary of weights $(k+ \tfrac{m}{2}, \bar{k} + \tfrac{m}{2})$. We shall denote these operators as $\partial_u$-descendants.

The momentum of a massless particle can be parametrised as
\begin{equation}
    \label{eq:mompar31}
    p^\mu = \epsilon \om q^{\mu} (z, \bar z) \ ,
\end{equation} 
where $q^\mu(z, \bar z)$ is given in \eqref{q in terms of z bar z future}, $\omega >0$ is the energy, and $\e = + 1$ or $\epsilon = -1$ if the particle is outgoing or incoming, respectively. An $n$-point massless scattering amplitude in momentum space is denoted by $\mA_n \left(\left\lbrace \om_1, z_1, \zb_1\right\rbrace_{J_1}^{\epsilon_1}, \dots ,\left\lbrace \om_n, z_n, \zb_n\right\rbrace_{J_n}^{\epsilon_n}\right)$, where $J_i$ are the particle helicities. The associated position space amplitude at $\mathscr I$, obtained via Fourier transform,\footnote{The sign convention of the Fourier transforms in \eqref{eq:AtoC} is opposite to the one chosen in \cite{Mason:2023mti} but agrees with \cite{Donnay:2022wvx}.}
\begin{multline}
\label{eq:AtoC}
    \mc_n\left(\left\lbrace u_1, z_1, \zb_1\right\rbrace_{J_1}^{\epsilon_1}, \dots , \left\lbrace u_n, z_n, \zb_n\right\rbrace_{J_n}^{\epsilon_n}\right) \\
    =   \int_0^{+\infty} \prod_{i=1}^n \frac{d\om_i}{2\pi} \, e^{-i\e_i \om_i u_i}  \mA_n \left(\left\lbrace \om_1, z_1, \zb_1\right\rbrace_{J_1}^{\epsilon_1}, \dots , \left\lbrace \om_n, z_n, \zb_n\right\rbrace_{J_n}^{\epsilon_n}\right) \ , 
\end{multline} is referred to as Carrollian amplitude \cite{Donnay:2022wvx,Mason:2023mti}. It can be re-interpreted as a correlator of Carrollian primary operators at $\mathscr I$,
\begin{equation}
\label{carrollian identification}
    \mc_n\left(\left\lbrace u_1, z_1, \zb_1\right\rbrace_{J_1}^{\epsilon_1}, \dots , \left\lbrace u_n, z_n, \zb_n\right\rbrace_{J_n}^{\epsilon_n}\right) \equiv \langle \Phi^{\epsilon_1}_{(k_1,\bar{k}_1)} (u_1, z_1, \bz_1) \ldots \Phi^{\epsilon_n}_{(k_n,\bar{k}_n)} (u_n, z_n, \bz_n)  \rangle \ , 
\end{equation} where the Carrollian weights are fixed in terms of the helicity
\begin{equation}
    k_i = \frac{1+\epsilon_i J_i}{2} \ , \qquad \bar k_i = \frac{1- \epsilon_i J_i}{2} \ .
    \label{fiexed Carrollian weights}
\end{equation} 
In the following, it will also be convenient to compute correlators of $\partial_u$-descendants, 
\begin{multline}
\label{descendants amplitudes}
 \mathcal{C}_n^{m_1\ldots m_n} \left(\left\lbrace u_1, z_1, \zb_1\right\rbrace_{J_1}^{\epsilon_1}, \dots , \left\lbrace u_n, z_n, \zb_n\right\rbrace_{J_n}^{\epsilon_n}\right) 
   = \partial_{u_1}^{m_1} \ldots \partial_{u_n}^{m_n} \mc_n\left(\left\lbrace u_1, z_1, \zb_1\right\rbrace_{J_1}^{\epsilon_1}, \dots , \left\lbrace u_n, z_n, \zb_n\right\rbrace_{J_n}^{\epsilon_n}\right) \\
   =    \int_0^{+\infty} \prod_{i=1}^n \frac{d\om_i}{2\pi} \, (-i \epsilon_i \omega_i)^{m_i} e^{-i\e_i \om_i u_i}  \mA_n \left(\left\lbrace \om_1, z_1, \zb_1\right\rbrace_{J_1}^{\epsilon_1}, \dots , \left\lbrace \om_n, z_n, \zb_n\right\rbrace_{J_n}^{\epsilon_n}\right) \\
   = \langle \partial_{u_1}^{m_1}\Phi^{\epsilon_1}_{(k_1,\bar{k}_1)} (u_1, z_1, \bz_1)  \ldots \partial_{u_n}^{m_n}\Phi^{\epsilon_n}_{(k_n,\bar{k}_n)} (u_n, z_n, \bz_n)  \rangle \ , 
\end{multline} where $m_i \in \mathbb N$ for $i=1, \ldots, n$. Tree-level Carrollian amplitudes for gluons and gravitons have been discussed in detail in \cite{Mason:2023mti, Banerjee:2019prz}. In the following, we will restrict ourselves to the case of scalars ($J=0$), which is sufficient to illustrate our prescription of the flat limit from AdS. For instance, the two-point Carrollian amplitudes can be obtained by applying \eqref{descendants amplitudes} to $\mathcal{A}_2 = \kappa_2 \frac{\delta (\omega_{12})}{\omega_1} \delta^{(2)} (z_{12}) \delta_{\e_1, -\e_2}$, which leads to 
\begin{align}
\label{twopointCarrollian}
    \mc_2^{m_1 m_2}  \left(\left\lbrace u_1, z_1, \zb_1\right\rbrace_{0}^{\epsilon_1}, \left\lbrace u_2, z_2, \zb_2\right\rbrace_{0}^{-\epsilon_1}\right)= \frac{\kappa_2}{4\pi^2}\frac{(-1)^{m_1}\Gamma\left(m_1+m_2\right)}{\left(u_{12}-i\varepsilon \, \e_1\right)^{m_1+m_2}}\delta^{(2)}\left(z_{12}\right) \ .
\end{align}
Scalar three-point amplitudes are non-zero only in the collinear limit in Lorentzian signature. It is common to work in Klein space (a spacetime with $(2,2)$ signature) to get around this. In this case, the celestial sphere is replaced by the celestial torus, there is only one copy of $\mathscr I$, $z$ and $\bar z$ are two independent real coordinates, $\epsilon = \pm 1$ in \eqref{eq:mompar31} now label the two Poincar\'e patches on the celestial torus, and the parametrization of the momentum \eqref{q in terms of z bar z future} becomes $p^\mu = \frac{\e \om}{\sqrt{2}} \left(1+z_i \zb_i, z_i + \zb_i, z_i - \zb_i , 1 - z_i \zb_i \right)$. See e.g. \cite{Atanasov:2021oyu, Jorge-Diaz:2022dmy} for more details. The three-point Carrollian amplitude is then obtained by applying \eqref{descendants amplitudes} to $\mathcal{A}_3 = \kappa_3 \delta^{(4)} (p_1+p_2+p_3)$, which gives\footnote{The parametrization used for momenta in this paper (both in Lorentzian and Klein) differs from that of \cite{Mason:2023mti} by a factor of $\frac{1}{\sqrt{2}}$.}
\begin{multline}
\label{eq:threeptcarr}
     {\mc}^{m_1 m_2 m_3}_3 = - \frac{i\kappa_3\, \e_1 \e_2 \e_3}{ \left(2\pi\right)^3 } \left(z_{12}\right)^{m_1-1}\left(z_{13}\right)^{m_2-1}\left(z_{23}\right)^{m_3-1} \delta\left(\zb_{12}\right)\delta\left(\zb_{23}\right)\\
     \times  \Theta\left(-\frac{z_{13}}{z_{23}}\e_1\e_2\right)\Theta\left(\frac{z_{12}}{z_{23}}\e_1\e_3\right) \frac{\Gamma\left(\sum_{i=1}^3 m_i -1\right)}{\left(z_{23}u_1 + z_{31}u_2+z_{12}u_3 - i \varepsilon \e_1\, \text{sign} (z_{23}) \right)^{\sum_{i=1}^3 m_i -1}}  \ , 
\end{multline} where $\Theta(x)$ denotes the step function. In Lorentzian signature, the four-point contact diagram $\mA_{4,c} = \kappa_4 \delta^{(4)} (p_1+p_2+p_3+p_4)$ expressed in position space at $\mathscr I$ reads 
\begin{multline}
\label{eq:4ptcontactcarr}
   \mc_{4,c}^{m_1 m_2 m_3 m_4} = \frac{\kappa_4}{\left(2\pi\right)^4}\delta\left(z-\zb\right)  \Theta\left(-z \e_1 \e_4\right) \Theta\left(\left(1-z\right)z\e_2 \e_4 \right) \Theta\left( (z-1) \e_3 \e_4 \right) \left(-1\right)^{m_1+m_3}\\ 
  \frac{\left|z_{14}\right|^{2m_3}\left|z_{24}\right|^{2m_1-2}\left|z_{34}\right|^{2m_2}}{\left|z_{12}\right|^{2m_1}\left|z_{13}\right|^{2m_3+2}\left|z_{23}\right|^{2m_2}} \frac{ z^{m_1-m_2} \left(1-z\right)^{m_2-m_3} \Gamma\left(\sum_{i=1}^4 m_i\right)}{\left( u_4 - u_1 z \left|\frac{z_{24}}{z_{12}}\right|^2+u_2 \frac{1-z}{z}\left|\frac{z_{34}}{z_{23}}\right|^2 - u_3\frac{1}{1-z}\left|\frac{z_{14}}{z_{13}}\right|^2  \right)^{\sum_{i=1}^4 m_i}} \ ,
\end{multline}
where $z=\frac{z_{12}z_{34}}{z_{13}z_{24}}$ is the two-dimensional cross-ratio, $\left|z_{ij}\right|^2 = z_{ij} \zb_{ij} $ and $m=\sum_{i=1}^4 m_i$. We can proceed in a similar way for the four-point exchange diagram given by $\mathcal{A}_{4,e} =  \kappa_3^2 \left(\frac{1}{s}+ \frac{1}{t} + \frac{1}{u}\right) \delta^{(4)}( p_1 + p _2 + p_3)$, where $s=(p_1 + p_2)^2$, $t=(p_1 + p_3)^2$ and $u=(p_1 + p_4)^2$ are the Mandelstam variables. Expressing the $s$-channel term of $\mA_{4,e}$ in position space at $\mathscr I$, we get
\begin{multline}
\label{eq:4ptexchangecarr}
    \mc_{4,e}^{m_1m_2 m_3 m_4} =\frac{1}{2} \frac{\kappa_3^2}{\left(2\pi\right)^4}\delta\left(z-\zb\right)  \Theta\left(-z \e_1 \e_4\right) \Theta\left(\left(1-z\right)z\e_2 \e_4 \right) \Theta\left( (z-1) \e_3 \e_4 \right) \left(-1\right)^{m_1+m_3-1}\\ 
  \frac{\left|z_{14}\right|^{2m_3}\left|z_{24}\right|^{2m_1-4}\left|z_{34}\right|^{2m_2-2}}{\left|z_{12}\right|^{2m_1}\left|z_{13}\right|^{2m_3+2}\left|z_{23}\right|^{2m_2-2}} \frac{ z^{m_1-m_2} \left(1-z\right)^{m_2-m_3-1} \Gamma\left(\sum_{i=1}^4 m_i-2\right)}{\left( u_4 - u_1 z \left|\frac{z_{24}}{z_{12}}\right|^2+u_2 \frac{1-z}{z}\left|\frac{z_{34}}{z_{23}}\right|^2 - u_3\frac{1}{1-z}\left|\frac{z_{14}}{z_{13}}\right|^2  \right)^{\sum_{i=1}^4 m_i-2}} \ .
\end{multline} The corresponding expressions for the $t$- and $u$-channels can easily be deduced. As we will explain in Section \ref{sec:Carrollian limit of boundary CFT correlators}, these explicit expressions for the Carrollian correlators can be obtained from a Carrollian limit ($c\to 0$) of CFT correlators, without the need to refer to the bulk.

\subsection{Bulk-to-boundary propagators}
\label{sec:Bulk-to-boundary propagatorsflat}

In this section, we derive an explicit expression for the bulk-to-boundary propagators in Minkowski space by taking a large radius limit of the bulk-to-bulk propagator. In flat spacetime, the bulk-to-bulk propagator for a massless scalar field, which solves the sourced Klein-Gordon equation 
\begin{equation}
 \Box_{X_{1}} \mathcal{G}_{BB}^{Flat}(X_1,X_2) = \frac{\delta^{(4)}(X_1-X_2)}{\sqrt{-g}} \ , \qquad \Box= \Big[  \frac{2}{r^2} \partial_z {\partial}_{\bar z} - \frac{2}{r} \partial_u - 2 \partial_u \partial_r   \Big]
 \label{box equation flat}
\end{equation} and is relevant for scattering is given by the Feynman propagator
\begin{equation}
    \mathcal{G}_{BB}^{Flat}(X_1,X_2)  = \frac{-i}{\left(2\pi\right)^2} \frac{1}{\xi^{Flat}_{12} + i \varepsilon}
    \label{bulktobulkflat} \ ,
\end{equation} where
\begin{equation}
    \xi_{12}^{Flat} = -2  r_{12} u_{12} + 2 r_1 r_2 |z_1 - z_2|^2 
    \label{distance flat}
\end{equation} is the spacetime interval between $X^\mu_1 (u_1, r_1, z_1, \bar z_1)$ and $X^\mu_2 (u_2, r_2, z_2, \bar z_2)$, i.e., $\xi_{12}^{Flat} = (X_{12})^\mu (X_{12})_\mu$ with $X_{12}^\mu = X_1^\mu - X_2^\mu$. The infinitesimal regulator $\varepsilon >0$ in position space ensures that the solution is well-defined on the physical spacetime and can be obtained by analytic continuation in Cartesian coordinates from Euclidean signature \cite{Duffin}. The bulk-to-boundary propagator admits an integral representation using Fourier transform:
\begin{equation}
\begin{split}
     \mathcal{G}_{BB}^{Flat}(X_1,X_2)   &= -\int \frac{d^4p}{\left(2\pi\right)^4} \frac{e^{-i p \cdot X_{12}}}{p^2 - i \varepsilon_p} \ .
    \end{split}
    \label{fourierBulktoBulk}
\end{equation} In Appendix \ref{sec:Propagators in flat space}, we provide more details on the convention of regulators and show explicitly that the integral expression \eqref{fourierBulktoBulk} agrees with the position space expression \eqref{bulktobulkflat}.

The bulk-to-boundary propagator is obtained from the bulk-to-bulk propagator \eqref{bulktobulkflat} by sending one point to infinity and rescaling appropriately to extract the finite piece. As discussed in Appendix \ref{sec:Propagators in flat space}, the outgoing bulk-to-boundary propagator associated with positive energy particle is defined as 
\begin{equation}
    \mathcal{G}_{Bb,+}^{Flat}(x_1;X_2)  \equiv \lim_{r_1 \to + \infty } r_1 \mathcal{G}_{BB}^{Flat}(X_1,X_2) =  \frac{-i}{2\left(2\pi\right)^2} \frac{1}{-u_{1}-q_1 \cdot X_2 + i \varepsilon } \ ,
    \label{Bbout}
\end{equation} where we used $q_1 \cdot X_2 = -u_2-r_2|z_{12}|^2$ in the parametrization \eqref{Retarded flat BMS coordinates} and \eqref{q in terms of z bar z future} to write the last expression. Similarly, the incoming bulk-to-boundary propagator associated with positive energy particle is defined as
\begin{equation}
\mathcal{G}_{Bb,-}^{Flat}(x_2;X_1)  \equiv \lim_{r_2 \to - \infty } r_1 \mathcal{G}_{BB}^{Flat}(X_1,X_2) =  \frac{-i}{2\left(2\pi\right)^2} \frac{1}{-u_{2}-q_2 \cdot X_1 - i \varepsilon } \ ,
\label{Bbin}
\end{equation} where we rescaled $-\varepsilon / r_2 \to \varepsilon$ to keep $\varepsilon > 0$. These bulk-to-boundary propagators can be nicely rewritten as 
\begin{equation}
\mathcal{G}_{Bb,\epsilon}^{Flat}(x;X) = \frac{-\epsilon}{2\left(2\pi\right)^2} \int_{0}^{+\infty} {d\omega} e^{-\varepsilon \omega} e^{-i \epsilon \omega u_x } e^{-i \epsilon \om q^\mu_x X_\mu} = \frac{-i}{2\left(2\pi\right)^2} \frac{1}{-u_x - q_x \cdot X +i \epsilon \varepsilon} \ ,
\label{bulktoboundaryflat}
\end{equation} which corresponds to the Fourier transform of a plane wave. 

As discussed in \cite{Donnay:2022wvx}, the boundary value at $\mathscr I$ of a bulk field with radiative falloffs is identified with a Carrollian primary operator in the putative Carrollian CFT. For a bulk scalar field $\phi (X)$, we have explicitly
\begin{equation}
    \Phi^{\epsilon =\pm1}_{(\frac{1}{2}, \frac{1}{2})}(x)  =  \lim_{r\to \pm \infty} r \, \phi (X) \ ,
    \label{boundary values}
\end{equation} where the Carrollian weights have been fixed according to \eqref{fiexed Carrollian weights} with $J = 0$. The Feynman propagator is defined as the time-ordered correlator 
\begin{equation}
     \mathcal{G}_{BB}^{Flat}(X_1,X_2) = \left\langle T (\phi (X_1) \phi (X_2) )\right\rangle \ .
\end{equation} It is natural to rewrite the bulk-to-boundary propagators \eqref{Bbout} and \eqref{Bbin} as 
\begin{equation}
\mathcal{G}_{Bb,+1}^{Flat}(x;X) = \left\langle \Phi^{\epsilon =+1}_{(\frac{1}{2}, \frac{1}{2})}(x) \phi (X) \right\rangle \ , \qquad \mathcal{G}_{Bb,-1}^{Flat}(x;X) = \left\langle \phi (X) \Phi^{\epsilon =-1}_{(\frac{1}{2}, \frac{1}{2})}(x)  \right\rangle \ .
\end{equation} In particular, as discussed in \cite{Mason:2023mti} (see also Equation \eqref{descendants amplitudes}), it is sometimes useful to consider correlators with $\partial_u$-descendants 
\begin{align}
\label{descendantsbB}
&\mathcal{G}_{Bb,+1}^{Flat,m}(x;X) = \left\langle \partial_u^m \Phi^{\epsilon =+1}_{(\frac{1}{2}, \frac{1}{2})}(x) \phi (X) \right\rangle \ , \qquad \mathcal{G}_{Bb,-1}^{Flat,m}(x;X) = \left\langle  \phi (X) \partial_u^m \Phi^{\epsilon =-1}_{(\frac{1}{2}, \frac{1}{2})}(x)  \right\rangle \ ,
  \\
&\mathcal{G}_{Bb,\epsilon}^{Flat,m}(x;X) = \frac{-\epsilon}{2\left(2\pi\right)^2} \int_{0}^{+\infty} {d\omega} (-i \epsilon \omega)^m e^{-\varepsilon \omega} e^{-i \epsilon \omega u } e^{-i \epsilon \om q^\mu X_\mu} = \frac{-i}{2\left(2\pi\right)^2} \frac{\Gamma (m+1)}{(-u - q \cdot X +i \epsilon \varepsilon)^{m+1}} \ , \nonumber
\end{align} which reproduce the bulk-to-boundary correlators considered in \cite{Bagchi:2023fbj,Bagchi:2023cen}. As a consistency check, the Carrollian two-point function \eqref{twopointCarrollian} with normalization $\kappa_2 = - i \pi$ can be obtained from \eqref{descendantsbB} by simply taking $\epsilon = \pm 1$, $r \to \mp \infty$, and using the stationary phase approximation.

\subsection{Feynman diagrams for Carrollian amplitudes}
\label{sec:Feynman diagrams for Carrollian amplitudes}

The bulk-to-bulk and bulk-to-boundary propagators introduced in the previous section allow us to express the Carrollian amplitudes directly in terms of Feynman rules \cite{Liu:2024nfc}. These expressions will be the flat space analogues of the AdS Witten diagrams, and in fact, they will be those on which we will land when taking the flat limit inside of the integrals involved in the computation of AdS Witten diagrams, see next section. 

The three-point Carrollian amplitude \eqref{eq:threeptcarr} can be expressed as a contact diagram in position space for which the external propagators are attached to points at $\mathscr I$: 
\begin{equation}
    \mathcal{C}_3 (x^{\epsilon_1}_1,x^{\epsilon_2}_2,x^{\epsilon_3}_3) = \beta_3 \kappa_3 \int_{Flat} d^4 X \mathcal{G}^{Flat}_{Bb,\epsilon_1}(x_1,X) \mathcal{G}^{Flat}_{Bb,\epsilon_2} (x_2,X)\mathcal{G}^{Flat}_{Bb,\epsilon_3}(x_3,X) \ ,
\label{2pointdiagram}
\end{equation} with $\beta_3 =  \frac{-4  \epsilon_1 \epsilon_2 \epsilon_3 }{\pi}$.\footnote{The constants $\beta_3$, $\beta_4^c$ and $\beta_4^e$ introduced in this section are consistent with our definition of Carrollian amplitude \eqref{eq:AtoC} and the normalization of the bulk-to-boundary propagator \eqref{bulktoboundaryflat}.\label{footnote}}  Here the integration is performed over Minkowski spacetime. Using the integral representation of the bulk-to-boundary propagator in \eqref{bulktoboundaryflat}, we obtain 
\begin{equation}
\begin{split}
    \mathcal{C}_3 (x^{\epsilon_1}_1,x^{\epsilon_2}_2,x^{\epsilon_3}_3)  &= \frac{\kappa_3}{(2\pi)^4} \int_{0}^{+\infty}  \prod_{i=1}^3 \frac{d\omega_i}{2 \pi} e^{-i \epsilon_i \omega_i u_i }  \int_{Flat} d^4 X   e^{-i \sum_{i=1}^3 p^\mu_i X_\mu} \\
    &= \kappa_3  \int_{0}^{+\infty} \prod_{i=1}^3  \frac{d\omega_i}{2 \pi} e^{-i \epsilon_i \omega_i u_i }\delta^{(4)}(p_1 + p _2 + p_3) \ .
\end{split}
\end{equation}  By comparing this expression to \eqref{eq:AtoC}, this corresponds to the definition of the three-point Carrollian amplitude associated with the momentum space amplitude $\mathcal{A}_3 = \kappa_3 \delta^{(4)}(p_1 + p _2 + p_3)$. This discussion can be easily extended to the three-point correlator of $\partial_u$-descendants using the bulk-to-boundary propagators \eqref{descendantsbB}: 
\begin{equation}
    \mathcal{C}^{m_1 m_2 m_3}_3 (x^{\epsilon_1}_1,x^{\epsilon_2}_2,x^{\epsilon_3}_3) = \beta_3 \kappa_3  \int_{Flat} d^4 X \mathcal{G}^{Flat,m_1}_{Bb,\epsilon_1}(x_1,X) \mathcal{G}^{Flat,m_2}_{Bb,\epsilon_2} (x_2,X)\mathcal{G}^{Flat,m_3}_{Bb,\epsilon_3}(x_3,X) \ .
    \label{3ptdiagramflat}
\end{equation} The last expression matches with \eqref{descendants amplitudes} for $n=3$. The above example can be easily extended to any $n$-point contact diagram. For completeness, we also display the correlator of $\partial_u$-Carrollian descendants associated with the four-point contact diagram: 
\begin{multline}
\label{4pointcontact}
    \mathcal{C}^{m_1 m_2 m_3 m_4}_{4,c}(x^{\epsilon_1}_1,x^{\epsilon_2}_2,x^{\epsilon_3}_3) \\
    =  \beta_4^c \kappa_4 \int_{Flat} d^4X \mathcal{G}^{Flat,m_1}_{Bb,\epsilon_1}(x_1,X) \mathcal{G}^{Flat,m_2}_{Bb,\epsilon_2} (x_2,X)\mathcal{G}^{Flat,m_3}_{Bb,\epsilon_3}(x_3,X) \mathcal{G}^{Flat,m_4}_{Bb,\epsilon_4}(x_4,X)
\end{multline} with $\beta_4^c = 16 \epsilon_1 \epsilon_2 \epsilon_3 \epsilon_4$. The four-point exchange diagram coming from the cubic interaction can also be expressed by
\begin{align}
\label{flat space4ptexch}
&\mathcal{C}^{m_1m_2m_3m_4}_{4,e} (x_1^{\epsilon_1}, x_2^{\epsilon_2}, x_3^{\epsilon_3}, x_4^{\epsilon_4}) \\ &= \beta_4^e \kappa_3^2 \int_{Flat} d^4 X d^4Y \mathcal{G}^{Flat,m_1}_{Bb,\epsilon_1} (x_1;X) \mathcal{G}^{Flat,m_2}_{Bb,\epsilon_2} (x_2;X) \mathcal{G}^{Flat}_{BB}(X,Y) \mathcal{G}^{Flat,m_3}_{Bb,\epsilon_3} (x_3;Y) 
    \mathcal{G}^{Flat,m_4}_{Bb,\epsilon_4} (x_4;Y) \nonumber 
\end{align} where $\beta_4^e = -16  \epsilon_1 \epsilon_2 \epsilon_3 \epsilon_4$. Using the integral representation of the bulk-to-bulk \eqref{fourierBulktoBulk} and the bulk-to-boundary \eqref{bulktoboundaryflat} propagators, we can rewrite this expression as
\begin{align}
&\mathcal{C}^{m_1m_2m_3m_4}_{4,e} (x_1^{\epsilon_1}, x_2^{\epsilon_2}, x_3^{\epsilon_3}, x_4^{\epsilon_4}) \nonumber\\
     &= \frac{\kappa_3^2}{(2\pi)^4}  \int_{0}^{+\infty}  \prod_{i=1}^4 \frac{d\omega_i}{2 \pi} (-i\epsilon_i \omega_i)^{m_i} e^{-i \epsilon_i \omega_i u_i } \int \frac{d^4 p}{(2\pi)^4} \int d^4 X d^4Y \frac{e^{-ip^\mu (X-Y)_\mu}}{p^2} e^{-i (p_1 + p_2)^\mu X_\mu} e^{-i (p_3 + p_4)^\mu Y_\mu} \nonumber \\
     &=\kappa_3^2  \int_{0}^{+\infty} \prod_{i=1}^4 \frac{d\omega_i}{2 \pi} (-i\epsilon_i \omega_i)^{m_i} e^{i \epsilon_i \omega_i u_i }  \int d^4 p \frac{1}{p^2} \delta^{(4)} (p+ p_1 + p_2) \delta^{(4)} (p- p_3 - p_4) \nonumber \\
     &= \kappa_3^2  \int_{0}^{+\infty} \prod_{i=1}^4 \frac{d\omega_i}{2 \pi}(-i\epsilon_i \omega_i)^{m_i} e^{i \epsilon_i \omega_i u_i } \frac{1}{(p_1 +p_2)^2} \delta^{(4)} \Big(\sum_{i=1}^4 p_i \Big) \ .
\end{align}
Comparing with \eqref{descendants amplitudes}, this corresponds to the definition of the three-point Carrollian amplitude associated with the $s$-channel exchange diagram $\mathcal{A}_{4,e} =  \frac{\kappa_3^2}{s} \delta^{(4)}( p_1 + p _2 + p_3)$. The general proof of the representation of Carrollian amplitudes in terms of Feynman diagrams can be found in \cite{Liu:2024nfc}.

\section{Flat limit of AdS Witten diagrams in Bondi coordinates}
\label{sec:Flat limit of AdS Witten diagrams in Bondi coordinates}

In this section, we discuss the Bondi coordinates in AdS. We then write the building blocks of the AdS Witten diagrams in these coordinates and demonstrate that they admit a smooth limit when taking $\ell \to \infty$. Indeed, we show that the bulk-to-bulk and bulk-to-boundary propagators reduce precisely to those of flat space derived in the previous section. Putting all together, we confirm that the limit inside of the integrals is smooth and gives back the Carrollian amplitudes in their form presented in Section \ref{sec:Feynman diagrams for Carrollian amplitudes}.

\subsection{Bondi coordinates in AdS}

The (planar) Bondi coordinates $(u,r,z, \bar z)$, with $u,r\in \mathbb R, z \in \mathbb C$, originally introduced in flat spacetime \cite{Bondi:1962px,Sachs:1962zza}, also exist in AdS \cite{Barnich:2012aw,Poole:2018koa,Compere:2019bua,Compere:2020lrt}. In these coordinates, the AdS metric takes the form
\begin{equation}
    ds^2_{AdS} = - \frac{r^2}{\ell^2} du^2 - 2 du dr + 2 r^2 dz d\bar{z} \ ,
    \label{AdSBondi}
\end{equation} where the dimensions of length are $\ell \sim L$, $u \sim L$, $r \sim L$, $z \sim L^0$, $\bar z \sim L^0$. One can indeed check explicitly that the metric has constant curvature, i.e. $R_{\mu\nu\rho\sigma} = -\frac{1}{\ell^2}( g_{\mu\rho}g_{\nu\sigma}- g_{\mu\sigma} g_{\nu\rho})$. Rescaling this metric by $1/r^2$ and taking $r\to \pm \infty$ yields the boundary metric
\begin{equation}
    ds^2_{\partial AdS} =  - \frac{1}{\ell^2} du^2 + 2 dz d\bar{z} \ .
\label{boundary metric AdS}
\end{equation} The main advantage of Bondi coordinates is that they admit a well-defined flat limit. Indeed, taking $\ell \to \infty$ in \eqref{AdSBondi} gives back the flat metric \eqref{Minkowski metric}, $ds^2_{AdS} \to ds^2_{Flat}$. Moreover, implementing this limit at the level of the boundary metric \eqref{boundary metric AdS} yields the degenerate metric, $ds^2_{\partial AdS} \to 0du^2 + 2 dz d\bar z$, which is part of the Carrollian structure at the boundary of flat space \cite{Henneaux:1979vn,1977asst.conf....1G,Duval:2014uva,Ashtekar:2014zsa}. Indeed, this is exactly the metric one would have obtained by taking the Carrollian limit ($c\to 0$) of the Minkowski line element in three dimensions $-c^2 du^2 +2 dz d\bar z$. Hence, we see by this simple geometric observation that the flat limit in the bulk implies the Carrollian limit at the boundary via the formal identification $c_{boundary} \equiv \frac{1}{\ell_{bulk}} \to 0$.

We now introduce the embedding space coordinates which will be very useful in the next sections, and discuss their relation with intrinsic coordinate systems: global, Poincaré and Bondi coordinates. \textit{Lorentzian} AdS$_4$ can be described as a hyperboloid embedded in $\mathbb{R}^{3,2}$ with coordinates 
$X^I = \left(X^+, X^-, X^0, X^1, X^2\right)$ and metric
\begin{equation}
    \label{eq:embeddingmetric}
    G_{IJ} dX^I dX^J =  -dX^+dX^- -(dX^0)^2 + (dX^1)^2 + (dX^2)^2 \ .
\end{equation}
Indeed, AdS${}_{4}$ is the hyperboloid defined by the universal cover of 
\begin{equation}
    \label{eq:LAdSdef}
    X \cdot X = -X^{+}X^{-} -(X^0)^2 + (X^1)^2 + (X^2)^2 = -\ell^2 \ . 
\end{equation} We typically work with the universal cover of this quadric since the latter has closed timelike loops. This can be seen by temporarily switching from light-cone coordinates to ordinary ones via $X^{\pm} = X^3 \pm X^4$. This gives
\begin{equation}
    (X^4)^2+ (X^0)^2+\ell^2 = (X^1)^2+(X^2)^2+(X^3)^2 \ .
\end{equation}
We will also use \textit{Euclidean} AdS${}_4$ which is defined as the quadric
\begin{equation}
    \label{eq:EAdSdef}
    X \cdot X = -X^{+}X^{-} +(X^0)^2 + (X^1)^2 + (X^2)^2 = -\ell^2 \ , \qquad X^+ + X^- >0 \ ,
\end{equation}
in the embedding space $\mathbb{R}^{4,1}$. This is related to the Lorentzian one by analytic continuation.  In the rest of this section, we will work with Lorentzian AdS$_4$. We introduce three intrinsic coordinate systems.

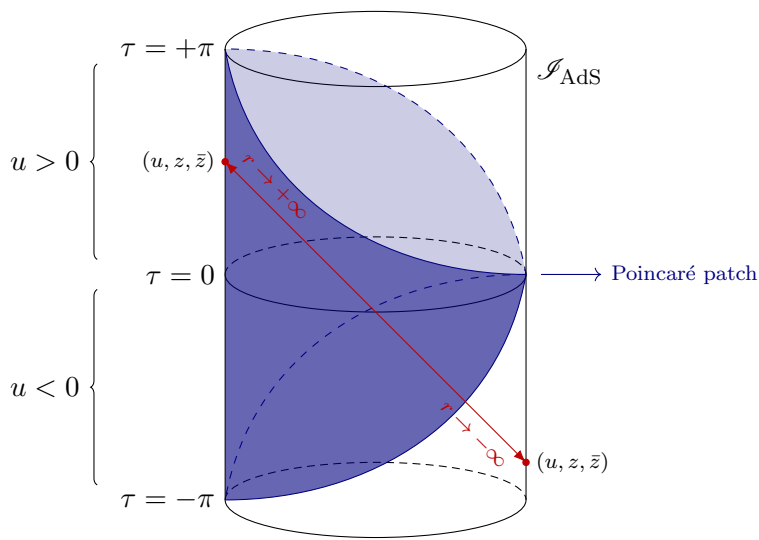
\begin{figure}[ht!]
\centering
\begin{tikzpicture}
	\def\h{3};\def\l{2};\def\dy{0.5};\def\ofx{1.7};\def\ofy{0.2};
	\coordinate (bl) at (-\l,-\h);
	\coordinate (br) at ( \l,-\h);
	\coordinate (cl) at (-\l,  0);
	\coordinate (cr) at ( \l,  0);
	\coordinate (cc) at (  0,  0);
	\coordinate (tl) at (-\l, \h);
	\coordinate (tr) at ( \l, \h);
	\coordinate (al) at ($(cl)+(0,1.5)$);
	\coordinate (ar) at ($(al)+(2*\l,-2*\l)$);
	\coordinate (bracetopfin) at ($(tl)+(-\ofx,-\ofy)$);
	\coordinate (bracetopini) at ($(cl)+(-\ofx, \ofy)$);
	\coordinate (bracebotini) at ($(bl)+(-\ofx, \ofy)$);
	\coordinate (bracebotfin) at ($(cl)+(-\ofx,-\ofy)$);
	\draw[red,opacity=0] ($(bl)+(-3.1,-1)$) -- ($(tl)+(-3.1,1)$) -- ($(tr)+(3.1,1)$) -- ($(br)+(3.1,-1)$) -- cycle;
	\draw[] (bl)node[left]{$\tau=-\pi$} -- (cl)node[left]{$\tau=0$} -- (tl)node[left]{$\tau=+\pi$};
	\draw[] (br) -- (tr)node[anchor=north west]{$\mathscr I_{\text{AdS}}$};
	\draw[] (bl) arc[x radius=\l, y radius=\dy, start angle=-180, end angle=0];
	\draw[densely dashed] (bl) arc[x radius=\l, y radius=\dy, start angle=180, end angle=0];
	\draw[] (cl) arc[x radius=\l, y radius=\dy, start angle=-180, end angle=0];
	\draw[densely dashed] (cl) arc[x radius=\l, y radius=\dy, start angle=180, end angle=0];
	\draw[] (tl) arc[x radius=\l, y radius=\dy, start angle=-180, end angle=180];
	\node[left] at (al) {\scriptsize $(u,z,\bar z)$};
	\node[right] at (ar) {\scriptsize $(u,z,\bar z)$};
	\draw[decorate, decoration={brace}] (bracetopini) -- (bracetopfin);
	\draw[decorate, decoration={brace}] (bracebotini) --(bracebotfin);
	\node[left,outer sep=3pt] at ($(bracetopini)!0.5!(bracetopfin)$) {$u>0$};
	\node[left,outer sep=3pt] at ($(bracebotini)!0.5!(bracebotfin)$) {$u<0$};
	\draw[navyblue] (cr) to[out=180,in=-80] (tl);
	\draw[navyblue,dashed] (cr) to[out=100,in=0] (tl);
	\draw[navyblue,dashed] (cr) to[out=180,in=80] (bl);
	\draw[navyblue] (cr) to[out=-100,in=0] (bl);
	\fill[navyblue,opacity=0.2] (tl) to[in=100,out=0] (cr) to[out=-100,in=0] (bl);
	\fill[navyblue,opacity=0.5] (tl) to[in=180,out=-80]  (cr) to[out=-100,in=0] (bl);
	\draw[->,navyblue] ($(cr)+(0.2,0)$) -- ($(cr)+(1,0)$)node[right]{\scriptsize Poincaré patch};
	\node[rotate=-45,anchor=south west,red!75!black] at (al) {\scriptsize $r\to+\infty$};
	\node[rotate=-45,anchor=north east,red!75!black] at (ar) {\scriptsize $r\to-\infty\,$};
	\draw[Latex-Latex,red!75!black] (al)node[circle,fill,inner sep=1pt]{} -- (ar)node[circle,fill,inner sep=1pt]{};
\end{tikzpicture}
\caption{The blue region represents a Poincar\'e patch ($\rho>0$) in AdS. To cover the other half of the cylinder, we can use a second Poincar\'e patch corresponding to $\rho <0$. In particular, this naturally divides the boundary into two regions. The Bondi coordinates also cover the region outside of the patch (this follows from \eqref{eq:ptob} and $r\in \mathbb R$). In red is a null ray defined by $(u,z,\bar z) = \text{constant}$ in Bondi coordinates. The point obtained by taking $r\to +\infty$ is inside of the patch, while the point $r\to -\infty$ is outside of the patch.} \label{Fig:Poincare}
\end{figure}

\begin{itemize}
\item Global coordinates in AdS are obtained by choosing the following parametrization: 
\begin{equation}
    (X^+, X^-) = \frac{\ell}{\cos R} \left(  \sin \tau , \cos \tau \right) \ , \qquad (X^0 , X^1, X^2) = \ell \tan R \left( \frac{z + \bar z}{1+ z \bar z} ,  \frac{-i(z-\bar z)}{1 + z \bar z}, \frac{1 - z \bar z}{1 + z \bar z} \right) \ .
\end{equation} The metric reads 
\begin{equation}
    ds^2 = \frac{\ell^2}{\cos^2 R} [-d\tau^2 + d\rho^2 + \sin^2 \rho d\Omega_{S} ] \ ,
    \label{global coordinate}
\end{equation} where $d\Omega_{S} = \frac{4dz d\bar z}{(1+ z\bar z)^2}$ is the round sphere metric. From this embedding, the range of coordinates is $\tau \in [-\pi, \pi]$ and $R \in [0,\frac{\pi}{2}]$, see Figure \ref{Fig:Poincare}. As mentioned above, to get the universal cover of the quadric, we take instead $\tau \in \mathbb R$.

    \item Poincar{\' e} coordinates $(\rho,x^\mu)$ are obtained by choosing the following parametrization:
\begin{equation}
\label{eq:Poincare}
    X = \frac{\ell}{\rho}\left(1, \rho^2+ x \cdot x, x^0, x^1, x^2\right) \ ,
\end{equation}
where $x \cdot x = \eta_{\mu \nu} x^{\mu} x^{\nu}$ with $\eta$ being the metric on $\mathbb{R}^{1,2}$. Here we chose $\rho, x^\mu$ to be dimensionless and $\ell$ to have dimensions of length. The metric in these coordinates is 
\begin{equation}
    ds^2_{AdS} = \frac{\ell^2}{\rho^2} \Big(d\rho^2 - (dx^0)^2 +(dx^1)^2 + (dx^2)^2 \Big) \ .
\end{equation} 
The conformal boundary is located at $\rho = 0$ and the region where $\rho > 0$ describes a Poincar{\'e} patch, see Figure \ref{Fig:Poincare} (the other Poincar\'e patch corresponds to the region where $\rho < 0$). Boundary Poincar{\'e} coordinates are denoted by
\begin{equation}
    x = \left(x^0, x^1, x^2 \right)
\end{equation}
and the boundary metric is 
\begin{equation}
    ds^2_{\partial AdS} = -(dx^0)^2 +(dx^1)^2 + (dx^2)^2  \ .
\end{equation} 
\item Bondi coordinates $(u,r,z,\bar z)$ are obtained via the parametrization
\begin{equation}
    X = r\left(1, \frac{2u}{r}-\frac{u^2}{\ell^2}+2 w \bar w, -\frac{\ell}{r}+\frac{u}{\ell}, \frac{w+\bar w}{\sqrt{2}},  \frac{w-\bar w}{\sqrt{2}i}\right) \ .
\end{equation}
The relation between Poincar{\'e} and Bondi coordinates is 
\begin{equation}
    \label{eq:ptob}
    \rho = \frac{\ell}{r} \ , \quad x^0 = -\frac{\ell}{r}+\frac{u}{\ell} \ , \quad x^1 = \frac{w+\bar w}{\sqrt{2}} \ , \quad x^2 = \frac{w-\bar w}{\sqrt{2}i} \ .
\end{equation}
The interplay between Bondi coordinates and Poincar\'e patch is discussed in Figure \ref{Fig:Poincare}. One can check that the metric in Bondi coordinates is indeed given by \eqref{AdSBondi}. The conformal boundary is reached by letting $r \to \pm \infty$. After a suitable rescaling, the boundary coordinates are
\begin{equation}
      \left(1, -\frac{u^2}{\ell^2}+2 w \bar w, \frac{u}{\ell}, \frac{w+\bar w}{\sqrt{2}},  \frac{w-\bar w}{\sqrt{2}i}\right) \ ,
\end{equation}
and the boundary metric reproduces \eqref{boundary metric AdS}.  
\end{itemize}

The chordal distance (or geodesic distance) between two points in AdS${}_4$ appears frequently in computations. This is defined as
\begin{equation}
    \xi_{12}^{AdS} = (X_1 - X_2)\cdot (X_1 - X_2) = \frac{\ell^2}{\rho_1 \rho_2} \Big(\rho^2_{12}+x_{12}^2 \Big) \label{eq:chordalp} \ ,
\end{equation} 
where $x_{12}^2 = \eta_{\mu \nu} (x_1 - x_2)^{\mu} (x_1 - x_2 )^{\nu}$ and the first expression for $\xi^{AdS}_{12}$ is in terms of the embedding coordinates while the second in terms of Poincar{\'e} coordinates. This can also be written in Bondi coordinates as 
\begin{equation}
     \xi_{12}^{AdS} = - \frac{1}{\ell^2} r_1 r_2 u_{12}^2  -2  r_{12} u_{12} + 2 r_1 r_2 |z_{12}|^2  = \xi^{Flat}_{12} - \frac{1}{\ell^2}r_1 r_2 u_{12}^2 \ . \label{eq:chordalb} 
\end{equation}
A virtue of Bondi coordinates is that the chordal distance between two points reduces to the flat space distance \eqref{distance flat} in the limit $\ell \to \infty$.

\subsection{Bulk-to-bulk propagator} 

The two key ingredients in the computation of AdS Witten diagrams are the bulk-to-bulk and bulk-to-boundary propagators. For scalars, the bulk-to-bulk propagator $\mathcal{G}^{AdS,\Delta}_{BB}$ can be obtained by solving the sourced Klein-Gordon equation in AdS$_4$ 
\begin{equation}
    (\Box_{X_1} + m^2) \mathcal{G}^{AdS,\Delta}_{BB}(X_1,X_2) = \frac{1}{\sqrt{-g}}\delta^{(4)} \left(X_{12}\right) \ ,
\label{KG equation AdS}
\end{equation} with $m^2 \ell^2 = \Delta (3-\Delta)$. In Bondi coordinates, the d'Alembert operator reads explicitly as
\begin{equation}
\label{eq:wavebondi}
    \Box = \Big[  \frac{2}{r^2} \partial_z {\partial}_{\bar z} + \frac{r}{\ell^2} (4 \partial_r + r \partial_r^2) - \frac{2}{r} \partial_u - 2 \partial_u \partial_r   \Big]
\end{equation} and Equation \eqref{KG equation AdS} consistently reduces to the flat space sourced Klein-Gordon Equation \eqref{box equation flat} when $\ell \to \infty$. Note that for our purposes, $\D$ does not rescale with $\ell$ as we take the flat space limit. AdS isometries imply that the solution of \eqref{KG equation AdS} is a function of the chordal distance \eqref{eq:chordalb} only. It will be convenient to use the following dimensionless variable 
\begin{equation}
    \chi_{12} = -\frac{4\ell^2}{\xi^{AdS}_{12}} \ .\label{eq:chidef}
\end{equation} Note that $\chi_{12} \to 0$ as either point approaches the boundary, i.e. if $r_1 \to \infty$  or $r_2 \to \infty$. On the other hand $\chi_{12} \to \infty$ as $\ell \to \infty$. In terms of $\chi_{12}$ \eqref{KG equation AdS} becomes 
\begin{equation}
    \D (\D-3)\mathcal{G}^{AdS,\Delta}_{BB}+2 \chi_{12}\, (\mathcal{G}^{AdS}_{BB})' + (\chi_{12} - 1)\, \chi_{12}^2 \,(\mathcal{G}^{AdS,\Delta}_{BB})'' = \frac{\delta^{(4)}(X_{12})}{\sqrt{-g}} \ .
    \label{boxdelta equation at all order}
\end{equation}
In Euclidean signature, the general solution is 
\begin{equation}
     \label{eq:wavesol1}
    \mathcal{G}^{AdS,\Delta}_{BB}  =C_1 (\Delta) \, \chi_{12}^{3-\D}\, \hypf \left(2-\D, 3-\D, 4-2\D; \chi_{12} \right) + C_2(\Delta)  \,\chi_{12}^{\D} \,\hypf \left(-1+\D,\D, -2+2\D; \chi_{12} \right) \ .
\end{equation} The solution in Lorentzian signature can be obtained by analytically continuing the embedding space coordinate $X^0$ following the prescription of \cite{Duffin}, which for the time-ordered propagator implies $\xi_{12}^{AdS} \to \xi_{12}^{AdS} + i \varepsilon $. From now on, we will only consider the solution multiplying $C_2(\Delta)$ and drop the other one, which corresponds to the usual choice obtained by imposing regularity in the interior \cite{Witten:1998qj}. As we will see, this choice reproduces the expected behaviour for the bulk-to-bulk propagator in the flat limit. Requiring \eqref{boxdelta equation at all order} to be satisfied implies\footnote{This normalization is almost the same as the one in (2.6) of \cite{DHoker:1999mqo}. The extra factor of $i$ arises due to Wick rotation and the extra factor of $(-1)$ is due to the different convention for the Green function (compare \eqref{KG equation AdS} and (2.7) of \cite{DHoker:1999mqo}) }
\begin{align}
\label{eq:c2def}
   C_2(\Delta) = \begin{cases}
        -\frac{i(-1)^{\D}\Gamma \left(\D\right)}{2 \times 4^{\D} \ell^2 \pi^{\frac{3}{2}}\Gamma\left(\D-\frac{1}{2}\right)} \qquad &\D >1 \\
        \frac{i}{16 \pi^2 \ell^2} \qquad &\D = 1
    \end{cases}  \ .
\end{align} 
In the limit $\chi_{12} \to \infty$, corresponding to the flat limit (see below \eqref{eq:chidef}), the bulk-to-bulk propagator behaves as
\begin{align}
    \label{eq:wavesold1} 
    \mathcal{G}^{AdS,\Delta}_{BB}(X_1,X_2) \xrightarrow[]{\chi_{12} \to \infty}  &\frac{i}{16 \ell^2 \pi^2}  \chi_{12} \nonumber \\
\overset{\text{Poincar{\'e}}}{=}& \frac{-i}{(2\pi)^2} \frac{\rho_1 \rho_2}{\ell^2(\rho^2_{12}+x_{12}^2)+ i \varepsilon} \\ 
    \overset{\text{Bondi}}{=}& \frac{-i}{(2\pi)^2}  \frac{1}{-\frac{1}{\ell^2} r_1 r_2 u_{12}^2  -2  r_{12} u_{12} + 2 r_1 r_2 |z_{12}|^2 + i \varepsilon} = \frac{-i}{(2\pi)^2}\frac{1}{\xi^{AdS}_{12}+ i \varepsilon}  \nonumber \ .
\end{align}  We identified the chordal distance \eqref{eq:chordalb} in the last equality. As we can see from these expressions, 
the naive flat space limit of the propagator ($\ell \to \infty$) in Poincar{\'e} coordinates does not lead to the expected result. By contrast, this limit can be easily taken in Bondi coordinates where it yields 
\begin{equation}
   \mathcal{G}^{AdS,\Delta}_{BB}(X_1,X_2)  \xrightarrow[\text{Bondi}]{\ell \to \infty} 
  \frac{-i}{(2\pi)^2} \frac{1}{-2  r_{12} u_{12} + 2 r_1 r_2 |z_{12}|^2+ i \varepsilon}  =  \frac{-i}{(2\pi)^2} \frac{1}{\xi^{Flat}_{12}+i\varepsilon} \ .
\label{flatlimitBB}
\end{equation} Hence, we see that in Bondi coordinates, the AdS massive bulk-to-bulk propagator straightforwardly reduces to the massless Feynman propagator propagator \eqref{bulktobulkflat} in the limit $\ell \to \infty$. In particular, the parameter $\Delta$ completely disappears in the limit. Of course, this could have been anticipated from the observation made below \eqref{eq:wavebondi} where we noticed that the massive Klein-Gordon equation in AdS exactly reduces to the massless one in flat space.

\subsection{Bulk-to-boundary propagator}
\label{sec:Bulk-to-boundary propagator}

Analogously to the flat space discussion in Section \ref{sec:Bulk-to-boundary propagatorsflat}, the bulk-to-boundary propagator in AdS can be obtained from the bulk-to-bulk propagator discussed in the previous section by sending one point to infinity, which amounts to take $\chi_{12} \to 0$. In this limit, the bulk-to-bulk propagator \eqref{eq:wavesol1} (with $C_1 = 0)$ behaves as
\begin{equation}
\begin{split}
     \mathcal{G}^{AdS,\Delta}_{BB} (X_1,X_2)    \xrightarrow[]{\chi_{12} \to 0}& C_2(\Delta) \chi_{12}^{\D} \\ \overset{\text{Poincar{\' e}}}{=}& C_2(\Delta) \rho_1^{\D} \Big(\frac{-4\rho_2}{\rho^2_{2}+x_{12}^2 + i\varepsilon} \Big)^{\D} \\
      \overset{\text{Bondi}}{=}& C_2(\Delta) \left(\frac{\ell}{r_1}\right)^{\D} \Big(\frac{-4\ell}{- \frac{1}{\ell^2} r_2 u_{12}^2  -2 u_{12} + 2 r_2 |z_{12}|^2+ i\varepsilon}\Big)^{\D} \ .
\end{split} 
\label{eq:bblimit}
\end{equation} The bulk-to-boundary propagator is then defined as 
\begin{align}
    \label{eq:bboundary}
   \mathcal{G}_{Bb}^{AdS,\Delta} \left(x_1;X_2\right)  \overset{\text{Poincar{\' e}}}{\equiv }& \ell \lim_{\rho_1 \to 0}\rho_1^{-\D} \, \mathcal{G}_{BB}^{AdS,\Delta} (X_1,X_2) = \ell C_2 (\Delta) \Big(\frac{-4\rho_2}{\rho^2_{2}+x_{12}^2 + i \epsilon \varepsilon } \Big)^{\D}  \\
   \overset{\text{Bondi}}{\equiv}& \ell \lim_{r_1 \to \infty} \left(\frac{r_1}{\ell}\right)^{\D} \,\mathcal{G}_{BB}^{AdS,\Delta} (X_1,X_2) = \ell C_2 (\Delta)   \Big(\frac{-4\ell}{- \frac{1}{\ell^2} r_2 u_{12}^2  -2 u_{12} + 2 r_2 |z_{12}|^2 + i \epsilon \varepsilon }\Big)^{\D} \nonumber \ ,
\end{align} where the factors of $\rho$ and $r$ allow us to extract the finite piece, and the factors of $\ell$ are fixed by consistency with the units (see e.g. \cite{Penedones:2010ue}). As in flat space, in Bondi coordinates, we can distinguish between outgoing and incoming bulk-to-boundary propagators, corresponding to $r_1 \to +\infty$ (the case above) and $r_2 \to - \infty$, respectively (see Figure \ref{fig:flatlimit}). Hence the outgoing ($\epsilon = +1$) and incoming ($\epsilon = -1$) bulk-to-boundary propagators are given by 
\begin{equation}
\mathcal{G}_{Bb,\epsilon}^{AdS,\Delta} \left(x;X\right) = \ell C_2 (\Delta)   \Big(\frac{-4\ell}{- \frac{1}{\ell^2} r_2 u_{xX}^2  -2 u_x - 2 q_x \cdot X + i \epsilon \varepsilon }\Big)^{\D} \ .
\end{equation} 
In the flat limit, these bulk-to-boundary propagators reproduce those of Minkowski spacetime written in Equation \eqref{descendantsbB},
\begin{equation}
   \label{flat limit Bb}\mathcal{G}^{AdS,\Delta}_{Bb,\epsilon}(x;X)  \xrightarrow[\text{Bondi}]{\ell \to \infty} 
  \frac{-i \alpha (\Delta)}{2(2\pi)^2} \frac{\Gamma (\Delta)}{(- u - q \cdot X + i \epsilon \varepsilon)^{\Delta}}  =  \alpha (\Delta) \mathcal{G}^{Flat,m}_{Bb}(x;X) \ ,
\end{equation} after identifying $\Delta \equiv m +1$. The prefactor is given by  
    \begin{align}
\label{eq:alphadef}
  \alpha (\Delta) = i (2\pi )^2 \ell^{\Delta +1} (-1)^\Delta 2^{\Delta + 1} \frac{C_2 (\Delta)}{\Gamma (\Delta)}  = \begin{cases}
        \frac{\pi^{\frac{1}{2}} \ell^{\Delta - 1}}{2^{\Delta-2}\Gamma (\Delta - \frac{1}{2})} \qquad &\D >1 \\
        1 \qquad &\D = 1
    \end{cases} \ .
\end{align} For the case $\Delta = 1$, the expression for the bulk-to-boundary propagator reduces exactly to the one in Minkowski space \eqref{bulktoboundaryflat} without extra factor. The other values of $\Delta$ do not provide any new data on flat space, they simply count the number of $\partial_u$-derivatives taken on Carrollian primaries in the flat space correlators, see \eqref{descendantsbB}. This is consistent with the observation made earlier for the bulk-to-bulk propagator where the information carried by $\Delta$ was somehow redundant in the limit.

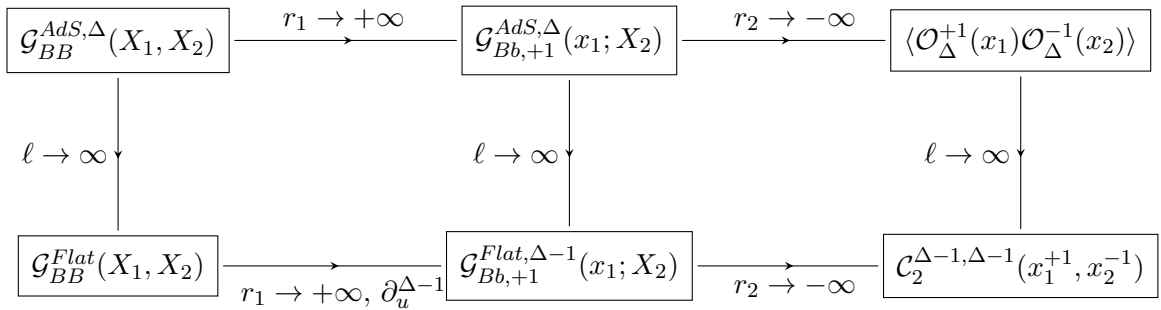
\begin{figure}[h!]
\begin{center}
\begin{tikzpicture}[scale=2.0]
    \tikzset{blok/.style={draw,outer sep=2pt,inner sep=5pt}};
    \tikzset{->-/.style={decoration={markings, mark=at position .53 with {\arrow{stealth}}}, postaction={decorate}}};
    \def\l{3};
    \def\shrink{2}; % Shrink factor for vertical arrows
    \begin{scope}
        \node[blok] (topleft) at (0,\l) {$\mathcal{G}_{BB}^{AdS,\Delta}(X_1,X_2)$ };
        \node[blok] (botleft) at (0,\l/\shrink) {$\mathcal{G}_{BB}^{Flat}(X_1,X_2)$};
        \node[blok] (topcent) at (\l,\l) {$\mathcal{G}_{Bb,+1}^{AdS,\Delta}(x_1;X_2)$};
        \node[blok] (botcent) at (\l,\l/\shrink) {$\mathcal{G}_{Bb,+1}^{Flat,\Delta - 1}(x_1;X_2)$};
        \node[blok] (toprigh) at (2*\l,\l) {$\langle \mathcal{O}^{+1}_{\Delta} (x_1) \mathcal{O}^{-1}_{\Delta} (x_2)  \rangle$};
        \node[blok] (botrigh) at (2*\l,\l/\shrink) {$\mathcal{C}_2^{\Delta - 1,\Delta - 1}(x_1^{+1}, x_2^{-1})$};
        \draw[->-] (topleft) -- (botleft);
        \draw[->-] (topcent) -- (botcent);
        \draw[->-] (toprigh) -- (botrigh);
        \draw[->-] (topleft) -- (topcent);
        \draw[->-] (topcent) -- (toprigh);
        \draw[->-] (botleft) -- (botcent);
        \draw[->-] (botcent) -- (botrigh);
        \node[left] at ($(topleft)!.5!(botleft)$) {$\ell\to\infty$};
        \node[left] at ($(topcent)!.5!(botcent)$) {$\ell\to\infty$};
        \node[left] at ($(toprigh)!.5!(botrigh)$) {$\ell\to\infty$};
        \node[above] at ($(topleft)!.5!(topcent)$) {$r_1\to+\infty$};
        \node[above] at ($(topcent)!.5!(toprigh)$) {$r_2\to-\infty$};
        \node[below] at ($(botleft)!.5!(botcent)$) {$r_1\to+\infty$, 
  $\partial^{\Delta-1}_u$};
        \node[below] at ($(botcent)!.5!(botrigh)$) {$r_2\to-\infty$};
    \end{scope} 
\end{tikzpicture}
\caption{Relations between the propagators and boundary correlators in AdS and Minkowski spacetime. The diagram commutes perfectly for the case $\Delta = 1$, without extra factor (see \eqref{eq:c2def}).}
    \label{fig:limits}
\end{center}
\end{figure}

In Figure \ref{fig:limits}, we summarise the connection between the various propagators in AdS and flat space. We also anticipate the results of Sections \ref{sec:AdS Witten diagram} and \ref{sec:Two point function} by making the link with the two-point functions.

%\begin{figure}[h!]
%\begin{center}
%\begin{tikzpicture}[scale=2.0]
%	\tikzset{blok/.style={draw,outer sep=2pt,inner sep=5pt}};
%	\tikzset{->-/.style={decoration={markings, mark=at position .53 with {\arrow{stealth}}}, postaction={decorate}}};
%	\def\l{3};
%	\begin{scope}
%		\node[blok] (topleft) at (0,\l) {$\mathcal{G}_{BB}^{AdS,\Delta}(X_1,X_2)$ };
%		\node[blok] (botleft) at (0,0) {$\mathcal{G}_{BB}^{Flat}(X_1,X_2)$};
%		\node[blok] (topcent) at (\l,\l) {$\mathcal{G}_{Bb,+1}^{AdS,\Delta}(x_1;X_2)$};
%		\node[blok] (botcent) at (\l,0) {$\mathcal{G}_{Bb}^{Flat,\Delta - 1}(x_1;X_2)$};
%		\node[blok] (toprigh) at (2*\l,\l) {$\langle \mathcal{O}^{+1}_{\Delta} (x_1) \mathcal{O}^{-1}_{\Delta} (x_2)  \rangle$};
%		\node[blok] (botrigh) at (2*\l,0) {$\mathcal{C}_2^{\Delta - 1,\Delta - 1}(x_1^{+1}, x_2^{-1})$};
%		\draw[->-] (topleft) -- (botleft);
%		\draw[->-] (topcent) -- (botcent);
%		\draw[->-] (toprigh) -- (botrigh);
%		\draw[->-] (topleft) -- (topcent);
%		\draw[->-] (topcent) -- (toprigh);
%		\draw[->-] (botleft) -- (botcent);
%		\draw[->-] (botcent) -- (botrigh);
%		\node[left] at ($(topleft)!.5!(botleft)$) {$\ell\to\infty$};
%		\node[left] at ($(topcent)!.5!(botcent)$) {$\ell\to\infty$};
%		\node[left] at ($(toprigh)!.5!(botrigh)$) {$\ell\to\infty$};
%		\node[above] at ($(topleft)!.5!(topcent)$) {$r_1\to+\infty$};
%		\node[above] at ($(topcent)!.5!(toprigh)$) {$r_2\to-\infty$};
%		\node[below] at ($(botleft)!.5!(botcent)$) {$r_1\to+\infty$, 
%  $\partial^{\Delta-1}_u$};
%		\node[below] at ($(botcent)!.5!(botrigh)$) {$r_2\to-\infty$};
%	\end{scope} 
%\end{tikzpicture}

\subsection{AdS Witten diagram}
\label{sec:AdS Witten diagram}

\begin{figure}[ht!]
\centering
\begin{tikzpicture}[scale=.9]
	\def\h{3};\def\l{2};\def\dy{0.5};\def\ofx{1.7};\def\ofy{0.2};
	\coordinate (bl) at (-\l,-\h);
	\coordinate (br) at ( \l,-\h);
	\coordinate (cl) at (-\l,  0);
	\coordinate (cr) at ( \l,  0);
	\coordinate (cc) at (  0,  0);
	\coordinate (tl) at (-\l, \h);
	\coordinate (tr) at ( \l, \h);
	\coordinate (bbl) at (-\l,-2*\h);
	\coordinate (bbr) at ( \l,-2*\h);
	\coordinate (al) at ($(cl)+(0,1.5)$);
	\coordinate (ar) at ($(al)+(2*\l,-2*\l)$);
	\coordinate (bracetopfin) at ($(tl)+(-\ofx,-\ofy)$);
	\coordinate (bracetopini) at ($(cl)+(-\ofx, \ofy)$);
	\coordinate (bracebotini) at ($(bl)+(-\ofx, \ofy)$);
	\coordinate (bracebotfin) at ($(cl)+(-\ofx,-\ofy)$);
	\draw[red,opacity=0] ($(bl)+(-1.5,-4)$) -- ($(tl)+(-1.5,1)$) -- ($(tr)+(13,1)$) -- ($(br)+(13,-4)$) -- cycle;
	\draw[] (bl) -- (cl) -- (tl);
	\draw[] (br) -- (tr)node[anchor=north west]{$\mathscr I_{\text{AdS}}$};
	\draw[] (bl) arc[x radius=\l, y radius=\dy, start angle=-180, end angle=0];
	\draw[densely dashed] (bl) arc[x radius=\l, y radius=\dy, start angle=180, end angle=0];
	\draw[] (cl) arc[x radius=\l, y radius=\dy, start angle=-180, end angle=0];
	\draw[densely dashed] (cl) arc[x radius=\l, y radius=\dy, start angle=180, end angle=0];
	\draw[] (tl) arc[x radius=\l, y radius=\dy, start angle=-180, end angle=180];
	\draw[navyblue] (cr) to[out=180,in=-80] (tl);
	\draw[navyblue,dashed] (cr) to[out=100,in=0] (tl);
	\draw[navyblue,dashed] (cr) to[out=180,in=80] (bl);
	\draw[navyblue] (cr) to[out=-100,in=0] (bl);
	\fill[navyblue,opacity=0.2] (tl) to[in=100,out=0] (cr) to[out=-100,in=0] (bl);
	\fill[navyblue,opacity=0.5] (tl) to[in=180,out=-80]  (cr) to[out=-100,in=0] (bl);
	\draw (bl) -- (bbl);
	\draw (br) -- (bbr);
	\draw[densely dashed] (bbl) arc[x radius=\l, y radius=\dy, start angle=180, end angle=0];
	\draw (bbl) arc[x radius=\l, y radius=\dy, start angle=-180, end angle=0];
	\draw[orange,dashed] (bbr) to[out=100,in=0] (bl);
	\draw[orange] (bbr) to[out=180,in=-80] (bl);
	\fill[orange,opacity=0.5] (bbr) to[in=-80,out=180]  (bl) to[out=0,in=-100] (cr);
	\fill[orange,opacity=0.2] (bbr) to[in=-80,out=180]  (bl) to[out=0,in=100] (bbr);
	\node[navyblue,rotate=90,anchor=south east] at (cl) {out $(\epsilon=+1)$};
	\node[orange,rotate=-90,anchor=south west] at (br) {in $(\epsilon=-1)$};
	\def\del{1.5};
	\coordinate (nexus) at ($(cr)!0.5!(br)$);
	\coordinate (inarr) at ($(nexus)+(\del,0)$);
	\coordinate (fiarr) at ($(inarr)+(2*\del,0)$);
	\draw[very thick,black!70,-Latex] (inarr) -- (fiarr);
	\node[black!70,above,outer sep=4pt] at ($(inarr)!0.5!(fiarr)$) {$\ell\to+\infty$};
	\def\are{3};
	\coordinate (L) at ($(fiarr)+(\del,0)$);
	\coordinate (B) at ($(L)+(\are,-\are)$);
	\coordinate (R) at ($(B)+(\are, \are)$);
	\coordinate (T) at ($(R)-(\are,-\are)$);
	\draw (L) -- (B) -- (R) -- (T) -- cycle;
	\draw[] (L) to[out=-20,in=-160] (R);
	\draw[densely dashed] (L) to[out=20,in=160] (R);
	\fill[orange,opacity=0.5] (L) to[out=20,in=160] (R) -- (B) -- (L);
	\fill[navyblue,opacity=0.5] (L) to[out=-20,in=-160] (R) -- (T) -- (L);
	\node[anchor=south west] at ($(T)!0.5!(R)$) {$\mathscr I^+$};
	\node[anchor=north west] at ($(B)!0.5!(R)$) {$\mathscr I^-$};
	\node[below] at (B) {$i^-$};
	\node[above] at (T) {$i^+$};
	\node[right] at (R) {$i^0$};
	\node[left] at (L) {$i^0$};
	\node[circle,inner sep=1.5pt,fill] at (B) {};
	\node[circle,inner sep=1.5pt,fill] at (T) {};
	\node[circle,inner sep=1.5pt,fill] at (R) {};
	\node[circle,inner sep=1.5pt,fill] at (L) {};
	\node[navyblue,rotate=45,above] at ($(T)!0.5!(L)$) {out $(\epsilon=+1)$};
	\node[orange,rotate=-45,below] at ($(B)!0.5!(L)$) {in $(\epsilon=-1)$};
	\node[anchor=south west,outer sep=10pt] (toparrowL) at ($(cr)!0.3!(tr)$) {};
	\node[anchor=south east,outer sep=10pt] (toparrowR) at ($(T)!0.5!(L)$) {};
	\draw[line width=3pt,navyblue!20,-Latex] (toparrowL) to[out=20,in=145] (toparrowR);
	\node[anchor=north west,outer sep=10pt] (downarrowL) at ($(br)!0.3!(bbr)$) {};
	\node[anchor=north east,outer sep=10pt] (downarrowR) at ($(B)!0.5!(L)$) {};
	\draw[line width=3pt,orange!20,-Latex] (downarrowL) to[out=-20,in=-145] (downarrowR);
	\node[rotate=-45,anchor=south west,red!75!black] at (al) {\scriptsize $r\to+\infty$};
	\node[rotate=-45,anchor=north east,red!75!black] at (ar) {\scriptsize $r\to-\infty\,$};
	\draw[red!75!black,Latex-Latex] (al)node[circle,fill,inner sep=1pt]{} -- (ar)node[circle,fill,inner sep=1pt]{};
	\node[left] at (al) {\scriptsize $(u,z,\bar z)$};
	\node[right] at (ar) {\scriptsize $(u,z,\bar z)$};
	\coordinate (tipr) at ($(L)!0.3!(T)$);
	\coordinate (orir) at ($(B)!0.3!(R)$);
	\draw[red!75!black,Latex-Latex] (tipr)node[circle,fill,inner sep=1pt]{} -- (orir)node[circle,fill,inner sep=1pt]{};
	\node[below,rotate=45] at (orir) {\scriptsize $(u,z,\bar z)$};
	\node[rotate=-45,anchor=south west,red!75!black] at (tipr) {\scriptsize $r\to+\infty$};
	\node[rotate=-45,anchor=north east,red!75!black] at (orir) {\scriptsize $r\to-\infty\,$};
\end{tikzpicture}
\caption{The left figure represents two Poincar\'e patches in AdS: the blue one is covered by Bondi coordinates with $r>0$, the orange one is covered by Bondi coordinates with $r<0$ (see also Figure \ref{Fig:Poincare}). These two patches provide a natural division of the AdS boundary into two regions characterised by $r\to +\infty$ and $r\to -\infty$, respectively. These two regions will be the locus for the insertion of operators creating outgoing ($\epsilon = +1$) and incoming ($\epsilon = -1$) propagators. Notice that this separation is conformally invariant (hence, the number $\epsilon$ is non-ambiguous from the dual CFT perspective). The right figure corresponds to flat space in Bondi coordinates (see also Figure \ref{FigureBondi}). The complete figure summarizes the effect of taking the flat limit $\ell \to \infty$ in the bulk: the boundary of the blue patch is sent to $\mathscr I^+$ and the boundary of the orange region is sent to $\mathscr I^-$. In particular, there is no kinematic restriction before taking the limit (the operators can be inserted anywhere on the two patches). This picture is very different from the one consisting of zooming in the center of AdS (see Figure \ref{fig:centerAdS}).} \label{fig:flatlimit}
\end{figure}
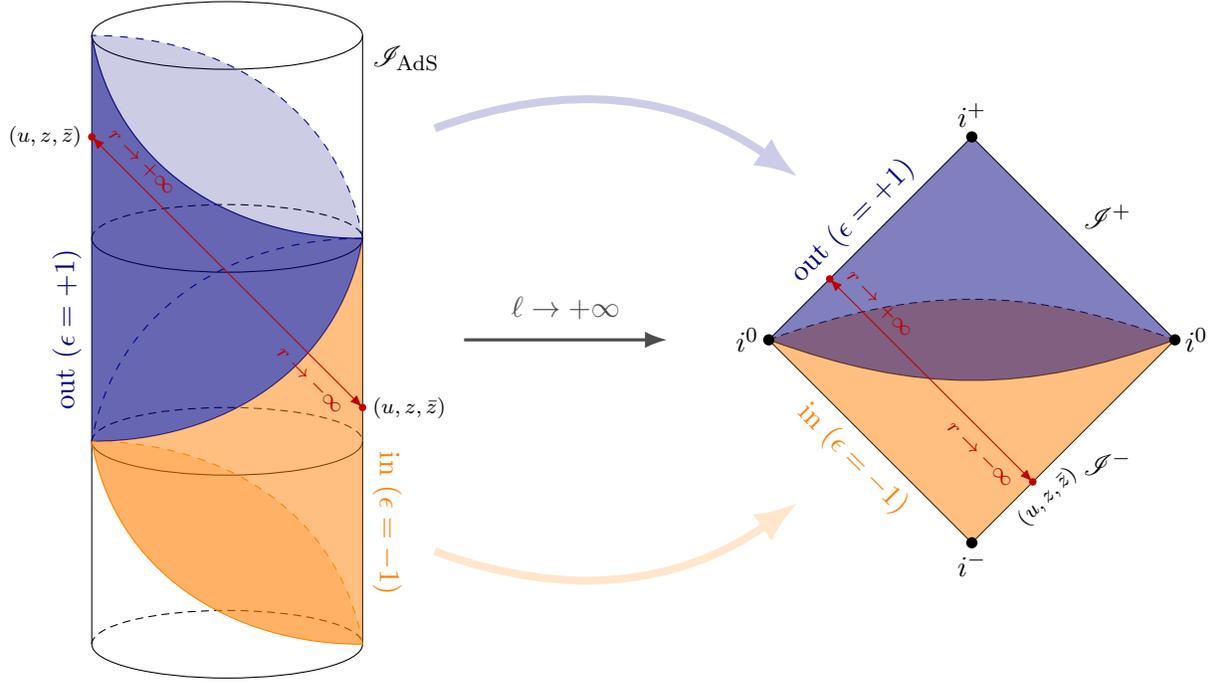

In the previous section, we showed that the massive bulk-to-bulk and bulk-to-boundary propagators reduce respectively to the massless bulk-to-bulk and bulk-to-boundary propagators in Minkowski spacetime in the flat limit $\ell \to \infty$. These propagators in AdS being the building blocks of the Witten diagrams, we are somehow guaranteed that the latter will also have a well-defined limit. In this section, we show that this is indeed the case and that we directly end up on the flat space Carrollian amplitudes expressed in terms of Feynman rules in Section \ref{sec:Feynman diagrams for Carrollian amplitudes}. By contrast with the approach taken recently in \cite{deGioia:2022fcn,deGioia:2023cbd,deGioia:2024yne,Bagchi:2023fbj,Bagchi:2023cen,Penedones:2010ue}, we do not assume any constraint on the inserted operators (compare Figures \ref{fig:centerAdS} and \ref{fig:flatlimit}), which is one of the virtues of Bondi coordinates. We will come back on this in Section \ref{sec:Carrollian limit of boundary CFT correlators} where this limit will be re-interpreted as a Carrollian limit at the boundary of AdS.

Correlators of operators inserted at the boundary of AdS$_4$ can be computed explicitly using the AdS Witten diagrams \cite{Witten:1998qj,Freedman:1998tz,Liu:1998bu,DHoker:1998ecp,DHoker:1998bqu}. By consistency, these correlators satisfy the conformal Ward identities in three dimensions and are therefore naturally interpreted as correlators of a CFT in a bottom-up approach of AdS/CFT. We denote by $\mathcal{O}^\epsilon_\Delta (x)$ the boundary operators corresponding to CFT primaries of spin $s=0$ and conformal dimension $\Delta$. The extra index $\epsilon = \pm 1$ labels operators inserted inside/outside the causal diamond at the boundary determined by the Poincar\'e patch (see Figure \ref{fig:flatlimit}), and will be associated with outgoing/incoming operators.

The two-point correlation function can be computed by starting from $\mathcal{G}_{Bb,+1}^{AdS,\Delta}$ in \eqref{eq:bboundary} and taking $r_2 \to -\infty$. With the appropriate rescaling, we find
\begin{equation}
      \langle \mathcal{O}^{+1}_{\Delta} (x_1) \mathcal{O}^{-1}_{\Delta} (x_2) \rangle = \ell \lim_{r_2 \to - \infty } \left(\frac{r_2}{\ell}\right)^{\D}\mathcal{G}_{Bb,+1}^{AdS,\Delta}(x_1,X_2) = \tilde C_2 (\Delta)  \frac{1}{(-\frac{1}{\ell^2} u_{12} + 2 |z_{12}|^2 - i \varepsilon)^\Delta} \ ,
\label{2ptAdSWitten}
\end{equation} where $\tilde C_2 = (-4)^\Delta \ell^2 C_2 (\Delta)$ with $C_2 (\Delta)$ defined in \eqref{eq:c2def}. The flat limit of this two-point function will be investigated in detail in the next section and re-interpreted as a Carrollian limit at the boundary. The three-point correlator is computed through 
\begin{equation}
    \langle \mathcal{O}^{\epsilon_1}_{\Delta_1} (x_1) \mathcal{O}^{\epsilon_2}_{\Delta_2} (x_2) \mathcal{O}^{\epsilon_3}_{\Delta_3} (x_3) \rangle =\beta_3 \kappa_3  \int_{AdS} d^4X \mathcal{G}^{AdS,\Delta_1}_{Bb,\epsilon_1}(x_1,X) \mathcal{G}^{AdS,\Delta_2}_{Bb,\epsilon_2} (x_2,X)\mathcal{G}^{AdS,\Delta_3}_{Bb,\epsilon_3}(x_3,X) \ ,
\label{3ptWd}
\end{equation} where the conventional factor $\beta_3$ is the same as the one appearing in Equation \eqref{2pointdiagram} (see also Footnote \ref{footnote}). Using Equation \eqref{flat limit Bb}, we see that \eqref{3ptWd} reduces precisely to \eqref{3ptdiagramflat} provided the CFT primaries are related to the Carrollian CFT primaries at $\mathscr I$ defined in Section \ref{sec:Definition and properties} via
\begin{equation}
    \mathcal{O}_{\Delta} (x) \equiv \alpha (\Delta) \partial_u^{\Delta-1}\Phi (x) \ , \qquad  \Delta = m+1 \ , \qquad m = 0, 1, 2, \ldots 
\label{operator scaling}
\end{equation} In the left-hand side, $x= (u,z, \bar z)$ corresponds to a point at the boundary of AdS, $\mathscr{I}_{AdS}$, while in the right-hand side, this corresponds to a point at the boundary of flat space, $\mathscr I$. The identification $\Delta = m+1$ was anticipated in \eqref{flat limit Bb} and shows that the conformal dimension $\Delta$ in AdS becomes redundant in the flat limit as it counts the number of $\partial_u$-derivatives acting on Carrollian primaries. Notice that the conformal Carrollian dimension of $\partial_u^m \Phi^\epsilon$ (which is also a Carrollian primary of weights $(\frac{m+1}{2}, \frac{m+1}{2})$, see below \eqref{carrollian primary} together with \eqref{fiexed Carrollian weights}) is given by $\Delta_c = k + \bar k = m+1$, which coincides with the conformal dimension of the corresponding CFT primary! As we shall explain in the next section, this is consistent with the fact that the dilatation operator in the CFT is not affected by the Carrollian limit induced at the boundary. Moreover, the specific scaling of $\alpha (\Delta)$ in $\ell$ is such that \eqref{operator scaling} corresponds to an electric Carrollian limit \cite{Chen:2021xkw,deBoer:2021jej,Henneaux:2021yzg,Baiguera:2022lsw,Rivera-Betancour:2022lkc,deBoer:2023fnj} at the boundary.

Similarly, the AdS correlator associated with the four-point contact diagram reads as 
\begin{multline}
\label{fourpointAdS}
    \langle \mathcal{O}^{\epsilon_1}_{\Delta_1} (x_1) \mathcal{O}^{\epsilon_2}_{\Delta_2} (x_2) \mathcal{O}^{\epsilon_3}_{\Delta_3} (x_3) \mathcal{O}^{\epsilon_4}_{\Delta_3} (x_4) \rangle_c \\
    =  \beta^c_4 \kappa_4  \int_{AdS} d^4X \mathcal{G}^{AdS,\Delta_1}_{Bb,\epsilon_1}(x_1,X) \mathcal{G}^{AdS,\Delta_2}_{Bb,\epsilon_2} (x_2,X)\mathcal{G}^{AdS,\Delta_3}_{Bb,\epsilon_3}(x_3,X) \mathcal{G}^{AdS,\Delta_4}_{Bb,\epsilon_4}(x_4,X) 
\end{multline} and reduces to \eqref{4pointcontact} after the identification \eqref{flat limit Bb}. Finally, the four-point exchange diagram can be computed as 
\begin{align}
&\langle \mathcal{O}^{\epsilon_1}_{\Delta_1} (x_1) \mathcal{O}^{\epsilon_2}_{\Delta_2} (x_2) \mathcal{O}^{\epsilon_3}_{\Delta_3} (x_3) \mathcal{O}^{\epsilon_4}_{\Delta_4} (x_4) \rangle_e \\ &\qquad= \beta^e_4 \kappa_3^2 \int_{AdS} d^4 X d^4Y \mathcal{G}^{AdS,\Delta_1}_{Bb,\epsilon_1} (x_1;X) \mathcal{G}^{AdS,\Delta_2}_{Bb,\epsilon_2} (x_2;X) \mathcal{G}^{AdS,\Delta}_{BB}(X,Y) \mathcal{G}^{AdS,\Delta_3}_{Bb,\epsilon_3} (x_3;Y) 
    \mathcal{G}^{AdS,\Delta_4}_{Bb,\epsilon_4} (x_4;Y) \nonumber
\end{align} and maps perfectly to the corresponding flat space Carrollian amplitude \eqref{flat space4ptexch} after taking \eqref{flatlimitBB} and \eqref{flat limit Bb} into account, as well as the identification \eqref{operator scaling}.

\section{Carrollian limit of boundary CFT correlators}
\label{sec:Carrollian limit of boundary CFT correlators}

\subsection{General procedure}

The results of Section \ref{sec:Flat limit of AdS Witten diagrams in Bondi coordinates} show that in the flat limit $\ell \to \infty$, the integral representation of boundary correlators in AdS$_4$ directly reduces to the integral representation of Carrollian amplitudes (similar results were obtained in \cite{deGioia:2022fcn,deGioia:2023cbd,deGioia:2024yne,Bagchi:2023fbj,Bagchi:2023cen} using a different prescription for the flat limit, see Figure \ref{fig:centerAdS}). This can be dubbed the ``bulk perspective'' since these formulas make it obvious that the correlators arise from local interactions in a bulk spacetime. It is then natural to wonder how the flat limit would be implemented from a ``boundary perspective''. A closely related question is how the fingerprint of bulk locality is encoded in the boundary correlators. The two- and three-point functions are completely fixed by symmetries and are thus agnostic to the presence or absence of a bulk spacetime. Four-point correlators, on the other hand, display a \textit{bulk point singularity} which is a manifestation of bulk locality \cite{Gary:2009ae,Maldacena:2015iua}. Moreover, the residue on this singularity is related to the flat space scattering amplitude. Here, we take a holographic perspective and ask the question: ``What operation in the dual CFT corresponds to taking the flat limit in the bulk?''. 

As suggested in the previous sections, the answer to this question is related to a Carrollian limit at the level of the boundary correlators. A glance at \eqref{boundary metric AdS} shows that $\frac{1}{\ell}$ plays the role of the speed of light at the boundary. Formally identifying $c \leftrightarrow \frac{1}{\ell}$, we see that the flat limit in the bulk $(\ell \to \infty)$ corresponds to a Carrollian limit $c \to 0$ at the boundary. In this section, we show that the Carrollian limit is dominated by singularities of the correlators. For two- and three-point correlators these are the light-cone singularities and at four points it is the bulk point singularity. The Carrollian limit thus naturally ``zooms in'' on these singularities. Hence, by contrast with most of the previous approaches, these singularities naturally emerge within the Carrollian limit. Furthermore, in the Carrollian limit, where $\mathbb{R}^{1,2} \to \scri$, these singular configurations can be identified with precisely the configurations of points on $\scri$ which allow scattering to occur while respecting momentum conservation. This interpretation is supported by the fact that the leading terms in the Carrollian limit of AdS correlators coincide with Carrollian amplitudes.

We will adopt the following convention for this section. Euclidean (time ordered) correlators will be denoted by $\an{\dots}_E$. Lorentzian correlators will not carry any subscripts. Time-ordered Lorentzian correlators will be denoted by $\an{T\left\lbrace \dots \right\rbrace}$. We will use $\left(c\, t, z, \zb \right)$ as Euclidean coordinates with metric $ds^2_E = c^2 dt^2 + 2 dz d\zb$. Here $t$ is the Euclidean time coordinate which we will analytically continue to the Lorentzian time $u$. The vector containing all the coordinates of a point will be denoted by $x$ with its Lorentzian or Euclidean nature being inferred from context.

\subsection{Two-point function}
\label{sec:Two point function}

The two-point function (of primaries) is non zero only when both operators have the same conformal dimension. For two scalars, each having conformal dimension $\D$, it is given by 
\begin{equation}
\label{eq:2ptfuncE}
    \an{\mo_{\D}\left(x_1\right)\mo_{\D}\left(x_2\right)}_E = \frac{\tilde{C}_2\left(\D\right) }{\left(c^2 t_{12}^2+2|z_{12}|^2\right)^{\D}} \ ,
\end{equation}
where $\tilde{C}_2\left(\D\right) = C_2\left(\D\right)c^{-2} (-4)^{\D}$ with $C_2\left(\D\right)$ defined in \eqref{eq:c2def} and we formally identified $\ell = \frac{1}{c}$. The normalization of the $2$-point function is consistent with our conventions for the bulk-to-bulk and bulk-to-boundary propagators, see Equation \eqref{2ptAdSWitten}. In order to analytically continue to Lorentzian signature \cite{Duffin}, we first choose an Euclidean Wightman function. We then set $t_i = i u_i + \varepsilon_i$ and take the limit $\varepsilon_i \to 0$ assuming $\varepsilon = \varepsilon_1-\varepsilon_2>0$. This leads to two Lorentzian Wightman functions 
\begin{align}
 \label{eq:2ptfuncs}
\an{\mo_{\D}\left(x_1\right)\mo_{\D}\left(x_2\right)} &= \frac{\tilde{C}_2\left(\D\right) }{\left(-c^2 u_{12}^2+2|z_{12}|^2+i \varepsilon u_{12}\right)^{\D}} \ , \\
\an{\mo_{\D}\left(x_2\right)\mo_{\D}\left(x_1\right)} &= \frac{\tilde{C}_2\left(\D\right) }{\left(-c^2 u_{12}^2+2|z_{12}|^2 -i \varepsilon u_{12}\right)^{\D}} \nonumber
\end{align}
and the time-ordered two-point function
\begin{align}
    \label{eq:to2ptfunc}
    \an{T\left\lbrace\mo_{\D}\left(x_1\right)\mo_{\D}\left(x_2\right)\right\rbrace} &\equiv \Theta \left(u_{12}\right)  \an{\mo_{\D}\left(x_1\right)\mo_{\D}\left(x_2\right)} +\Theta \left(-u_{12}\right)  \an{\mo_{\D}\left(x_2\right)\mo_{\D}\left(x_1\right)} \\
     &=  \frac{\tilde{C}_2\left(\D\right)}{\left(-c^2 u_{12}^2 + 2 |z_{12}|^2+i\varepsilon \right)^{\D}} \ . \nonumber
\end{align}
We can now compute the Carrollian limits of the various two-point correlators. Since they only differ by $i \varepsilon$ prescriptions, for concreteness, we will focus on \eqref{eq:to2ptfunc}. In the conventions of Section \ref{sec:AdS Witten diagram}, this would correspond to a configuration where the first boundary operator $\mathcal{O}_\Delta (x_1)$ is inserted inside the causal diamond determined by the Poincar\'e patch (blue region in left figure \ref{fig:flatlimit}, $\epsilon = +1$) while the second operator $\mathcal{O}_\Delta (x_2)$ is inserted outside the causal diamond (orange region in the left figure \ref{fig:flatlimit}, $\epsilon = -1$). We will not write the $\epsilon$ indices explicitly in this section. We start by noting that we can have two qualitatively different behaviors in the Carrollian limit. It is finite if $\left|z_{12}\right|^2 \neq 0$ whereas it diverges as $c^{-2\D}$ when $\left|z_{12}\right|^2 = 0$. We can express this schematically as 
\begin{align}
    \label{eq:EClimit}
   \an{T\left\lbrace\mo_{\D}\left(x_1\right)\mo_{\D}\left(x_2\right)\right\rbrace}\xrightarrow[]{c \to 0} \tilde{C}_2\left(\D\right)  \frac{1}{2^{\D} |z_{12}|^{2\D} } + \frac{1}{c^{\alpha}}g\left(u_{12}, \D\right) \delta^{(2)}\left(z_{12}\right) \ ,
\end{align}
with $g\left(u_{12}, \D\right)$, $\alpha$ to be determined shortly and $\delta^{(2)}\left(z_{12}\right) \equiv \delta\left(z_{12}\right)\delta\left(\zb_{12}\right)$. We would like to emphasize that this is only meant to hold heuristically. The mathematically rigorous statements that follow from this are
\begin{align}
    \label{eq:mathstatements}
    \lim_{c \to 0} c^{\alpha} \an{T\left\lbrace\mo_{\D}\left(x_1\right)\mo_{\D}\left(x_2\right)\right\rbrace} =  
    \begin{cases}
      \frac{  \tilde{C}_2\left(\D\right) }{2^{\D} |z_{12}|^{2\D} } \qquad &\alpha = 0 \ ,\\ 
      \frac{\tilde{C}_2\left(\D\right)}{2\left(\D-1\right)} \frac{\delta^{(2)}\left(z_{12}\right) }{\left(-u_{12}^2+i \varepsilon \right)^{\D-1}}  \qquad &\alpha = 2\D-2 \ .
    \end{cases} 
\end{align}
The function $g\left(u_{12}, \D\right)$ was determined by multiplying both sides of \eqref{eq:EClimit} by $c^{\alpha}$ and integrating $z_1$ over the complex plane. The two-point functions \eqref{eq:mathstatements} are the standard two-point Carrollian CFT correlators in three dimensions: the first is the magnetic branch and the second is the electric branch (see e.g. \cite{Chen:2021xkw,deBoer:2021jej,Henneaux:2021yzg,Baiguera:2022lsw,Rivera-Betancour:2022lkc,deBoer:2023fnj}). Before taking the limit, the configuration $z_{12} = 0$ is timelike. Thus the electric branch can only be reached when the two operators are timelike separated and in this case, it is the \textit{dominant} contribution (for $Re(\D)>0$) as $c \to 0$. Given two timelike-separated points and the constraint $z_{12} = \zb_{12} = 0$ arising in the Carrollian limit, we can construct the following ``$1 \to 1$ scattering'' process in flat space: 
\begin{align}
    &\text{Particle 1: Outgoing at} \left(u_1, z_1, \zb_1\right) \text{with momentum } p_1 = \om \left(1+z_1 \zb_1, z_1+\zb_1, -i (z_1-\zb_1), 1 - z_1 \zb_1 \right) \nonumber .\\ 
    &\text{Particle 2: Incoming at} \left(u_2, z_2, \zb_2\right) \text{with momentum } p_2 = -\om \left(1+z_2 \zb_2, z_2+\zb_2, -i (z_2-\zb_2), 1 - z_2 \zb_2 \right). \nonumber
\end{align}
Both particles have energy $\om >0$ and momentum is conserved ($p_1+p_2 = 0$) on the support of $z_{12}= \zb_{12} = 0$.

This link is buttressed by the observation
\begin{align}
     \lim_{c \to 0} c^{2\D-2} \an{T\left\lbrace\mo_{\D}\left(x_1\right)\mo_{\D}\left(x_2\right)\right\rbrace} =  \frac{8i\pi^2 \tilde{C}_2\left(\D\right)}{\Gamma\left(2\D-1\right)} \mc_2^{\D-1,\D-1} \ ,
\end{align}
where $\mc_2^{\D-1,\D-1}$ is the $2$-point Carrollian amplitude defined in \eqref{twopointCarrollian} with $\kappa_2 = -i\pi$ (see below \eqref{descendantsbB}), $\e_1 = 1$ and $m_1=m_2=\Delta-1$ (by contrast with the CFT case, Carrollian correlators with $m_1 \neq m_2$ can be obtained in flat space by just taking derivatives with respect to $u_1$ and $u_2$, and this is compatible with the Carrollian Ward identities \cite{Donnay:2022wvx}). Based on this, we can define the electric and magnetic Carrollian operators ($\Phi$ and $\Psi$ respectively) as
\begin{equation}
\label{eq:e&mexp}
  \alpha\left(\D\right) \partial_u^{\D-1}  \Phi\left(x\right) =\mo_{\D}\left(x\right) \ , \qquad \Psi_{\D}\left(x\right) = \sqrt{\frac{1}{\tilde{C}_2\left(\D\right)}}\mo_{\D}\left(x\right) \ ,
\end{equation}
with $\alpha\left(\D\right)$ being defined in \eqref{eq:alphadef} (with the formal identification $c\leftrightarrow \frac{1}{\ell}$) and both operators having Carrollian weights $\left(\frac{\D}{2}, \frac{\D}{2}\right)$. This ensures that the conformal dimension of the Carrollian CFT primaries is the same as the dimension of the CFT primaries, which is consistent with the fact that the dilatation operator is not affected by the Carrollian limit. Notice that the definition of the electric Carrollian operator $\Phi(x)$ from the CFT primary $\mathcal{O}_\Delta(x)$ is identical to the one found in \eqref{operator scaling} using the bulk perspective on the flat limit.

The correlation functions of the operators \eqref{eq:e&mexp} are
\begin{align}
    \label{eq:e&mcorrelators}
   &\an{\partial_u^{\D-1}\Phi\left(x_1\right) \partial_u^{\D-1}\Phi\left(x_2\right)} = \mc_2^{\D-1,\D-1} \ , \quad \an{\Psi_{\D}\left(x_1\right) \Psi_{\D}\left(x_2\right)} =  \frac{1}{2^{\D}\left|z_{12}\right|^{2\D}}
\end{align}
and the electric branch reproduces exactly the $2$-point Carrollian amplitude.\footnote{The interpretation of the magnetic branch in the scattering theory in flat space has not yet been completely unveiled, but it is most likely related to soft energy modes, see e.g. \cite{Himwich:2020rro,Pasterski:2021dqe,Freidel:2022skz,Fiorucci:2023lpb} where such time-independent correlation functions appear naturally.} Hence there is a perfect agreement between the flat limit in the bulk and the Carrollian limit at the boundary, upon identifying $c\leftrightarrow \frac{1}{\ell}$. As we will see in the next few sections, the electric rescaling in \eqref{eq:e&mexp} also produces finite $3$- and $4$-point correlators which are related to the Carrollian amplitudes.

\subsection{Three-point function}

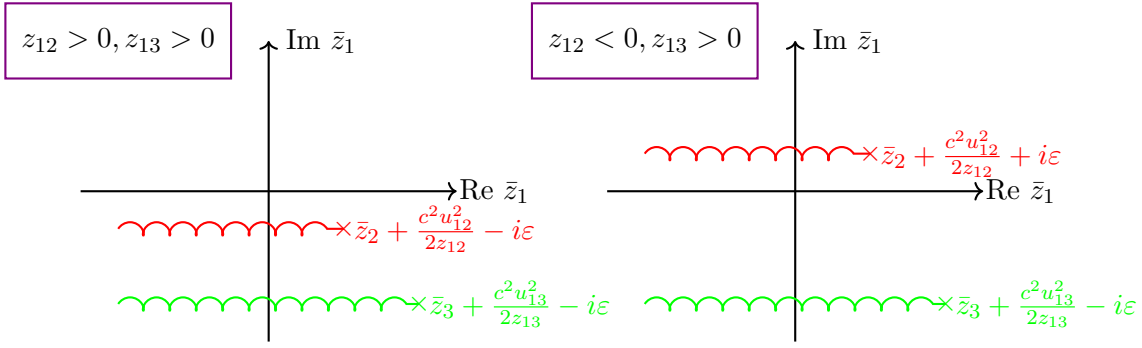
\begin{figure}
    \centering
\begin{tikzpicture}
    % Axes for left plot
    \draw[thick, ->] (-6,0) -- (-1,0);
    \draw[thick, ->] (-3.5,-2) -- (-3.5,2);
    
    % Axes for right plot
    \draw[thick, ->] (1,0) -- (6,0);
    \draw[thick, ->] (3.5,-2) -- (3.5,2);
    
    % Inequalities boxes
    \draw[thick, violet] (-7, 1.5) rectangle (-4, 2.5);
    \node at (-5.5, 2) {${z}_{12} > 0 , {z}_{13} > 0$};
    
    \draw[thick, violet] (0, 1.5) rectangle (3, 2.5);
    \node at (1.5, 2) {${z}_{12} < 0 , {z}_{13} > 0$};
    
    % Horizontal colored lines and labels for the left figure
    \draw[red, thick, decorate, decoration={coil,aspect=0.8}] (-5.5, -0.5) -- (-2.5, -0.5);
    \node[red, right] at (-2.5, -0.5) {$\bar{z}_2 + \frac{c^2 u_{12}^2}{2{z}_{12}} - i\varepsilon$};
    \node[red] at (-2.5, -0.5) {$\times$};
    
    \draw[green, thick, decorate, decoration={coil,aspect=0.8}] (-5.5, -1.5) -- (-1.5, -1.5);
    \node[green, right] at (-1.5, -1.5) {$\bar{z}_3 + \frac{c^2 u_{13}^2}{2{z}_{13}} - i\varepsilon$};
    \node[green] at (-1.5, -1.5) {$\times$};
    
    % Horizontal colored lines and labels for the right figure
    \draw[red, thick, decorate, decoration={coil,aspect=0.8}] (1.5, 0.5) -- (4.5, 0.5);
    \node[red, right] at (4.5, 0.5) {$\bar{z}_2 + \frac{c^2 u_{12}^2}{2{z}_{12}} + i\varepsilon$};
    \node[red] at (4.5, 0.5) {$\times$};
    
    \draw[green, thick, decorate, decoration={coil,aspect=0.8}] (1.5, -1.5) -- (5.5, -1.5);
    \node[green, right] at (5.5, -1.5) {$\bar{z}_3 + \frac{c^2 u_{13}^2}{2{z}_{13}} - i\varepsilon$};
    \node[green] at (5.5, -1.5) {$\times$};
    
    % Axes labels
    \node at (-2.8, 2) {Im $\bar{z}_1$};
    \node at (-0.5, 0) {Re $\bar{z}_1$};
    \node at (6.5, 0) {Re $\bar{z}_1$};
     \node at (4.2, 2) {Im $\bar{z}_1$};

\end{tikzpicture}
\caption{Two of four possible configurations of branch points. The other two are just reflections of the above in the real $\zb_1$ axis.}
    \label{fig:branchpoints}
\end{figure}

The boundary correlator of three scalars in Euclidean AdS${}_4$ interacting via a cubic interaction with coupling constant $\kappa_3$ can be computed explicitly via AdS Witten diagrams (the Euclidean version of \eqref{3ptWd}, see e.g. \cite{Penedones:2010ue}). In our normalization conventions, we have
\begin{equation}
    \label{eq:3ptfuncE}
\an{\mo_{\D_1}\left(x_1\right)\mo_{\D_2}\left(x_2\right)\mo_{\D_3}\left(x_3\right)}_{E} =\frac{\kappa_3}{c}\frac{ 
 \, \mathcal{N}_{\D_1, \D_2, \D_3}}{\left(x_{12}^2\right)^{\D_{12}} \left(x_{23}^2\right)^{\D_{23}} \left(x_{13}^2\right)^{\D_{13}}} \ ,
\end{equation}
where  $x_{ij}^2 = c^2 t_{ij}^2 + 2\left|z_{ij}\right|^2, \, \D_{ij} = \D_i + \D_j - \frac{1}{2}\sum_{k=1}^3 \D_k$  and
\begin{align}
\mathcal{N}_{\D_1, \D_2, \D_3} = \frac{\pi^{\frac{3}{2}}}{2}\Gamma\left(\frac{\sum_{i=1}^3 \D_i -3}{2}\right)\frac{\Gamma\left(\D_{12}\right)\Gamma\left(\D_{23}\right)\Gamma\left(\D_{13}\right)}{\Gamma\left(\D_1\right)\Gamma\left(\D_2\right)\Gamma\left(\D_3\right)} \beta_3 \tilde{C}_2\left(\D_1\right)\tilde{C}_2\left(\D_2\right)\tilde{C}_3\left(\D_1\right) \ .
\end{align}
The factor of $\frac{1}{c}$ follows from the identification $\ell \leftrightarrow \frac{1}{c}$ and $\beta_3$ is the factor that was introduced in \eqref{2pointdiagram}. We will focus on analysing the Carrollian limit of the $3$-point function arising in a $(2,2)$ signature (or Kleinian) spacetime where the Carrollian $3$-point amplitudes are non-trivial. Kleinian correlators can be obtained from Euclidean ones by first analytically continuing to Lorentzian signature following the same procedure outlined for the $2$-point function and then treating $z_i, \zb_i$ as independent variables. For concreteness, we will focus on the Kleinian correlators obtained from the Lorentzian time-ordered correlators which we denote with the subscript $K$
\begin{align}
    \label{eq:3ptklein}
     \an{\mo_{\D_1}\left(x_1\right)\mo_{\D_2}\left(x_2\right)\mo_{\D_3}\left(x_3\right)}_K = \frac{\kappa_3}{c}\frac{ 
 \, \mathcal{N}_{\D_1, \D_2, \D_3}}{\left(x_{12}^2 + i \varepsilon\right)^{\D_{12}} \left(x_{23}^2 + i \varepsilon\right)^{\D_{23}} \left(x_{13}^2+ i \varepsilon\right)^{\D_{13}}} \ .
\end{align}
We will restrict our attention to the Carrollian limit on the support of the configuration $\bar z_{12} = \bar z_{13} = 0$ which is intimately tied to the Carrollian amplitude. We leave an exhaustive analysis of various branches arising in the limit to future work. Heuristically, we have 
\begin{align}
    \label{eq:3pt22structure}
\an{\mo_{\D_1}\left(x_1\right)\mo_{\D_2}\left(x_2\right)\mo_{\D_3}\left(x_3\right)}_K \xrightarrow[]{c \to 0} \frac{\mathcal{G}\left(u_i, z_i\right)}{c^{\alpha}} \delta\left(\zb_{12}\right)\delta\left(\zb_{13}\right) \ .
\end{align}
 We can determine the coefficient by integrating with respect to $\zb_1, \zb_2$.
\begin{align}
\label{eq:22deltacoeff}
    \mathcal{G} &= c^{\alpha} \int_{-\infty}^{\infty} d\zb_1 \int_{-\infty}^{\infty} d\zb_2 \, \an{\mo_{\D_1}\left(x_1\right)\mo_{\D_2}\left(x_2\right)\mo_{\D_3}\left(x_3\right)}_K \nonumber \\
    &= \frac{\kappa_3}{c}\frac{c^{\alpha}\mathcal{N}_{\D_1, \D_2, \D_3}}{\left(2z_{12}\right)^{\D_{12}} \left(2z_{13}\right)^{\D_{13}}}  \int_{-\infty}^{\infty} d\zb_2 \, \frac{I_3^{(\zb_1)}}{\left(x_{23}^2 +i \varepsilon\right)^{\D_{23}}} \ ,
\end{align}
where we have defined
\begin{equation}
    I_3^{(\zb_1)} \equiv  \int_{-\infty}^{\infty} d\zb_1 \frac{1}{\left(\zb_{12}- \frac{c^2 u_{12}^2}{2z_{12}} + i \,\text{sign}\left(z_{12}\right) \varepsilon\right)^{\D_{12}}  \left(\zb_{13}- \frac{c^2 u_{13}^2}{2z_{13}} + i \,\text{sign}\left(z_{13}\right) \varepsilon\right)^{\D_{13}}} \ .
\end{equation}
When $\D_1 \geq 2$, there is no branch point at $\infty$ and we can close the contour either in the upper or lower half planes. The positions of the two branches thus dictate whether or not the integral vanishes. Two of the four possible configurations are shown in Figure \ref{fig:branchpoints}. Clearly, the branch points must lie on either side of the real $\zb_1$ axis for the integral to be non vanishing.

In this case, we can close the contour on either side and deform it to wrap around one of the branch cuts. Assuming $z_{12}<0$, $z_{13}>0$, closing the contour in the upper half plane and deforming it to wrap around the branch cut from $-\infty$ to $\zb_2+\frac{c^2 u_{12}^2}{2z_{12}}$ we get
\begin{align}
     I_3^{(\zb_1)} =\int_{-\infty}^{\zb_2+\frac{c^2 u_{12}^2}{2z_{12}}} d\zb_1 \frac{ \Theta \left(z_{12}z_{31} \right)\left(e^{i \pi \D_{12}} - e^{-i \pi \D_{12}}\right) }{\left|\zb_{12}-\frac{c^2 u_{12}^2}{2z_{12}} \right|^{\D_{12}}  \left(\zb_{13} - \frac{c^2 u_{13}^2}{2z_{13}} + i \,\text{sign}\left(z_{13}\right) \epsilon\right)^{\D_{13}}} \ .
\end{align}
This integral is easily performed by changing variables to $\bar{w} = -\zb_1+\zb_2+\frac{c^2 u_{12}^2}{2z_{12}}$
\begin{align}
    I_3^{(\zb_1)} &=2i\sin{\pi \D_{12}}\Theta \left(z_{12}z_{31} \right)\int_{0}^{\infty} d\bar{w} \frac{1}{\bar{w}^{\D_{12}}  \left(-\bar{w} +\zb_{23} +\frac{c^2}{2}\left(\frac{u_{12}^2}{z_{12}} - \frac{u_{13}^2}{z_{13}}\right) +i \,\text{sign}\left(z_{13}\right) \epsilon\right)^{\D_{13}}}\nonumber \\
 &=\frac{(-1)^{\D_1-1+\D_{13}}2i\sin{\pi \D_{12}}\Theta \left(z_{12}z_{31} \right) B\left(1-\D_{12},\D_1-1\right)}{\left(\zb_{23}+\frac{c^2}{2}\left(\frac{u_{12}^2}{z_{12}} - \frac{u_{13}^2}{z_{13}}\right) +i \,\text{sign}\left(z_{13}\right) \epsilon\right)^{\D_1-1}} \ .
\end{align}
Plugging this back into \eqref{eq:22deltacoeff} and using the identity $\Gamma(x)\Gamma(1-x) = \frac{\pi}{\sin(x)}$ we get
\begin{align}
    \mathcal{G} &= \kappa_3c^{\alpha-1} \left[\frac{2\pi i\mathcal{N}_{\D_1, \D_2, \D_3}(-1)^{\D_{13}+\D_1-1}\Gamma\left(\D_1-1\right)}{\Gamma\left(\D_{12}\right)\Gamma\left(\D_{13}\right)}\right] \left[\frac{\Theta \left(z_{12}z_{31} \right)}{\left(2z_{12}\right)^{\D_{12}} \left(2z_{13}\right)^{\D_{13}}\left(2z_{23}\right)^{\D_{23}}}  \right]\\
    &\nonumber\qquad \times  \int_{-\infty}^{\infty} d\zb_2 \, \frac{1}{\left(-\frac{c^2 u_{23}^2}{2z_{23}}+\zb_{23}+i \text{sign}\left(z_{23}\right)\epsilon\right)^{\D_{23}}} \, \frac{1}{\left(\zb_{23}+\frac{c^2}{2}\left(\frac{u_{12}^2}{z_{12}} - \frac{u_{13}^2}{z_{13}}\right) +i \,\text{sign}\left(z_{13}\right) \epsilon\right)^{\D_1-1}} \ .
\end{align}
The condition for the two branch points of this final integral to be on opposite sides of the real $\zb_2$ axis is $z_{23}z_{13}<0$. Assuming $\D_1+\D_2+\D_3 \geq 6$, closing the contour in the upper half plane and deforming it to wrap around the branch cut from $-\infty$ to $\zb_3 + \frac{c^2 u_{23}^2}{2z_{23}}$ (for $z_{23}<0, z_{13}>0$), we get
\begin{equation}
    \begin{split}
\label{eq:gfinal}
    \mathcal{G} &= \kappa_3c^{\alpha-1} \left[\frac{2\pi i\mathcal{N}_{\D_1, \D_2, \D_3}(-1)^{\D_{13}+\D_{23}\D_1-1}\Gamma\left(\D_1-1\right)}{\Gamma\left(\D_{12}\right)\Gamma\left(\D_{13}\right)}\right] \left[\frac{\Theta \left(z_{12}z_{31} \right)\Theta \left(z_{23}z_{31} \right)}{\left(2z_{12}\right)^{\D_{12}} \left(2z_{13}\right)^{\D_{13}}\left(2z_{23}\right)^{\D_{23}}}  \right]\\
    &\qquad \times  2i \,\sin{\pi \D_{23}}\int_{0}^{\infty} d\zb_2 \, \frac{1}{\zb_2^{\D_1-1}\left(\zb_{2}-\frac{c^2}{2}\frac{\left(u_1 z_{23}+u_2 z_{31}+u_3 z_{12}\right)^2}{z_{12}z_{23}z_{13}} +i \,\text{sign}\left(z_{23}\right) \epsilon\right)^{\D_{23}}}\\
    & = \frac{\kappa_3\pi^{2} \mathcal{N}_{\D_1, \D_2, \D_3} \Gamma\left(-2+\frac{\sum_{i=1}^3 \D_i}{2}\right)\Theta \left(z_{12}z_{31} \right)\Theta\left(z_{13}z_{23}\right)\left(z_{12}\right)^{\D_{3}-2} \left(z_{13}\right)^{\D_{2}-2}\left(z_{23}\right)^{\D_{1}-2}}{c^{-\alpha-3+\sum_{i=1}^3 \D_i} \Gamma\left(\D_{12}\right)\Gamma\left(\D_{23}\right)\Gamma\left(\D_{13}\right)\left(u_1 z_{23}+u_2 z_{31}+u_3 z_{12}+i \varepsilon \text{sign}z_{23}\right)^{-4+\sum_{i=1}^3 \D_i}} \ .
    \end{split}
\end{equation}
In arriving at the second equality, we have used the identity $\Gamma(x)\Gamma(1-x) = \frac{\pi}{\sin(x)}$. We must have $\alpha = \sum_{i=1}^3 \D_i -3$ for the limit to be finite and non-zero. Finally, the Legendre duplication formula for the Gamma function implies
\begin{align}
   \frac{ \mathcal{N}_{\D_1, \D_2, \D_3} \Gamma\left(-2+\frac{\sum_{i=1}^3 \D_i}{2}\right)}{\Gamma\left(\D_{12}\right)\Gamma\left(\D_{23}\right)\Gamma\left(\D_{13}\right)} =\left(2\pi\right)^4 \beta_3
 \Gamma\left(-4+\sum_{i=1}^4 \D_i \right)\prod_{i=1}^3 \frac{\tilde{C}_2\left(\D_i\right)}{2^{\D_i}\Gamma\left(\D_i\right)} \ .
\end{align}
Plugging in the expression for $\mathcal{G}$ from \eqref{eq:gfinal} into \eqref{eq:3pt22structure} and using the value $\alpha = \sum_{i=1}^3 \D_i - 3$ determined above, we arrive at the following compact formula for the Carrollian limit:
\begin{equation}
    \begin{split}
    \lim_{c \to 0}\frac{\an{\mo_{\D_1}\left(x_1\right)\mo_{\D_2}\left(x_2\right)\mo_{\D_3}\left(x_3\right)}_K }{\alpha\left(\D_1\right)\alpha\left(\D_2\right)\alpha\left(\D_3\right)} &= \an{\partial_u^{\D_1-1}\Phi\left(x_1\right) \partial_u^{\D_2-1}\Phi\left(x_2\right) \partial_u^{\D_3-1}\Phi\left(x_3\right)}_K \\
   & = \mc_3^{\D_1-1,\D_2-1,\D_3-1} \ .
    \end{split}
\end{equation}
In the first equality, we have recognised the rescaled operators as the electric Carrollian operators defined in \eqref{eq:e&mexp} and in the second equality we have expressed $\mathcal{G}$ in terms of the three-point Carrollian amplitude  \eqref{eq:threeptcarr} with $\e_1 = -\e_2 = -\e_3 = 1$. Notice the perfect agreement of the numerical factors in the limit, resulting from a consistent choice of conventions between AdS and flat space. It would be interesting to obtain correlators of the magnetic Carrollian operators as well as mixed ones as limits of Lorentzian/Kleinian operators. We leave a systematic analysis of all possible limits to future work while also referring the reader to \cite{deGioia:2024yne, Bagchi:2023fbj, Bagchi:2023cen} for some results in Lorentzian signature.

\subsection{Four-point function}

Four-point functions are not fixed by conformal symmetry and contain dynamical information about the theory. We are interested in analysing the Carrollian limits of various CFT$_3$ four-point functions arising from local interaction in an AdS${}_4$ bulk. We consider both contact and scalar exchange interactions.

\subsubsection{Euclidean signature expressions}

\paragraph{Contact diagrams} We consider a $\phi^4$ interaction in the AdS bulk with coupling $\kappa_4$. The four-point Euclidean correlator is given by 
\begin{align}
\label{eq:4ptgencorrelator}
\an{\mo_{\D_1}\left(x_1\right)\mo_{\D_2}\left(x_2\right)\mo_{\D_3}\left(x_3\right)\mo_{\D_4}\left(x_4\right)}_E^c =\kappa_4 \beta_4^c  \int_{AdS_4} d^4X  \, \prod_{i=1}^4 \, G^{\D_i}_{\partial B}\left(x_i, X\right) \ .
\end{align}
The superscript $c$ serves as a reminder that we are analysing a contact interaction. The factor $\beta_4^c$ is introduced by consistency with our conventions in \eqref{fourpointAdS}. The integral expression \eqref{eq:4ptgencorrelator} cannot be expressed in terms of elementary functions for arbitrary values of $\D_i$ and is usually taken to be the definition of the $D$-function (see e.g. \cite{Bissi:2022mrs} for a comprehensive overview). Hence we write
\begin{align}
    \label{eq:Dfunc}
\an{\mo_{\D_1}\left(x_1\right)\mo_{\D_2}\left(x_2\right)\mo_{\D_3}\left(x_3\right)\mo_{\D_4}\left(x_4\right)}_E^c &\equiv \kappa_4 \beta_4^c D_{\D_1, \D_2, \D_3, \D_4} \left(x_i\right)\\
&\nonumber \equiv \kappa_4 \beta_4^c \mathcal{N}_4 \frac{\left(x_{14}^2\right)^{\frac{1}{2}\Sigma_{\D}-\D_1-\D_4}\left(x_{34}^2\right)^{\frac{1}{2}\Sigma_{\D}-\D_3-\D_4}}{\left(x_{13}^2\right)^{\frac{1}{2}\Sigma_{\D}-\D_4}\left(x_{24}^2\right)^{\D_2}}\bar{D}_{\D_1, \D_2, \D_3, \D_4}\left(U,V\right) \ .
\end{align}
In the second equality above, we have expressed the $D$-function in terms of another one which is just a function of the conformal cross ratios. The definitions of all the quantities entering the above equation are 
\begin{align}
\label{eq:4ptbasicdefs}
    \mathcal{N}_4 = \frac{\pi^{\frac{3}{2}}}{2}\beta_4^c\Gamma\left(\frac{\Sigma_{D} -3}{2}\right)\prod_{i=1}^4\frac{\tilde{C}_2\left(\D_i\right)}{\Gamma\left(\D_i\right)} \ , \qquad \Sigma_{\D} = \sum_{i=1}^4 \D_i \ , \qquad U = \frac{x_{12}^2x_{34}^2}{x_{13}^2 x_{24}^2} \ , \qquad V = \frac{x_{23}^2x_{14}^2}{x_{13}^2x_{24}^2} \ .
\end{align}
However, for special values of $\D_i$, the $D$-function can be expressed in terms of elementary functions. For instance, we have
\begin{align}
     &\bar{D}_{1,1,1,1} \left(U, V\right) = \frac{1}{Z-\Zb}\left[2\,\text{Li}_2(Z) - 2\, \text{Li}_2(\Zb) +\log Z \Zb \log \left(\frac{1-Z}{1-\Zb}\right)\right] \label{eq:dbar1111} \ ,\\
     &\bar{D}_{1,1,1,2} \left(U, V\right) = \frac{\pi^{\frac{3}{2}}}{1-\sqrt{U}-\sqrt{V}} \label{eq:dbar1112} \ .
\end{align}
$\bar{D}_{1,1,1,1}$ is expressed in terms of variables $Z, \Zb$ which are defined by 
\begin{align}
    \label{eq:ZZbdefs}
    U = Z \Zb \ , \qquad V = (1-Z)(1-\Zb) \ .
\end{align}
The derivation of \eqref{eq:dbar1111} can be found in \cite{Denner:1991qq, Usyukina:1992jd} and we present a derivation of \eqref{eq:dbar1112} in Appendix \ref{sec:dbar1112}. The $\bar{D}$-functions with different values of conformal dimensions satisfy well-known recursion relations. Some of these are 
\begin{align}
\label{eq:dbarrecursions}
    &\bar{D}_{\D_1+1,\D_2+1,\D_3,\D_4} = -\partial_{U}\bar{D}_{\D_1,\D_2,\D_3,\D_4} = \left[\frac{1-Z}{Z-\Zb}\partial_{Z} -\frac{1-Z}{Z-\Zb}\partial_{\Zb}\right]\bar{D}_{\D_1,\D_2,\D_3,\D_4} \ , \\
    &\bar{D}_{\Delta_{1}, \Delta_{2}+1, \Delta_{3}+1, \Delta_{4}}=-\partial_{V} \bar{D}_{\Delta_{1}, \Delta_{2}, \Delta_{3}, \Delta_{4}}= -\left[\frac{\bar{Z} \partial_{\bar{Z}}-Z \partial_{Z}}{Z-\bar{Z}}  \right]\bar{D}_{\Delta_{1}, \Delta_{2}, \Delta_{3}, \Delta_{4}} \ , \\
    &\bar{D}_{\D_1, \D_2, \D_3+1, \D_4+1} = \left(\D_3+\D_4-\frac{1}{2}\Sigma_{\D} -U \partial_U \right) \bar{D}_{\D_1, \D_2,\D_3,\D_4} \ , \\
    &\hspace{3cm} = \left(\D_3+\D_4-\frac{1}{2}\Sigma_{\D} +Z\Zb \left[\frac{1-Z}{Z-\Zb}\partial_{Z} -\frac{1-Z}{Z-\Zb}\partial_{\Zb}\right] \right) \bar{D}_{\D_1, \D_2,\D_3,\D_4}\nonumber \ .
\end{align}
These can be used to compute
\begin{align}
    \label{eq:dbar2222}
    \bar{D}_{2,2,2,2} &= \partial_U \bar{D}_{1,1,1,1} + U \partial_U^2 \bar{D}_{1,1,1,1}\\
    &= \frac{1}{\left(Z-\Zb\right)^4}\left[\mathcal{P}_1\bar{D}_{1,1,1,1} +\mathcal{P}_2 \log Z \Zb +\mathcal{P}_3 \log (1-Z)(1-\Zb)\right]+\frac{2}{\left(Z-\Zb\right)^2} \ .
\end{align}
Here $\mathcal{P}_i$ are polynomials in $Z, \Zb$ which can be easily computed.

\paragraph{Exchange diagrams} The four-point correlator arising from a scalar exchange in the bulk is 
\begin{align}
\an{\mo_{\D_1}\left(x_1\right)\mo_{\D_2}\left(x_2\right)\mo_{\D_3}\left(x_3\right)\mo_{\D_4}\left(x_4\right)}_E^e =I_s + I_t + I_u \ ,
\end{align}
where $I_s$ corresponds to the $s$-channel exchange with the exchanged operator having conformal dimension $\D$,
\begin{align}
    I_s = \kappa_3^2 \beta_4^e  \int_{AdS_4} d^4X \, d^4Y  \, \prod_{i=1}^2 \, G^{\D_i}_{\partial B}\left(x_i, X\right)  \, \prod_{i=3}^4 \, G^{\D_i}_{\partial B}\left(x_i, Y\right) G^{\D}_{BB}\left(X,Y\right) \ ,
\end{align}
with $\beta_{4}^e$ being the factor introduced in \eqref{flat space4ptexch} and $I_t, I_u$ are given by the replacements $1 \leftrightarrow 3$ and $1 \leftrightarrow 4$ respectively. Exchange diagrams cannot be expressed in terms of elementary functions for generic values of conformal dimensions. However, they can be expressed as a finite sum of contact diagrams if $-\D+\D_3+\D_4 \in 2\mathbb{Z}^+$ when we have 
\begin{align}
\label{eq:finitesumexchange}
    I_s &= \kappa_3^2 \beta_4^e\sum_{k=k_{\text{min}}}^{k_{\text{max}}} a_k \left(x_{34}^2\right)^{k-\D_4} D_{\D_1, \D_2, \D_3-\D_4+k,k} \\
    &\nonumber = \frac{\left(x_{14}^2\right)^{\frac{1}{2}\Sigma_{\D}-\D_1-\D_4}\left(x_{34}^2\right)^{\frac{1}{2}\Sigma_{\D}-\D_3-\D_4}}{\left(x_{13}^2\right)^{\frac{1}{2}\Sigma_{\D}-\D_4}\left(x_{24}^2\right)^{\D_2}} \sum_{k=k_{\text{min}}}^{k_{\text{max}}} a_k \mathcal{N}'_4 \bar{D}_{\D_1, \D_2, \D_3-\D_4+k,k} \ ,
\end{align}
where $k_{\text{min}} = \frac{1}{2}\left(\D-\D_3+\D_4\right)$, $k_{\text{max}} = \D_4-1$, and
\begin{align}
    \mathcal{N}'_4 = \frac{\kappa_3^2 \beta_4^e \pi^{\frac{3}{2}} \ell^2 \Gamma\left(\frac{\Sigma_{D} -3}{2}+k-\D_4\right) \tilde{C}_2 \left(\D_1\right) \tilde{C}_2 \left(\D_2\right) \tilde{C}_2 \left(k+\D_3-\D_4\right) \tilde{C}_2 \left(k\right)}{2 \Gamma\left(\D_1\right) \Gamma\left(\D_2\right) \Gamma\left(k+\D_3-\D_4\right) \Gamma\left(k\right)} \ .
\end{align}
$\Sigma_D$ has the same meaning as in \eqref{eq:4ptbasicdefs} and $a_k$ satisfies the recursion relation\footnote{The value of $a_{k_{max}}$ differs from that of \cite{Bissi:2022mrs} due to differing normalizations for bulk-bulk and bulk-boundary propagators}
\begin{align}
    a_{k-1} = \frac{\left(k-\frac{\D}{2}+\frac{\D_3-\D_4}{2}\right)\left(k-\frac{3}{2}+\frac{\D}{2}+\frac{\D_3-\D_4}{2}\right)}{(k-1)(k-1+\D_3-\D_4)}a_k \ , \qquad a_{k_{\text{max}}} = \frac{1}{4\left(\D_3-\frac{3}{2}\right)\left(\D_4-\frac{3}{2}\right)} \ .
\end{align}
One case of interest is $\D_1 = \D_2 = \D_3 = \D_4 = \D = 2$, for which we have
\begin{align}
\label{eq:exchange2222s}
    I_s = \frac{\kappa_3^2 \beta^e_4 \ell^2}{4\pi^6\left(x_{13}^2\right)^2\left(x^2_{24}\right)^2} \bar{D}_{2,2,1,1} \ .
\end{align}
$I_u$ is given by permuting the labels $2, 3$ everywhere in \eqref{eq:exchange2222s} and $I_t$ by permuting $2, 4$.

\subsubsection{Analytic continuation to Lorentzian signature}
\label{sec:analyticcont}

Now we would like to analytically continue the $4$-point functions to Lorentzian signature. The procedure is very similar to those followed in \cite{Maldacena:2015iua, Gary:2009ae, Jain:2023fxc, Heemskerk:2009pn}. We review it here with minor modifications and highlight all the features necessary for understanding the Carrollian limit. For the sake of concreteness, we will illustrate this procedure with the simplest contact diagram,
\begin{equation}
    \label{eq:4ptall1}
\an{\mo_{1}\left(x_1\right)\mo_{1}\left(x_2\right)\mo_{1}\left(x_3\right)\mo_{1}\left(x_4\right)}_E^c = \frac{\pi^2}{2 x_{13}^2 x_{24}^2}\bar{D}_{1,1,1,1} \ ,
\end{equation}
with $\bar{D}_{1,1,1,1}$ being defined in \eqref{eq:dbar1111}. We can view it as a function defined on a Riemann surface with branch points at $Z, \Zb =0,1,\infty$. We want to analytically continue from the Euclidean sheet (on which $Z, \Zb$ are complex conjugates) to a different sheet by traversing the branch points in a specific way. This is accomplished by the following procedure.  
\begin{itemize}
    \item Start from a configuration in which all points are spacelike separated. By this we mean that the values of the coordinates $\left(t_i, z_i, \zb_i \right)$ are chosen such that $-c^2 t_{ij}^2+2\left|z_{ij}\right|^2 >0, \,  \forall i,j$.
    \item In Euclidean space, only time ordered correlators are well-defined. For concreteness, let us suppose $t_1<t_2<t_3<t_4$. We can analytically continue to Lorentzian time by setting $t_i = \varepsilon_i + i u_i$ and taking $\varepsilon_i$ to zero while respecting the ordering $\varepsilon_1<\varepsilon_2<\varepsilon_3<\varepsilon_4$. The resulting $i \varepsilon$ prescriptions completely define the analytic continuation to any Lorentzian configuration. This is the Euclidean sheet, $Z, \Zb$ are still complex conjugates and \eqref{eq:4ptall1} is single valued. 
    \item We now continuously deform the points to reach the desired causal configuration. This together with the $i\varepsilon$ prescriptions defines the path by $Z, \Zb$ to obtain the final configuration which might be on a different sheet of the Riemann surface. 
    \item An analysis of the path taken gives us a list of the branch points encircled and the monodromies that must be computed. 
\end{itemize}
The branch points encircled depend on the final configuration. We will analyse two distinct classes of configurations - \textit{scattering}  and \textit{non-scattering} configurations. As we shall explain in Section \ref{sec:contactcarrollian}, the scattering configuration will admit a well-defined and non-vanishing Carrollian limit for the electric falloffs in \eqref{eq:e&mexp},  which will lead to the Carrollian amplitudes described in Section \ref{sec:Definition and properties}. By contrast, the non-scattering configurations will not contribute in the electric Carrollian limit.

\begin{figure}[htb!]
    \centering
    \includegraphics[scale=0.8]{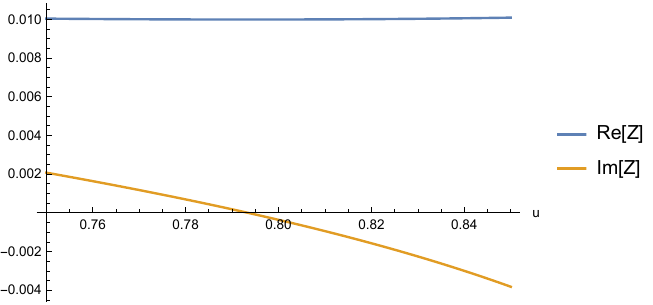}
     \includegraphics[scale=0.5]{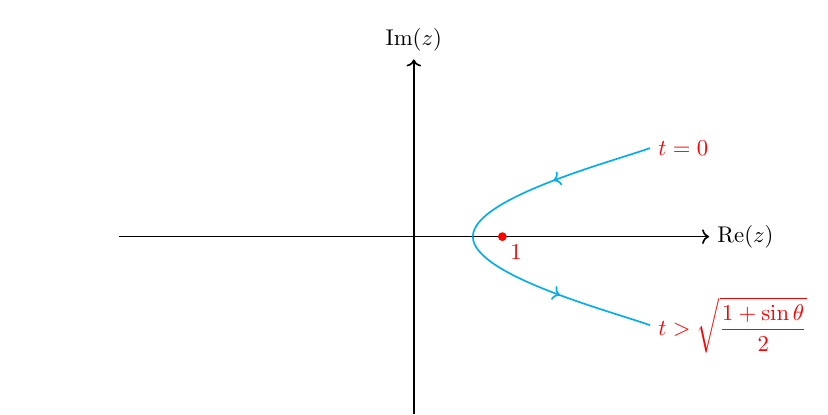}
    \caption{The figure on the left shows a plot of the real and imaginary parts of $Z$ for $\varepsilon = 10^{-5}, \theta = \frac{\pi}{12}$. The vertical red line is at $u = \sqrt{\frac{1+\sin \frac{\pi}{12}}{2}}$. Note that the imaginary part changes sign here indicating that the branch point is encircled. The figure on the right is a sketch of the trajectory of $Z$. }
    \label{fig:schannelz}
\end{figure}

\paragraph{Scattering configurations} These consist of two pairs of timelike separated points. This includes $x_1, x_2$ in the past of $x_3, x_4$ and other permutations. We can parametrize the path to this configuration from an initial spacelike one  by
\begin{align}
\label{eq:startconfig}
    x_1 = \left(-c \,t, 0, 1\right) \ , \quad x_2 = \left(-c\, t, 0, -1\right) \ , \quad x_3 = \left(c \,t, \cos \theta, \sin \theta\right) \ , \quad x_4 = \left(c \,t,  -\cos \theta, -\sin \theta \right) \ .
\end{align}
The spatial components of $x_i$ above are given in Cartesian coordinates and not in stereographic coordinates $z, \zb$. The initial configuration is at $t=0$ when all points are spacelike separated. The analytic continuation to Lorentzian time is given by $t = i u + \varepsilon$. In the process of moving $x_3, x_4$ to the future of $x_1, x_2$, we must necessarily cross the light cones $x_{13}^2=0$, $x_{23}^2 =0$, $x_{14}^2=0$, $x_{24}^2 = 0$. After analytic continuation, these are located at
\begin{align}
\label{eq:lightcones}
    &x_{13}^2 = x_{24}^2 = 2+4c^2t^2-2 \sin \theta = 2- 4c^2u^2-2 \sin \theta + i \varepsilon = 0 \implies c\, u = \pm\sqrt{\frac{1-\sin \theta}{2}} \pm i \varepsilon \ ,\\
    &\nonumber x_{14}^2 = x_{23}^2 = 2+4c^2t^2+2 \sin \theta = 2- 4c^2u^2+2 \sin \theta + i \varepsilon = 0  \implies c\, u =\pm \sqrt{\frac{1+\sin \theta}{2}} \pm i \varepsilon \ .
\end{align}
Next, $Z, \Zb$ are given in this parametrization by 
\begin{align}
    &Z = \frac{2-4 \cos \theta \sqrt{-c^2t^2(1+c^2t^2)}+2(-1-2c^2t^2)\sin \theta}{\left(-1-2c^2t^2+\sin \theta \right)^2} \ , \\
    &\Zb = \frac{2+4 \cos \theta \sqrt{-c^2t^2(1+c^2t^2)}-2(1+2c^2t^2)\sin \theta}{\left(-1-2c^2t^2+\sin \theta \right)^2} \ .\nonumber
\end{align}
The analytic continuation to Lorentzian signature along with the correct $i\varepsilon$ prescriptions is once again given by setting $t = \varepsilon + i u$.
\begin{figure}
    \centering
    \includegraphics[scale=0.8]{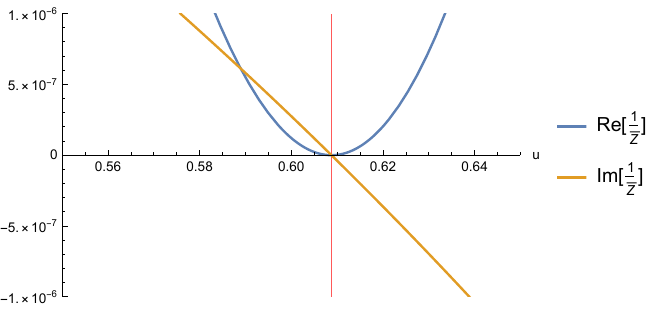}
     \includegraphics[scale=0.5]{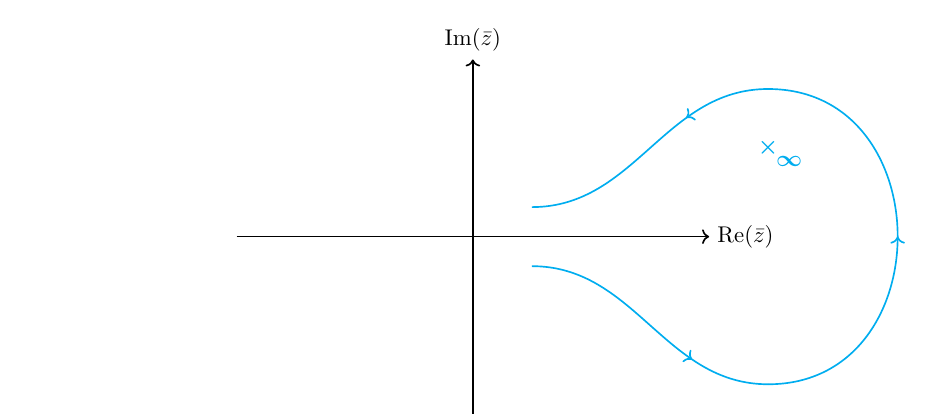}
    \caption{The figure on the left shows a plot of the real and imaginary parts of $\frac{1}{\Zb}$ for $\varepsilon = 10^{-5}, \theta = \frac{\pi}{12}$. The vertical red line is at $u = \sqrt{\frac{1-\sin \frac{\pi}{12}}{2}}$. Note that the imaginary part changes sign here indicating that the branch point is encircled. The figure on the right is a sketch of the trajectory of $\Zb$.}
    \label{fig:schannelzb}
\end{figure}
Near the points \eqref{eq:lightcones}, $Z, \Zb$ behave as
\begin{align}
&u \to \sqrt{\frac{1-\sin \theta}{2}} \ , \qquad Z \to \sec^2 \theta \ , \qquad \Zb \to \infty \implies \frac{1}{\Zb} \to 0 \ , \\
&u \to \sqrt{\frac{1+\sin \theta}{2}} \ , \qquad Z \to 1 \ , \qquad \Zb \to \csc^2 \theta \ . 
\end{align}
In the first case, $Z$ does not approach a branch point and $\Zb$ approaches $\infty$, while in the second case, $Z$ approaches $1$ and $\Zb$ does not approach a branch point. Figures \ref{fig:schannelzb} and \ref{fig:schannelz} show that the branch points at $\Zb =\infty$ and at $Z=1$ are traversed in a counterclockwise manner.
Monodromies of the polylogarithms and $\bar{D}_{1,1,1,1}$ around all branch points are well known (see e.g. \cite{Bourjaily:2020wvq}). Traversing $Z=1$ counterclockwise results in
\begin{equation}
\label{eq:Z1monodromy}
   \log \left(1-Z\right)  \to \log \left(1-Z\right)+2\pi i \ , \qquad \text{Li}_2 \left(Z\right)  \to \text{Li}_2 \left(Z\right) - 2\pi i\, \log Z \ , 
\end{equation}
while traversing $\Zb=\infty$ counterclockwise has the effect\footnote{We can use the identities $\text{Li}_2 \left(\Zb\right) = \text{Li}_2 \left(\frac{1}{\Zb}\right) -\frac{\pi^2}{6}-\frac{1}{2}\log^2 \Zb$, $\log \Zb = \log \frac{\frac{1}{\Zb}-1}{\frac{1}{\Zb}}$ to derive the monodromy around $\Zb=\infty$ using the monodromy around $\Zb=0$.}
\begin{align}
\label{eq:Zbinfmonodromy}
   &\log \left(\Zb\right)  \to \log \left(\Zb\right) - 2\pi i \ , \quad \log\left(1-\Zb\right) \to \log\left(1-\Zb\right)-2\pi i \ ,\\
   &\qquad\qquad\qquad\text{Li}_2 \left(\Zb\right)  \to \text{Li}_2 \left(\Zb\right) + 2\pi i\, \log \Zb+2\pi^2 \ .\nonumber
\end{align}
Putting all of this together, we get 
\begin{align}
    \label{eq:4ptl22config}
\bar{D}_{1,1,1,1} \to \bar{D}_{1,1,1,1} +  \frac{4\pi^2}{Z-\Zb} +\frac{2\pi i}{Z-\Zb} \log \frac{1-\Zb}{1-Z} \ .
\end{align}
There is now a singularity at $Z=\Zb$ and with the most singular term in an expansion around $Z=\Zb$ being $\hat{\Phi}_{1111}^{l s} = \frac{4\pi^2}{Z-\Zb}$. This is the bulk point singularity \cite{Maldacena:2015iua} and is absent in the Euclidean correlator. We stress that this analytic continuation is not unique. A different choice of path to reach the same final configuration can yield a different result. However, the presence of the singularity after analytic continuation is a universal feature. These formulas allow us to analytically continue other $\bar{D}$-functions. In particular, we have 
\begin{align}
    \label{eq:d2222l}
    \bar{D}_{2,2,2,2} \to \bar{D}_{2,2,2,2} + \frac{4 \pi^2 \mathcal{P}_1}{\left(Z-\Zb\right)^5} -\frac{2\pi i}{\left(Z-\Zb\right)^5}\left[\mathcal{P}_2 \left(Z-\Zb\right)+\mathcal{P}_1 \log \frac{1-Z}{1-\Zb}\right] \ .
\end{align}
This analytic continuation is also singular at $Z=\Zb$ with leading singularity $\hat{\Phi}_{2222}^{l s} = \frac{48\pi^2 \Zb^2 \left(1-\Zb\right)^2}{\left(Z-\Zb\right)^5}$. In arriving at this expression, we have used the explicit form of the polynomial $\mathcal{P}_1$ on the support of $Z=\Zb$. Finally, the other two scattering configurations can be reached by permuting the labels on the $x_i$ in \eqref{eq:startconfig}.  

\begin{figure}
    \centering
    \includegraphics[scale=0.7]{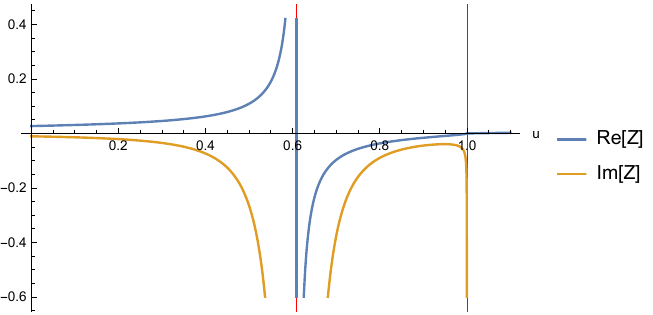}
    \includegraphics[scale=0.7]{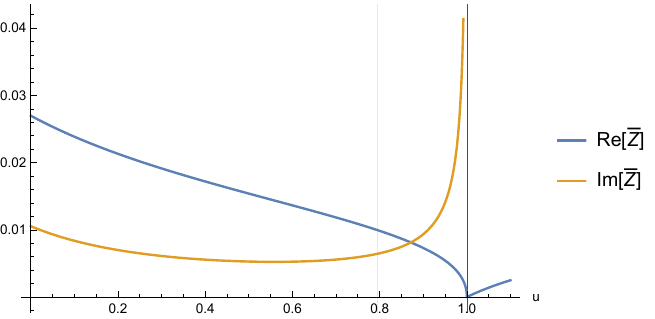}
    \caption{Plots of the real and imaginary parts of $Z$ (left) and $\Zb$ (right) for $\varepsilon = 10^{-5}, \theta = \frac{\pi}{12}$. The axes have been rescaled to highlight important features. The vertical red lines in the left figure are at $u=\sqrt{\frac{1-\sin \frac{\pi}{12}}{2}}, 1$ and for the right figure they are at $u = \sqrt{\frac{1+\sin \frac{\pi}{12}}{2}}$ and $u=1$. Note in particular that the imaginary part always remains negative.  }
    \label{fig:nonscat}
\end{figure}

\paragraph{Non-scattering configurations} These are configurations with only one point timelike separated from the rest. We will focus on the scenario in which $x_4$ is in the future of $x_1, x_2$ and $x_3$ which can be reached by starting from
\begin{align}
\label{eq:startconfig2}
    x_1 = \left(-t, 0, 1\right) \ , \quad x_2 = \left(-t, 0, -1\right) \ , \quad x_3 = \left(-t, \cos \theta, \sin \theta\right) \ , \quad x_4 = \left(t,  -\cos \theta, -\sin \theta \right) \ .
\end{align}
$x_4$ can be moved to the future of $x_1, x_2, x_3$ by crossing the branch cuts at  $x_{14}^2 =0$, $x_{24}^2 =0$, $x_{34}^2 = 0$, which occurs at $t = \frac{\sqrt{1-\sin \theta}}{\sqrt{2}}$, $t = \frac{\sqrt{1+\sin \theta}}{\sqrt{2}}$ and $t =1$. In this process, $Z, \Zb$ (after setting $t=\varepsilon + i u$) approach:
\begin{align}
   &u \to \sqrt{\frac{1-\sin \theta}{2}} \ , \qquad Z \to \infty \implies \frac{1}{Z} \to 0 \ , \qquad \Zb \to \frac{1}{1-\sin \theta} \ , \\
&u \to \sqrt{\frac{1+\sin \theta}{2}} \ , \qquad Z \to -\csc\theta \ , \qquad \Zb \to 1 \ , \nonumber\\
&u \to 1 \ , \qquad Z \to 0 \ , \qquad \Zb \to 0 \ . \nonumber 
\end{align}
However, as shown in Figure \ref{fig:nonscat}, the sign of the imaginary parts never changes. Since no branch points are encircled in this process, we do not pick up any monodromies and the function remains unchanged. In particular, it is not singular as $Z \to \Zb$.

\subsubsection{Leading singularities of contact diagrams} 
\label{sec:ls}

We have just seen that contact diagrams are singular as $Z \to \Zb$ after analytic continuation to scattering configuration. We would like to compute the leading singularity for general contact diagrams with arbitrary $\Delta_{i}$. This will turn out to be useful for obtaining the leading term in the Carrollian limit in Section \ref{sec:contactcarrollian}. These will have the form
\begin{equation}
\label{eq:lscontact}
\bar{D}_{\D_1, \D_2, \D_3, \D_4} \xrightarrow[]{Z \to \Zb}\hat{\Phi}_{\Delta_{1} \Delta_{2} \Delta_{3} \Delta_{4}}^{l s}=\frac{f_{\Delta_{1} \Delta_{2} \Delta_{3} \Delta_{4}}(Z)}{(Z-\bar{Z})^{\Sigma_{\Delta}-3}} \ ,
\end{equation}
with $\Sigma_{\D}$ being defined in \eqref{eq:4ptbasicdefs}. Note that this is the leading singular term of $\bar{D}_{\D_1, \D_2, \D_3, \D_4}$ and we can obtain the leading singularity of the four-point contact diagram from \eqref{eq:Dfunc}. Contact diagrams (both Euclidean and Lorentzian) satisfy well-known relations \eqref{eq:dbarrecursions}  which can be easily derived from the representation \eqref{eq:dbarnewrep}. From these, we can derive relations for the leading singularities. More precisely
\begin{align}
\hat{\Phi}_{\Delta_{1}+1, \Delta_{2}+1, \Delta_{3}, \Delta_{4}}^{l s}=\mathcal{C}_{1100} &\hat{\Phi}_{\Delta_{1}, \Delta_{2}, \Delta_{3}, \Delta_{4}}^{l s} \ , \qquad \hat{\Phi}_{\Delta_{1}, \Delta_{2}+1, \Delta_{3}+1, \Delta_{4}}^{l s}=\mathcal{C}_{0110} \hat{\Phi}_{\Delta_{1}, \Delta_{2}, \Delta_{3}, \Delta_{4}}^{l_{s}} \nonumber \ , \\
& \label{eq:lsrecursions}\hat{\Phi}_{\Delta_{1}, \Delta_{2}, \Delta_{3}+1, \Delta_{4}+1}^{l s}=\mathcal{C}_{0011} \hat{\Phi}_{\Delta_{1}, \Delta_{2}, \Delta_{3}, \Delta_{4}}^{l_{s}} \ ,
\end{align}
and so on, where $\mathcal{C}_{\text{mnpq}}$ can be derived by first recalling that the derivatives $\partial_{U}, \partial_{V}$ can be rewritten in terms of $\partial_{Z}, \partial_{\bar{Z}}$ as
\begin{equation}
\partial_{U}=\frac{(Z-1) \partial_{Z}+(1-\bar{Z}) \partial_{\bar{Z}}}{Z-\bar{Z}} \ , \quad \partial_{V}=\frac{\bar{Z} \partial_{\bar{Z}}-Z \partial_{Z}}{Z-\bar{Z}} \ ,
\end{equation}
by keeping in the operators \eqref{eq:dbarrecursions} only terms that raise the power of divergences as much as possible. When acting on \eqref{eq:lscontact} this leads to the following replacements
\begin{equation}
\partial_{U} \rightarrow \frac{2(1-Z)}{Z-\bar{Z}} \partial_{\bar{Z}} \ , \quad \partial_{V} \rightarrow \frac{2 Z}{Z-\bar{Z}} \partial_{\bar{Z}} \ .
\end{equation}
Furthermore, constant terms can be ignored, as they do not raise the singularity. All in all, we find
\begin{gather}
\label{eq:raisingops}
\mathcal{C}_{1100}=-\frac{2(1-Z)}{Z-\bar{Z}} \partial_{\bar{Z}} \ , \quad \mathcal{C}_{0110}=-\frac{2 Z}{Z-\bar{Z}} \partial_{\bar{Z}} \ , \quad \mathcal{C}_{0011}=-\frac{2 Z^{2}(1-Z)}{Z-\bar{Z}} \partial_{\bar{Z}} \ , \\
\mathcal{C}_{1001}=-\frac{2(1-Z)^{2} Z}{Z-\bar{Z}} \partial_{\bar{Z}} \ , \quad \mathcal{C}_{0101}=\frac{2 Z(1-Z)}{Z-\bar{Z}} \partial_{\bar{Z}} \ , \quad \mathcal{C}_{1010}=\frac{2 Z(1-Z)}{Z-\bar{Z}} \partial_{\bar{Z}} \nonumber \ .
\end{gather}
From here on, we will restrict the $\D_i$ to only take integer values and denote these by $n_i$ to emphasize this. The operators \eqref{eq:raisingops} allow us to find families of leading singularities, for instance
\begin{equation}
\hat{\Phi}_{\D, \D, 1,1}^{l s}=-\pi^{3 / 2}(-4)^{\D} \Gamma\left(\D-\frac{1}{2}\right) \frac{(1-Z)^{\D-1}}{(Z-\bar{Z})^{2 \D-1}} .
\end{equation}
In general, we obtain
\begin{equation}
\label{eq:lsgen}
\hat{\Phi}_{\D_{1} \D_{2} \D_{3} \D_{4}}^{l s}=\mathcal{K}_\D \frac{Z^{\D_{3}+\D_{4}-2}(1-Z)^{\D_{1}+\D_{4}-2}}{(Z-\bar{Z})^{\Sigma_{\Delta}-3}}, \qquad \mathcal{K}_\D = \pi^{3 / 2} 2^{\Sigma_{\Delta}-2} \Gamma\left(\frac{\Sigma_{\Delta}-3}{2}\right)(-1)^{\D_1+\D_3} \ .
\end{equation}
This was derived first assuming $\Sigma_{\Delta}=$ even, but the prefactor is adjusted such that this result agrees for any integer $\D_{i}$. Finally, note that we can compute
\begin{equation}
\left(\frac{1}{Z-\bar{Z}} \partial_{\bar{Z}}\right)^{n} \frac{1}{(Z-\bar{Z})^{\alpha}}=\frac{2^{n} \Gamma\left(n+\frac{\alpha}{2}\right)}{\Gamma\left(\frac{\alpha}{2}\right)} \frac{1}{(Z-\bar{Z})^{\alpha+2 n}} \ .
\end{equation}
This allows us to introduce
\begin{equation}
\left(\frac{1}{Z-\bar{Z}} \partial_{\bar{Z}}\right)^{1 / 2} \frac{1}{(Z-\bar{Z})^{\alpha}}\equiv \frac{\sqrt{2} \Gamma\left(\frac{1}{2}+\frac{\alpha}{2}\right)}{\Gamma\left(\frac{\alpha}{2}\right)} \frac{1}{(Z-\bar{Z})^{\alpha+1}} \ ,
\end{equation}
so that we can define an operator $\mathcal{C}\equiv\left(\frac{1}{Z-\bar{Z}} \partial_{\bar{Z}}\right)^{1 / 2}$ whose action is well-defined on leading singularities. In terms of this, the operators that raise an index are given by
\begin{equation}
\mathcal{C}_{1}=s_{1}(1-Z) \sqrt{2} \mathcal{C} \ , \quad \mathcal{C}_{2}=s_{2} \sqrt{2} \mathcal{C} \ , \quad \mathcal{C}_{3}=s_{3} Z \sqrt{2} \mathcal{C} \ , \quad \mathcal{C}_{4}=s_{4} Z(1-Z) \sqrt{2} \mathcal{C} \ ,
\end{equation}
such that the composition $\mathcal{C}_{1} \mathcal{C}_{2}=\mathcal{C}_{2} \mathcal{C}_{1}$ raises the first two indices, and so on. This gives
\begin{equation}
s_{1} s_{2}=-1, s_{1} s_{4}=-1, s_{2} s_{3}=-1, s_{2} s_{4}=1, s_{3} s_{4}=-1, s_{1} s_{3}=1 \ .
\end{equation}
Comparing with the leading singularity for $\hat{\Phi}_{1112}$ we see $s_{4}=1$ and hence $s_{1}=s_{3}=-1, s_{2}=$ $s_{4}=1$.

\subsubsection{Carrollian limit of contact diagrams}
\label{sec:contactcarrollian}
We are now in a position to discuss the Carrollian limits of four-point correlators corresponding to contact interactions. Once again the behaviour of the Carrollian limit when the boundary operators are inserted at special configurations will be starkly different from when they are at generic positions. The collinear configurations $z_{ij} = \zb_{ij} = 0$ could potentially yield interesting Carrollian limits. However, motivated by the fact that the four-point Carrollian amplitude \eqref{eq:4ptcontactcarr} has a $\delta$ function singularity on the circle $z=\zb$ (here $z, \zb$ are two dimensional cross ratios and not to be confused with $Z, \Zb$), we will restrict our attention to this configuration. We postpone an analysis of the other possibilities and their interpretations to future work. 
We are interested in obtaining the leading term as $c \to 0$ on the support of $z=\zb$. To this end, we note that 
\begin{align}
\label{eq:ZZbc0exp}
    \left(Z-\Zb\right)^2 &= \left((1+U-V)^2-4 U\right)\xrightarrow[]{c \to 0} \left(z-\zb\right)^2 + 2c^2 \,\mathcal{F} + \mo \left(c^4\right) \ .
\end{align}
On the support of $z-\zb=0$, we have $\left(Z-\Zb\right)^2 \propto c^2$ implying that the leading term as $c\to 0$ on the support of $z-\zb=0$ can be obtained from the leading term as $Z \to \Zb$. This implies that the \textit{scattering configurations} are precisely the ones which are dominant in this limit and we can utilize the leading singularities computed in Section \ref{sec:ls}. The explicit form of $\mathcal{F}$ is
\begin{align}
\label{eq:den}
   \mathcal{F} = \left[ (z+\zb)\left( U_1-V_1 \right)-2 \, U_1 \right] \ ,
\end{align}
where
\begin{align}
    \frac{2U_1}{\left|z\right|^2} = \left(\frac{u_{13}^2}{\left|z_{13}\right|^2}+\frac{u_{24}^2}{\left|z_{24}\right|^2}-\frac{u_{34}^2}{\left|z_{34}\right|^2}-\frac{u_{12}^2}{\left|z_{12}\right|^2}\right), \,\, \frac{2V_1}{\left|1-z\right|^2} = \left(\frac{u_{13}^2}{\left|z_{13}\right|^2}+\frac{u_{24}^2}{\left|z_{24}\right|^2}-\frac{u_{14}^2}{\left|z_{14}\right|^2}-\frac{u_{23}^2}{\left|z_{23}\right|^2}\right)
\end{align}
are the subleading terms in the $c \to 0$ expansion of the 4d cross ratios defined in \eqref{eq:4ptbasicdefs}.
We can further simplify \eqref{eq:den} on the support of $z=\zb$ to
\begin{align}
\label{eq:carrollden}
  \mathcal{F}\left|_{z=\zb} \right. = \frac{\left|z\right|^2 \left|z_{23}\right|^2}{\left|z_{34}\right|^2\left|z_{24}\right|^2} \left(u_4 - z \left|\frac{z_{24}}{z_{12}}\right|^2 u_1 +\frac{1-z}{z}\left|\frac{z_{34}}{z_{23}}\right|^2 u_2 - \frac{1}{1-z}\left|\frac{z_{14}}{z_{13}}\right|^2 u_3 \right)^2,
\end{align}
which is the denominator of the four point Carrollian amplitudes \eqref{eq:4ptcontactcarr} and \eqref{eq:4ptexchangecarr}. The task of computing the Carrollian limit of the analytic continuation of \eqref{eq:Dfunc} has now been reduced to that of computing the same of the leading singularity \eqref{eq:lsgen}. We start by making the ansatz
\begin{align}
    \label{eq:4ptcarransatz}
   \lim_{c \to 0} c^{\alpha} \hat{\Phi}^{ls}_{\D_1\D_2\D_3\D_4} =  \mathcal{R} \delta\left(z-\zb\right) .
\end{align}
In order to compute $\mathcal{R}$, we must evaluate the integral of $\hat{\Phi}^{ls}_{\D_1\D_2\D_3\D_4}$ over $\zb$,
\begin{align}
   \mathcal{R} =c^{\alpha} \int_{-\infty}^{\infty} d\zb\, \hat{\Phi}^{ls}_{\D_1\D_2\D_3\D_4} = \mathcal{K}_{\D} c^{\alpha} \left(1-z\right)^{\D_1+\D_4-2}z^{\D_3+\D_4-2}  \int_{-\infty}^{\infty} d\zb\, \frac{1}{\left(\left(z-\zb\right)^2+2c^2 \mathcal{F}\right)^{\Sigma_{\D}-3}} \ ,
\end{align}
where we have dropped the $\mo\left(c^4\right)$ terms in \eqref{eq:ZZbc0exp}. The integral is ostensibly ill-defined since there could be singularities on the real line. However, both singularities are complex. To see this, we first recall that since $\mathcal{F}$ depends on $z,\zb$, the equation $ \left(z-\zb\right)^2+2c^2 \mathcal{F} = 0$ is difficult to solve exactly. However perturbative (in $c$) solutions are easily found. They are given by
\begin{align}
    \label{eq:denroots}
    \zb = z \pm c \sqrt{-2 \mathcal{F}\left|_{z=\zb}\right.} \ .
\end{align}
\eqref{eq:carrollden} shows that $\mathcal{F}\left|_{z=\zb}\right. \geq 0$. The roots are thus complex conjugate pairs and the integral is well-defined and always non-zero. It is readily performed by closing the contour in the upper half plane and then deforming it to wrap around the branch cut along the negative real axis. Assuming $\Sigma_{\D} >4$, this yields
\begin{align}
    \mathcal{R}&= c^{\alpha} \mathcal{K}_{\D} \left(1-z\right)^{\D_1+\D_4-2}z^{\D_3+\D_4-2}  \int_{-\infty}^{z+ i c \sqrt{2\mathcal{F}\left|_{z=\zb}\right.}}\frac{ d\zb}{\left(\left(z-\zb\right)^2+2c^2 \mathcal{F}\right)^{\Sigma_{\D}-3}}  \\
    &\nonumber = c^{4-\Sigma_{\D}+\alpha} \mathcal{K} \,\left(\frac{\left|z_{23}\right|^2}{\left|z_{34}\right|^2\left|z_{24}\right|^2}\right)^{\frac{4-\Sigma_{\D}}{2}} \frac{z^{2-\D_1-\D_2}\left(1-z\right)^{\D_1+\D_4-2}}{\left(u_4 - z \left|\frac{z_{24}}{z_{12}}\right|^2 u_1 +\frac{1-z}{z}\left|\frac{z_{34}}{z_{23}}\right|^2 u_2 - \frac{1}{1-z}\left|\frac{z_{14}}{z_{13}}\right|^2 u_3 \right)^{\Sigma_{\D}-4}}  \ ,
\end{align}
where $\mathcal{K} =\mathcal{K}_{\D} \frac{2^{\frac{4-\Sigma_{\D}}{2}}}{\sqrt{\pi}}\cos \frac{\Sigma_{\D}\pi}{2}\Gamma\left(\frac{\Sigma_{\D}-4}{2}\right)\Gamma\left(\frac{5-\Sigma_{\D}}{2}\right)$ and we must have $\alpha = \Sigma_{\D}-4$ for the limit to be finite and non-zero. Finally, recalling that $\hat{\Phi}^{ls}_{\D_1\D_2\D_3\D_4}$ is the leading singularity of $\bar{D}_{\D_1, \D_2, \D_3, \D_4}$ and the actual contact diagram is related by a prefactor given by \eqref{eq:Dfunc}, it is now straightforward to compute the Carrollian limit of the Lorentzian correlator. After some simplification, it reads
\begin{align}
\label{eq:climit4ptfinal}
   &\lim_{c \to 0} \frac{\an{T\left\lbrace\mo_{1}\left(x_1\right)\mo_{1}\left(x_2\right)\mo_{1}\left(x_3\right)\mo_{1}\left(x_4\right)\right\rbrace}^c}{  \alpha\left(\D_1\right) \alpha\left(\D_2\right) \alpha\left(\D_3\right) \alpha\left(\D_4\right)} = \an{\partial_{u_1}^{\D_1-1}\Phi\left(x_1\right)\partial_{u_2}^{\D_2-1}\Phi\left(x_2\right)\partial_{u_3}^{\D_3-1}\Phi\left(x_3\right)\partial_{u_4}^{\D_4-1}\Phi\left(x_4\right)}^c\nonumber \\
   &=\kappa_4 \frac{(-1)^{\D_1+\D_3}}{(2\pi)^4}  \,\left(\frac{\left|z_{23}\right|^2}{\left|z_{34}\right|^2\left|z_{24}\right|^2}\right)^{\frac{4-\Sigma_{\D}}{2}} \frac{z^{2-\D_1-\D_2}\left(1-z\right)^{\D_1+\D_4-2} \delta\left(z-\zb\right)}{\left(u_4 - z \left|\frac{z_{24}}{z_{12}}\right|^2 u_1 +\frac{1-z}{z}\left|\frac{z_{34}}{z_{23}}\right|^2 u_2 - \frac{1}{1-z}\left|\frac{z_{14}}{z_{13}}\right|^2 u_3 \right)^{\Sigma_{\D}-4}} \ ,
\end{align}
where we have once again recognised the rescaled operators as electric Carrollian operators \eqref{eq:e&mcorrelators}. It is easy to check that the kinematic factors are in perfect agreement with those of \eqref{eq:4ptcontactcarr} on the support of $\delta\left(z-\zb\right)$. At first glance, it might appear that \eqref{eq:climit4ptfinal} is missing the $\Theta$ functions found in the Carrollian amplitude \eqref{eq:4ptcontactcarr}. However, we note that the leading singularity is the result of a specific analytic continuation where
\begin{align}
    Z \xrightarrow[]{c\to 0} z = \frac{2}{1-\sin \theta} \geq 1 \ ,
\end{align}
which corresponds to setting $\e_1 = \e_2 = -\e_3 = -\e_4 = 1$ in \eqref{eq:4ptcontactcarr}. The other in/out configurations can be obtained by performing the analytic continuation of Section \ref{sec:analyticcont} simply switching points $2\leftrightarrow 3$ and $2 \leftrightarrow 4$. The resulting leading singularity is identical. However, in the Carrollian limit,
\begin{equation}
    Z \xrightarrow[u \text{ channel}]{c \to 0} z_u = \frac{1-\sin \theta}{2}, \qquad Z \xrightarrow[t \text{ channel}]{c \to 0} z_t = -\frac{1+\sin \theta}{1-\sin \theta} \ ,
\end{equation}
and we have $0\leq z_u \leq 1$ and $z_t \leq 0$. Finally, the lack of a singularity when only one point is timelike separated from the others implies that configurations with 3 incoming and 1 outgoing (or vice versa) have a vanishing Carrollian limit. This is once again in agreement with the $\Theta$ functions in \eqref{eq:4ptcontactcarr}. Thus, we can write 
\begin{align}
\lim_{c \to 0} \frac{\an{T\left\lbrace\mo_{1}\left(x_1\right)\mo_{1}\left(x_2\right)\mo_{1}\left(x_3\right)\mo_{1}\left(x_4\right)\right\rbrace}^c}{  \alpha\left(\D_1\right) \alpha\left(\D_2\right) \alpha\left(\D_3\right) \alpha\left(\D_4\right)} &= \an{\partial_{u_1}^{\D_1-1}\Phi\left(x_1\right)\partial_{u_2}^{\D_2-1}\Phi\left(x_2\right)\partial_{u_3}^{\D_3-1}\Phi\left(x_3\right)\partial_{u_4}^{\D_4-1}\Phi\left(x_4\right)}^c\nonumber \\
   &=  \mc^{\D_1-1, \D_2-1, \D_3-1, \D_4-1}_{4,c} \ .
\end{align}
Hence, as in the three-point case, the limit reproduces exactly the Carrollian four-point function with the right normalization.

\subsubsection{Carrollian limit of exchange diagrams}
 It is apparent from \eqref{eq:bblimit} that the bulk-to-bulk propagator in AdS reduces to the Feynman propagator in flat space regardless of the value of $\D$ -- the conformal dimension of the exchanged operator. This property lets us adopt the following strategy to analyse the Carrollian limit of an arbitrary scalar exchange diagram. Focusing on the $s$-channel exchange for concreteness, for any value of the conformal dimension of the exchanged operator ($\D$), we choose the conformal dimensions of the external operators ($\D_i$) such that $\D_3+\D_4-\D \in 2\mathbb{Z}^+$. This ensures that the exchange diagram reduces to a finite sum of contact diagrams, see Equation  \eqref{eq:finitesumexchange}. For such configurations of $\D_i$, the problem of computing the Carrollian limit of the exchange diagram is reduced to that of computing the same limit of contact diagrams -- a problem solved in complete generality in Section \ref{sec:contactcarrollian}. We will find that the result agrees with the corresponding Carrollian amplitude. Furthermore, this result will be analytic in $\D_i $ and independent of $\D$. This allows us to relax the constraints on $\D_i$ thereby extending the result to generic exchange diagrams.

The computation of the  Carrollian limit of a finite sum of contact diagrams is guided by the logic laid out around \eqref{eq:ZZbc0exp} which implies that we focus on the leading term in an expansion around $Z = \Zb$. As $Z \to \Zb$ we have,
 \begin{align}
     I_s \xrightarrow[]{Z \to \Zb} \frac{\left(x_{14}^2\right)^{\frac{1}{2}\Sigma_{\D}-\D_1-\D_4}\left(x_{34}^2\right)^{\frac{1}{2}\Sigma_{\D}-\D_3-\D_4}}{\left(x_{13}^2\right)^{\frac{1}{2}\Sigma_{\D}-\D_4}\left(x_{24}^2\right)^{\D_2}} \sum_{k=k_{\text{min}}}^{k_{\text{max}}} a_k\, \mathcal{N}'_4\, \Phi^{ls}_{\D_1, \D_2, \D_3-\D_4+k,k} \ .
 \end{align}
 The leading singularity of $I_s$ is the term in the sum with $k = k_{max} = \D_4-1$ since $\Phi_{\D_1, \D_2, \D_3, \D_4}^{ls} \propto \left(Z-\Zb\right)^{3-\Sigma_{D}}$. The computation of the Carrollian limit proceeds exactly as before and we get 
 \begin{align}
     \lim_{c\to 0}\prod_{i=1}^4 \frac{1}{\alpha\left(\D_i\right)} I_s = \mc_{4,e}^{m_1, m_2, m_3, m_4} \ ,
 \end{align} where $\mc_{4,e}^{m_1, m_2, m_3, m_4}$ was defined in \eqref{eq:4ptexchangecarr}. This result is independent of $\Delta$ and analytic in $\D_i$ allowing us to relax the constraint on them thereby demonstrating that the Carrollian limit of a scalar exchange diagram is the corresponding Carrollian amplitude in full generality.

\section{Discussion}

In this work, we presented a flat limit procedure allowing to relate holographic correlators in AdS and Carrollian amplitudes in flat space. We focused on the scalar subsector of the AdS/CFT correspondence in four dimensions. The limit was taken in Bondi gauge, which allowed us to implement the Carrollian limit at the boundary of AdS. Importantly, we showed that the limit could be taken either at the level of the AdS Witten diagrams or directly at the level of the boundary correlators, yielding a perfect correspondence between flat limit and Carrollian limit via the identification $c_{boundary} = \frac{1}{\ell_{bulk}}$. In particular, the well-known kinematic constraints relevant for the flat limit, such as the bulk-point singularity, emerge naturally as a consequence of our way of taking the limit. This analysis opens the avenue to a systematic procedure to take the flat limit of AdS/CFT and obtain a top-down proposal for flat space holography. We conclude with comments and perspectives. 

In the flat limit considered in this paper, the conformal dimension does not scale with $\ell$. One consequence of this is that the AdS bulk-to-bulk propagator reduces to the massless propagator in flat space regardless of the value of its conformal dimension thus reducing any AdS exchange Witten diagram to the corresponding massless exchange diagram in flat space. It would be interesting to interpret this massless exchange from a Carrollian perspective and in particular understand which Carrollian operators are exchanged. This would amount to performing the analogue of a conformal block decomposition for this correlator and understanding the relation of this decomposition to the more standard 3d conformal block decomposition. We also hope to analyse this along with the flat limit in which $\D$ does scale with $\ell$ and its connection to Carrollian CFT in future work.

Let us emphasize that the Carrollian limit implemented in Section \ref{sec:Carrollian limit of boundary CFT correlators} on the CFT correlators offers a systematic procedure to construct Carrollian CFTs with time-dependent correlators. Hence we expect that these techniques might be useful beyond the context of holography in the study of Carrollian CFTs.

Here we restricted the analysis to scalar correlators. However, we believe that our flat limit procedure can be easily extended to correlators involving spinning particles such as those in \cite{Costa:2011mg, Costa:2014kfa}. Indeed, the expressions for Carrollian amplitudes are also known for gluons and gravitons \cite{Mason:2023mti, Banerjee:2019prz}. We leave this analysis for future work. 

It would be interesting to understand how the soft sector of scattering amplitudes in flat space, as well as the associated symmetries, emerge in the flat limit. In particular, it would be worth understanding in our set-up how the $Lw_{1+\infty}$ algebra uncovered from the collinear limit of scattering amplitudes in \cite{Guevara:2021abz,Strominger:2021mtt} are related to the AdS deformations studied in \cite{Taylor:2023ajd,Bittleston:2024rqe}.

\paragraph{Acknowledgements} We thank Kristan Jensen and Andreas Karch for useful correspondence and illuminating discussions. We also thank Shiraz Minwalla, Sabrina Pasterski, and Ana-Maria Raclariu for useful comments. We are grateful to Geoffrey Compère, Adrien Fiorucci, Lionel Mason, Marios Petropoulos for fruitful exchanges and collaborations on related topics, and Adrien Fiorucci again for his precious advice in the design of some of the figures. The work of LFA is supported by the European
Research Council (ERC) under the European Union’s Horizon 2020 research and innovation
programme (grant agreement No 787185). LFA is also supported in part by the STFC grant
ST/T000864/1. RR is supported by the Titchmarsh Research Fellowship at the Mathematical Institute and by the Walker Early Career Fellowship in Mathematical Physics at Balliol College. AYS is supported by the STFC grant ST/X000761/1. This work was supported by the Simons Collaboration on Celestial Holography.

\appendix

\section{Propagators in flat space}
\label{sec:Propagators in flat space}

In this appendix, we provide more details about the regulators of the Feynmann propagator in flat space and the derivation of the bulk-to-boundary propagator. We first check that the Fourier representation of the Feynman propagator \eqref{fourierBulktoBulk} is compatible with its position space expression \eqref{bulktobulkflat}, with the right normalization. For timelike-separated points, we have
\begin{equation}
\begin{split}
     \mathcal{G}_{BB}^{Flat}(X_1,X_2)   &= -\int \frac{d^4p}{\left(2\pi\right)^4} \frac{e^{-i p \cdot X_{12}}}{p^2 - i \varepsilon_p} = \int \frac{d^4p}{\left(2\pi\right)^4} \frac{e^{-i p \cdot X_{12}}}{\left(p^0 -\left|\vec{p}\right|+i\varepsilon_p\right)\left(p^0+\left|\vec{p}\right|-i\varepsilon_p\right)}\\
     &= -i \int \frac{d^3\vec{p}}{\left(2\pi\right)^3} \left[\Theta\left(X_{12}^0 \right)\frac{e^{-i X_{12}^0 \left|\vec{p}\right|-i \vec{p}\cdot \vec{X}_{12}}}{2\left|\vec{p}\right|}+\Theta\left(-X_{12}^0 \right)\frac{e^{i X_{12}^0 \left|\vec{p}\right|-i \vec{p}\cdot \vec{X}_{12}}}{2\left|\vec{p}\right|} \right] \\
    &= -i \int \frac{d^3\vec{p}}{\left(2\pi\right)^3} \frac{e^{-i |X_{12}^0| \left|\vec{p}\right|-i \vec{p} \cdot \vec{X}_{12}}}{2\left|\vec{p}\right|}  \ .
    \end{split}
    \label{interm}
\end{equation}
We have set the momentum space regulator $\varepsilon_p = 0$ after performing the $p^0$ integral. Working in the frame $\vec{X}_{12} = 0$, we get 
\begin{align}
\label{eq:topropfinal}
    \mathcal{G}_{BB}^{Flat}(X_1,X_2)   &=\frac{-i}{\left(2\pi\right)^2} \int_0^{+\infty} d\left|\vec{p}\right|  \left|\vec{p}\right| e^{-i \left|X_{12}^0\right| \left|\vec{p}\right|-\varepsilon \, \left|\vec{p}\right|} = \frac{-i}{\left(2\pi\right)^2} \frac{1}{\left(i \left|X_{12}^0\right|+\varepsilon\right)^2} = \frac{-i}{\left(2\pi\right)^2} \frac{1}{X_{12}^2+i\varepsilon} \ .
\end{align}
In the first step, we have introduced a regulator $\varepsilon$ in order to make the integral converge at $+\infty$. In the final expression, we have restored manifest Lorentz invariance and the result is valid for spacelike and null configurations as well. Note that this is exactly \eqref{bulktobulkflat}, obtained by analytically continuing the Euclidean Feynman propagator. It is instructive to repeat this computation in Bondi coordinates. Starting from \eqref{interm} and using the parametrization \eqref{Retarded flat BMS coordinates} and \eqref{q in terms of z bar z future}, we find 
\begin{align}
    \mathcal{G}_{BB}^{Flat}(X_1,X_2) =&-i \int \frac{\om \, d\om\, dw\, d\bar{w}}{\left(2\pi\right)^3} \left[\Theta\left(-X_{12}^0 \right)e^{-\varepsilon \om + i \om\left(u_{12}+r_1 \left|z_1-w\right|^2-r_2 \left|z_2-w\right|^2\right)}\right. \\
    &\left. \qquad +\Theta\left(X_{12}^0 \right)e^{-\varepsilon \om -i \om\left(u_{12} \left|w\right|^2+r_1\left|1+\bar{w}z_1\right|^2-r_2\left|1+\bar{w}z_2\right|^2\right)}\right]\nonumber\\
     &\hspace{-3.6cm}=\frac{-i}{\left(2\pi\right)^2} \left[\frac{\Theta\left(-X_{12}^0 \right)}{-2u_{12}r_{12}+2r_1 r_2\left|z_{12}\right|^2-i \varepsilon r_{12}}+\frac{\Theta\left(X_{12}^0 \right)}{-2u_{12}r_{12}+2r_1 r_2\left|z_{12}\right|^2+i \varepsilon \left(u_{12}+r_1\left|z_1\right|^2-r_2\left|z_2\right|^2 \right)}\right]\nonumber \ .
\end{align}
From here, we can use the following properties
\begin{equation}
    \begin{split}
\label{eq:causalitybondi}
&X_{12}^2 <0, \, X_{12}^0 >0 \implies \text{sign}\left(u_{12}+r_1\left|z_1\right|^2-r_2\left|z_2\right|^2 \right) = \text{sign}\left(r_{12}\right) = 1 \ , \\
    &X_{12}^2 <0, \, X_{12}^0 <0 \implies \text{sign}\left(u_{12}+r_1\left|z_1\right|^2-r_2\left|z_2\right|^2 \right) = \text{sign}\left(r_{12}\right) = -1 \ , 
\end{split}
\end{equation}
to arrive at \eqref{eq:topropfinal}. Equation \eqref{eq:causalitybondi} reveals an essential feature of the Feynman propagator in Bondi coordinates. If we take $r_1 \to +\infty$, we necessarily have $X_{12}^0 >0$ and this selects the positive energy solution ($\sim e^{-i \omega u_{12}}$) denoted by $(1)$. In contrast, taking $r_1 \to -\infty$ implies $X_{12}^0 <0$ picking out the negative energy solution ($\sim e^{+i \omega u_{12}}$) denoted by $(2)$. The opposite is true for $r_2$. Starting from \eqref{bulktobulkflat}, we can define four bulk-to-boundary propagators
\begin{equation}
    \begin{split}
 & \mathcal{G}_{Bb,+}^{Flat (1)} \left(x_1;X_2\right)\equiv \lim_{r_1 \to +\infty} r_1 \mathcal{G}_{BB}^{Flat}(X_1,X_2)  = \frac{-i}{2\left(2\pi\right)^2} \frac{1}{-u_{1}-q_1 \cdot X_2 + i \varepsilon } \ , \\    
& \mathcal{G}_{Bb,-}^{Flat (2)}  \left(x_1;X_2\right)\equiv \lim_{r_1 \to -\infty} r_1 \mathcal{G}_{BB}^{Flat}(X_1,X_2)  = \frac{-i}{2\left(2\pi\right)^2} \frac{1}{-u_{1}-q_1 \cdot X_2- i \varepsilon } \ ,\\
    & \mathcal{G}_{Bb,-}^{Flat (1)}  \left(x_2;X_1\right)\equiv \lim_{r_2 \to -\infty} r_2 \mathcal{G}_{BB}^{Flat}(X_1,X_2)  = \frac{-i}{2\left(2\pi\right)^2} \frac{1}{-u_{2}-q_2 \cdot X_1- i \varepsilon } \ ,\\
    & \mathcal{G}_{Bb,+}^{Flat (2)}  \left(x_2;X_1\right)\equiv \lim_{r_2 \to +\infty} r_2 \mathcal{G}_{BB}^{Flat}(X_1,X_2)  = \frac{-i}{2\left(2\pi\right)^2} \frac{1}{-u_{2}-q_2 \cdot X_1 + i \varepsilon } \ ,
\end{split}
\label{bB4soulutions}
\end{equation} where we used the relation $q^\mu (w,\bar w) X_\mu (u,r,z,\bar z) = -u-r|z-w|^2$ and rescaled the regulator to keep $\varepsilon>0$ in each case. The positive energy propagators correspond to the first and third expressions in \eqref{bB4soulutions}, which are the ones we consider in the main text. Hence, we define the bulk-to-boundary propagators in Minkowski space as in \eqref{bulktoboundaryflat}.

\section{\texorpdfstring{Computation of $\bar{D}_{1,1,1,2}$}{Computation of D (1,1,1,2)}}
\label{sec:dbar1112}

Contact terms in holographic CFTs are computed via Witten diagrams, which for scalar operators of external dimensions $\Delta_{i}$ are given in terms of $\bar{D}$-functions. In the Euclidean regime, these admit the following representation\cite{Alday:2016tll}
\begin{align}
\label{eq:dbarnewrep}
\bar{D}_{\Delta_{1} \Delta_{2} \Delta_{3} \Delta_{4}}(U, V)=\int_{-i \infty}^{i \infty} \frac{d j_{1} d j_{2}}{(2 \pi i)^{2}} U^{j_{1}} V^{j_{2}} \Gamma\left(j_{1}+j_{2}+\Delta_{2}\right) \Gamma\left(j_{1}+j_{2}+\Delta-\Delta_{4}\right)& \nonumber\\
\times \Gamma\left(-j_{1}\right) \Gamma\left(-j_{2}\right) \Gamma\left(-j_{1}-\Delta+\Delta_{3}+\Delta_{4}\right) \Gamma\left(-j_{2}+\Delta-\Delta_{2}-\Delta_{3}\right) & \ ,
\end{align}
where $2 \Delta=\sum_{i=1}^{4} \Delta_{i}$ and the integration contour is chosen such as to include all the poles generated by the second line of gamma functions, but none of the ones generated by the first. We are interested in the analytic continuation to the Lorentzian regime, given by \footnote{The expression below corresponds to the usual analytic continuation considered in \cite{Gary:2009ae}. The analytic continuation considered in the body of the paper corresponds instead to the extra factor $e^{2 \pi i\left(j_{1}+j_{2}\right)} \to e^{2 \pi i j_{1}}$.}
\begin{align}
\hat{\Phi}_{\Delta_{1} \Delta_{2} \Delta_{3} \Delta_{4}}(U, V)= & \int_{-i \infty}^{i \infty} \frac{d j_{1} d j_{2}}{(2 \pi i)^{2}} e^{2 \pi i\left(j_{1}+j_{2}\right)} U^{j_{1}} V^{j_{2}} \Gamma\left(j_{1}+j_{2}+\Delta_{2}\right) \Gamma\left(j_{1}+j_{2}+\Delta-\Delta_{4}\right) \\
& \times \Gamma\left(-j_{1}\right) \Gamma\left(-j_{2}\right) \Gamma\left(-j_{1}-\Delta+\Delta_{3}+\Delta_{4}\right) \Gamma\left(-j_{2}+\Delta-\Delta_{2}-\Delta_{3}\right) \ .
\end{align}
As an example let us consider $\Delta_{i}=1$. We need to sum over the poles at $j_{1}=0,1,2, \cdots$ and $j_{2}=0,1,2, \cdots$. Writing $U=Z \bar{Z}, V=(1-Z)(1-\bar{Z})$, it is possible to show that
\begin{equation}
\hat{\Phi}_{1111}(Z, \bar{Z})=\Phi(Z, \bar{Z})+\frac{2 \pi}{Z-\bar{Z}}(2 \pi+i \log (1-Z)-i \log Z-i \log (1-\bar{Z})+i \log \bar{Z}) \ ,
\end{equation}
where $\bar{D}_{1111}=\Phi(Z, \bar{Z})$ is the usual box function
\begin{equation}
\Phi(Z, \bar{Z})=\frac{\left(2 \mathrm{Li}_{2}(Z)-2 \mathrm{Li}_{2}(\bar{Z})\right)+\log \left(\frac{1-Z}{1-\bar{Z}}\right) \log (Z \bar{Z})}{Z-\bar{Z}} \ .
\end{equation}
Note that $\hat{\Phi}_{1111}(Z, \bar{Z})$ differs from the analytic continuation of $\bar{D}_{1111}$ to the Lorentzian regime obtained in \eqref{eq:4ptl22config}. However, as we show below, the leading singularities are in agreement. The Lorentzian contact diagram $\hat{\Phi}_{1111}(Z, \bar{Z})$ has the expected singularity as $\bar{Z} \rightarrow Z$ (where both $Z,\Zb \in (0,1)$)
\begin{equation}
\hat{\Phi}_{1111}^{l s}=\frac{4 \pi^{2}}{Z-\bar{Z}} \ .
\end{equation}
Another interesting example is $\hat{\Phi}_{1112}$. In this case we have poles at $j_{1}=m / 2, j_{2}=n / 2$ with $m, n=0,1, \cdots$. The final result is simply
\begin{equation}
\hat{\Phi}_{1112}(U, V)=\frac{\pi^{3 / 2}}{1-\sqrt{U}-\sqrt{V}} \ ,
\end{equation}
from which we can compute the leading singularity
\begin{equation}
\hat{\Phi}_{1112}^{l s}=\frac{8 \pi^{3 / 2} Z(1-Z)}{(Z-\bar{Z})^{2}} \ .
\end{equation}

\subsection*{References}
\bibliographystyle{style}
\renewcommand\refname{\vskip -1.3cm}%removes the title of the references
\bibliography{Draftv1}

\end{document}